\documentclass[10pt,aps,prb,amsmath,twocolumn,showpacs,hyperref]{revtex4-1}
\usepackage[utf8x]{inputenc}
\usepackage{braket}
\usepackage[normalem]{ulem}
\usepackage{graphicx}
\usepackage{caption}
\usepackage{subcaption}
\usepackage{mathtools}
\usepackage{dsfont}

\newcommand{\sgn}{\mathrm{sgn}}
\newcommand{\e}{\mathrm{e}}
\newcommand{\diff}{\mathrm{d}}
\newcommand{\tr}{\mathrm{tr}}
\newcommand{\ii}{\mathrm{i}}
\newcommand{\sinc}{\mathrm{sinc}}
\newcommand{\varvec}[1]{\mathbf{#1}}

\newcommand{\ddLambda}{\frac{\diff}{\diff\Lambda}}
\newcommand{\muchem}{\mu_{\text{chem}}}

\newcommand{\hTrunc}{Active Space Approximation}
\newcommand{\asa}{{ASA}}
\newcommand{\gowo}{{$\text{G}_0\text{W}_0$}} 
\newcommand{\gw}{{$\text{GW}$}} 
\newcommand{\gwbse}{{$\text{GW}{+}\text{BSE}$}}
\newcommand{\qpgw}{{$\text{qpGW}$}}
\newcommand{\frg}{{FRG}} 
\newcommand{\efrg}{{$\epsilon$FRG}}
\newcommand{\kfrg}{{$k$FRG}}

\newcommand{\ci}{\ii}

\newcommand{\OO}[1]{\ensuremath{\mathcal{O}\big(#1\big)}}

\makeatletter
\newcommand*{\balancecolsandclearpage}{%
  \close@column@grid
  \clearpage
  \twocolumngrid
}
\makeatother

\begin{document}

\title{A functional renormalization group approach to electronic structure calculations for systems without translational symmetry}
\date{May 14, 2016}
\author{Christian \surname{Seiler}}
\author{Ferdinand \surname{Evers}}
\affiliation{Institut f\"ur Theoretische Physik, Universit\"at Regensburg, D-93053 Regensburg}

\begin{abstract}
A formalism for electronic-structure calculations is presented that 
is based on the functional renormalization group (\frg). 
The traditional \frg{} has been formulated for systems that 
exhibit a translational symmetry with an associated Fermi surface, 
which can provide the organization principle for the
renormalization group 
(RG) procedure.
We here advance an alternative formulation,
where the RG-flow is organized in the energy-domain 
rather than in $k$-space. 
This has the advantage that it can also be applied
to inhomogeneous matter lacking a band-structure, 
such as disordered metals or molecules.
The energy-domain \frg{}  (\efrg) presented here 
accounts for Fermi-liquid corrections to 
quasi-particle energies and particle-hole excitations. 
It  goes beyond the state of the art \gw-BSE, because in \efrg{} the Bethe-Salpeter equation 
(BSE)is solved in a self-consistent manner.  
An efficient implementation of the approach 
that has been tested against exact diagonalization calculations and 
calculations based on the density matrix renormalization group is presented.

Similar to the conventional \frg{}, also the \efrg{} is able to 
signalize the vicinity of an instability of the Fermi-liquid fixed point  
via runaway flow of the corresponding interaction vertex. 
Embarking upon this fact, in an application of \efrg{} to the spinless 
disordered Hubbard model we calculate its phase-boundary in 
the plane spanned by the interaction and disorder strength. 
Finally, an extension of the approach to finite temperatures and 
spin $S{=}1/2$ is also given. 

\end{abstract}

\maketitle

\section{Introduction}

Correlation effects are the driving agent behind 
a great many of the phenomena that 
are comprising the contemporary physics of condensed 
matter systems. 
As long as interactions are not too strong, 
such correlation phenomena can be understood in terms 
of an effective single particle picture as it is provided, 
e.g., by the Fermi-liquid theory. 
In this weakly correlated limit, the density-functional theory 
(DFT) can yield useful, often quantitative results 
for the electronic structure of crystalline or molecular matter. 
Where DFT fails to be quantitative, post-DFT correction schemes 
have been introduced that can significantly improve the accuracy, 
in particular with respect to (charged) excitation energies. 
\cite{bookBechstedt15}
As a particularly successful example, we mention 
the \gw -approximation motivated by conventional diagrammatic perturbation theory.
\cite{HedinPaper,onida02,vanSetten15} 

At low enough temperatures most real materials undergo a transition 
into a correlated low-temperature phase, such as a magnet or a superconductor. 
Such phenomena are usually at the verge of applicability of perturbative methods. 
Still, perturbation theory can be very useful, because it often  signalizes 
the existence of such phase-transitions via divergent diagrams. 
In recent years a powerful method has been devised to deal 
with stronger correlations, 
the functional renormalization group (\frg), that has proven particularly 
successful in this respect. 
\cite{FrgReviewMetzner,SalmhoferBook}
It can be (roughly) thought of as a systematic extension of \gw-theory 
and its Bethe-Salpeter-type generalizations.
Because it monitors the RG-flow of a representative set of interaction vertices, 
\frg{} can predict in an unbiased way the leading Fermi liquid instabilities together 
with estimates for the corresponding phase boundaries. 

Beyond phase boundaries, the \frg{} is capable to predict 
a variety of other physical observables including
Luttinger-liquid parameters,\cite{FRGLuttinger1}
Fermi-liquid corrections,\cite{FRGFermiLiquidCorrection1,FRGFermiLiquidCorrection2}
and spin susceptibilities.\cite{FRGSpin,FRGSpin2,ClusterFRG}
Correspondingly, the \frg{} has been applied to a variety of systems, 
e.g., the Hubbard model in various parameter regimes,\cite{FRGHubbard,FRGHubbard2,FRGHubbard3,FRGHubbard4,FRGHubbardKatanin,FRGHubbard5,FRGAttractiveHubbbardSF}
single impurity models,\cite{FRGImpurity,FRGImpurity2}
spin--,\cite{FRGSpin,FRGSpin2,FRGSpin3,iqbal16,FRGSpin4,ClusterFRG}
and quantum critical systems,\cite{FRGAFCriticalPoint,FRGFermiSurfaceReconstruction,FRGDensityWaveMultiCritical}
and superfluids.\cite{FRGSuperfluid,FRGAttractiveHubbbardSF}
For an overview we direct the reader to Refs.~\onlinecite{FrgReviewMetzner,FrgReviewThomale}.

\subsection{Motivation underlying this work}
Good progress has been made in electronic structure calculations for 
real materials as well as for model Hamiltonians. 
Still, we believe that there is room for improvement. 
With an eye on ab-initio calculations, 
we observe that it is still very challenging to accurately calculate, e.g., 
ionization energies and electron affinities of small molecules or atom clusters.
Quantitative results from DFT can be obtained only via procedures, 
such as $\Delta$SCF, that rely on error cancellation. 
The \gowo-method in this respect seems more reliable; 
benchmarks for different implementations have recently become available. 
\cite{bruneval13,koerbel14,vanSetten15}
The \gowo-approximation is not selfconsistent, however,
and partly for this reason it comes in many flavors.
The development and testing of self-consistent and computationally 
affordable \gw-schemes is currently under way. 
\cite{rostgaard10,koerbel14,knight16,vanSetten16}
Even more challenging it is to calculate the dynamical 
response, e.g., the optical gap or the absorption spectrum. 
The traditional time-dependent DFT, such as TDLDA, 
tends to underestimate optical gaps in solids by $\sim eV$. 
Interestingly, it can quantitatively reproduce excitation gaps of small 
molecules when combined with long range 
functionals, especially if they are optimally tuned. 
\cite{kronik12,refaely15} 

In combination with \gw-theory one solves the Bethe-Salpeter equation to
find the optical properties. 
Due to the computational complexity, 
one usually keeps only the simplest non-trivial 
vertex corrections (\gwbse).
The approach yields results often with a typical  
accuracy of a few hundred meV, 
see Ref. \onlinecite{faber16} for a recent overview
and Ref. \onlinecite{bruneval15} for benchmarks. 
In some cases much larger deviations have been reported, 
however, calling for a further 
validation of \gwbse.\cite{hirose15} 
State of the art \gw-implementations can be found in 
many standard band structure codes, e.g., Refs. 
\onlinecite{friedrich10,fiesta,BGW,vasp,molgw}

(i) In this situation it seems advisable to go a step forward and 
explore more complete approximation schemes that in principle could go significantly 
beyond the lowest order 
BSE-technology by incorporating, e.g., a self-consistent evaluation of 
screening in the presence of vertex corrections. 
The extended scheme would thus provide a laboratory for testing  
the current BSE-technology against a more accurate higher order method. 
Our work is underlying the idea that the \frg{} could
be an interesting candidate for such a more advanced electronic 
structure theory. 

A certain limitation of the \frg{} in its current formulation 
is that it is applicable to homogeneous (clean) systems, only. 
It thus could form the basis for 
improved band structure calculations for crystalline matter,  
but it will be inapplicable to the more inhomogeneous systems
that we are mostly interested in, here. 
Specifically, the program lined out before in (i) cannot 
be followed within the present framework of \frg{} 
for molecules or disordered metals. 
From a methodological point of view, we therefore consider it 
an interesting challenge 
modifying the traditional $k$-space \frg{} (\kfrg) into a new tool 
-- energy-domain \frg{} (\efrg) -- 
that can also describe the phases 
and the corresponding transitions in  weakly correlated, inhomogeneous
matter.

(ii) To elaborate on the perspective for the \efrg{}, we mention two research fields with 
prospective applications. 
(1) Quantum chemistry calculations could benefit from \efrg{} 
in a range of system sizes where high-precision calculations, 
e.g. the couple-cluster approach, are computationally not affordable
any more. 
(2) The \efrg{} might prove a useful tool for investigating the 
effect that disorder has on those quantum phase transitions that 
have already been investigated in the clean limit. \cite{FrgReviewMetzner}
Conversely, there is the intriguing prospect to study 
the effect that weak interactions have on disordered 
systems with wavefunctions that are localized due to 
quantum interference.
\cite{evers2008} 

Motivated by (2), we here present an
implementation of an \efrg{} 
that can operate on disordered model Hamiltonians. 
Our goal is to explore the potential of the approach 
as a higher-order method 
for studies of weakly correlated fermions in 
generic environments lacking translational symmetries.

\subsection{FRG for systems without translational symmetries -- \efrg{}}

Consider a fermion system with a Hamiltonian 
that decomposes into a one-body and a two-body part,
\begin{equation}
 \hat H = \hat H_0 + \hat U.
\end{equation}
The non-interacting part, $\hat H_0$, includes a static potential. 
It is considered generic in the sense that 
it does not exhibit translational symmetries; 
its single-particle eigenstates $\ket{\alpha}, \alpha=1,\ldots,N$ are far from plane waves. 
They can be thought of as wavefunctions of a strongly disordered metal 
or as molecular orbitals, e.g., of a generic organic molecule.
We will leave the interacting part, $\hat U$, unspecified for the time being. 

\subsubsection{Excursion: Hedin's equations and \frg{}} 

As was recognized by L. Hedin, in order to compute physical observables 
in the presence of two-body interactions, 
one can solve a  set of self-consistent non-linear 
matrix equations for the exact (causal) Green's function, the corresponding 
self-energies and vertex-functions. 
\cite{bechstedtBook15,vignaleBook}
Unfortunately, 
Hedin's equations are impossible to solve exactly even with todays computational 
resourses for realistic system sizes. Difficulties arise because of 
(a) the complicated nature of the matrix-kernels and 
(b) the very large dimensions of the matrices involved, especially of 
the interaction vertex $\Gamma$. 
The ubiquitous approximation strategy therefore is truncating the matrix-equations
so that the kernels simplify and reducing the effective matrix size by 
grading the many-particle Hilbert space. 
Eventually, also the \frg{} relies on such a truncation scheme. 

However, even the truncated set of equations is very difficult to solve. 
Partially, this is because the requirement of the solution 
being self-consistent. 
Here the idea of the renormalization group (RG) with the 
corresponding flow-equation comes in. 
Speaking in a lose manner, 
what corresponds to an iteration cycle in conventional 
solutions of self-consistency problems is in the framework 
of \frg{} replaced by 
a consecutive integration of a differential equation that 
establishes the RG-flow.
The initializing guess of the iteration cycle  
corresponds to the initialization of the flow equation; 
the flow stops once the (self-consistent) fixed-point has been reached. 
Advantages of the RG-approach over self-consistency cycles are (a) that 
uncertainties related to the proper choice of the starting guess are removed 
and (b) there is a clear physical interpretation in terms of ``runaway flow'' 
even when the numerical integration breaks down, so the RG-flow cannot 
be followed all the way to the fixed-point. In contrast, the lack of 
convergence of a self-consistency cycle is much more difficult to 
interpret consistently.

\subsubsection{Mathematical challenges of \frg{}}

\begin{figure}[tbp]
   \includegraphics[width=.95\linewidth]{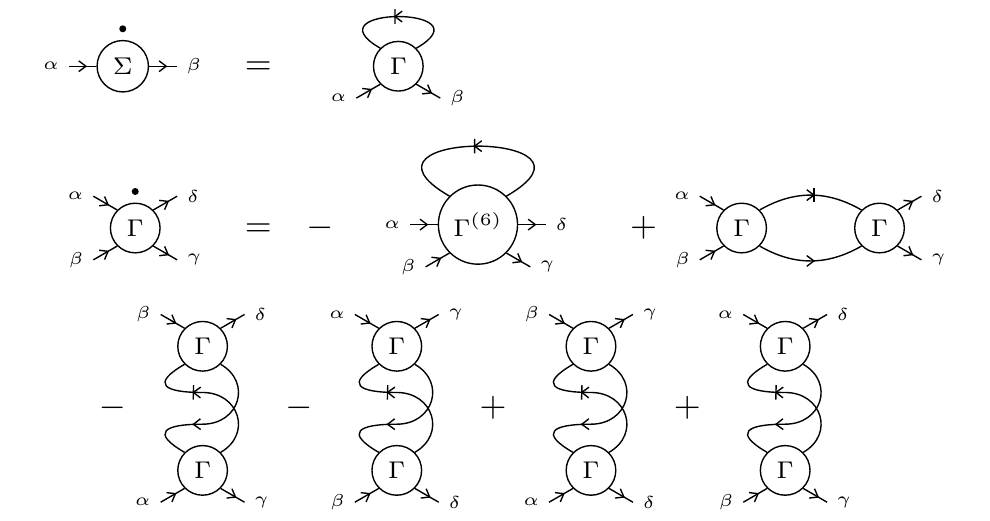}
   \caption[FRG flow equation diagrams]{Diagrammatic representation of the FRG flow equations for the
   self-energy $\Sigma^\Lambda$ and the vertex $\Gamma^\Lambda$. A vertical 
   bar denotes the single-scale propagator $\mathcal{S}^\Lambda$,
   the other propagators are $\mathcal{G}^\Lambda$. 
   As usual, external legs do not entail a propagator.}
   \label{fig:flow_equations}
\end{figure}

For the specific set of flow equations used in this work, 
we adopt the same truncation scheme for the RG-equations, Fig.~\ref{fig:flow_equations},
that also is underlying the traditional \frg{} for periodic systems (\kfrg). 
At this stage the only difference is that with \efrg{} each 
line represents a (Matsubara) Green's function deriving from a
resolvent $G=(\ci \omega -H_0)^{-1}$ that is not diagonal 
in momentum ($k-$) space.  
Fig. \ref{fig:flow_equations} gives a graphical representation of a 
set of nonlinear (integro-)differential equations that 
represent a typical initial-value problem; 
the flowing energy-cutoff $\Lambda$ plays a role analogous to 
a time. Ideally, after integrating the equations from $\Lambda=\infty$ 
to $\Lambda=0$ an exact solution of the (truncated)
vertex-equation has been found. 

As we already mentioned, solving the truncated set of flow equations, 
Fig.~\ref{fig:flow_equations}, still poses a problem of 
formidable computational complexity. The difficulty arises from the fact 
that the vertex function, $\Gamma(\Omega)$, 
is represented as very large family of matrices
with three continuous frequencies, $\Omega=(\omega_1,
\omega_2,\omega_3)$, acting as family parameters. 
In addition, each matrix has four indices, every one of which explores, 
in principle, the basis set of the full single-particle Hilbert space. 

\subsubsection{Established approximation strategies}

Two main simplification strategies can reduce the 
computational effort, making \frg{} feasible and competitive.
We offer a short overview.

\paragraph{Static (or adiabatic) approximation.} 
The frequency-dependence of the vertex function is neglected, 
$\Gamma(\omega_1,\omega_2;\omega_1')\rightarrow\Gamma(0)$.
This is analogous to the static screening approximation familiar from the traditional treatment of the BSE
imposed on top of \gowo. \cite{hybertsen86,strinati88,rohlfing98,rohlfing00}
In \frg{} one also ignores the frequency dependency of the self-energy, $\Sigma$.
As a consequence, $\Sigma_\text{\frg}$ turns into an energy-independent, hermitian correction to the reference 
Hamiltonian $\hat H_0$. The effective Hamiltonian matrix 
$H_\text{eff}=H_0+\Sigma_\text{\frg}$ defines the 
quasi-particle energies and wavefunctions. 
With respect to the static self-energy, the situation in \frg{} is completely analogous 
to the one in the quasi-particle self-consistent \gw-theory (\qpgw). 
\cite{vanSchilfgaarde06, brunval06prb,kotani07,shishkin:235102,vanSetten16}
The advantage of \frg{} over this theory is, that vertex corrections are accounted for 
in \frg{} in a {\em self-consistent} manner.

In the static approximation, the scaling of \frg{} with the dimension of the
single-particle Hilbert space, $N$, is formally $N^6$ if one does not consider
further symmetries such as translational invariance. It is thus roughly comparable
to the scaling of high-precision methods in quantum chemistry, like the coupled
cluster method (flavor CCSD).\cite{bartlett07}

\paragraph{Clean systems: Fermi-surface projection for \kfrg{}.} 
In the clean case,  $H_0$ exhibits a translational symmetry, so the number of independent matrix elements 
of $\Gamma(0)$ reduces significantly. Moreover, a Fermi-surface exists that helps to identify a hierarchical
structure within the matrix elements of $\Gamma(0)$. In many cases, only matrix elements with wavevectors
close to the Fermi surface dominate the physics of the system, so the vertex at momenta away from this surface
may be replaced by the vertex with momenta projected onto it, drastically simplifying the calculation.

\subsubsection{``Active-space'' approximation for \efrg{}.}
In the case of generic systems, there is no intrinsic symmetry guidance as towards how to simplify the matrix
structure of $\Gamma(0)$.  In particular, Fermi-surface projection is not feasible. The most important new
conceptual step in \efrg{} as compared to \kfrg{} will be to find an alternative to the common Fermi-surface projection. 
It should reduce the number of degrees of freedom that are kept explicit in the RG-calculation without invoking a
momentum-space concept. In this work we propose and test 
an ``active-space'' approximation that can achieve this goal.

\paragraph{How to choose the active space.}
Similar to the \frg{}-treatment of clean systems, we also 
work in the eigenstate basis $|\alpha\rangle$ of the non-interacting Hamiltonian $\hat H_0$.
Then the vertex function takes a matrix representation $\Gamma_{\alpha_1,\alpha_2,\alpha_3,\alpha_4}(0)$. 
To simplify the flow equations, we will approximate this matrix by the bare interaction vertex,
$U_{\alpha_1,\alpha_2,\alpha_3,\alpha_4}$, whenever one of the states $|\alpha_i\rangle, i=1,\ldots,4$ 
is outside a certain active space  $\mathcal{H}_M$ of the full single-particle 
Hilbert space, $\mathcal{H}$. A natural
choice of 
$\mathcal{H}_M$ corresponds to states with energy 
$\epsilon_\alpha$ in the vicinity of the chemical potential $\mu_\text{chem}$.
The index $M$ indicates  the size of the volume, which could be characterized by an energy scale or simply by
the number of states  that it contains. We will adopt the simplest choice associating $M$ with the number of
states kept in $\mathcal{H}_M$. 

\paragraph{Computational scaling.}
The important computational aspect of the active-space concept is that it brings the nominal scaling of \efrg{} down to $M^4N^2 + M^2N^3$.
The optimal choice of $M$ balances the computational effort against the required numerical accuracy of 
the calculational results. In our applications we found that typically $M=N/3$ is a reliable choice. It is 
implying a speedup of a factor $10^2$ for the applications that we have investigated. 
For the limit of large $N$, we argue that $M\sim N^{1/2}$ in two-dimensional systems, so that the net scaling
of  \efrg{} would be $N^4$.
It thus formally scales comparable to current implementations of the \gw-method.

\subsection{Application of \efrg{}: Disordered Hubbard model} 
As a first application of the new formalism 
and in order to demonstrate what can be achieved with it, 
we have studied the 2D-spinless, repulsive Hubbard 
model with on-site disorder
at half filling.
At zero disorder, $W{=}0$, the model exhibits a charge-density wave, 
while at zero interaction, $U{=}0$, the ground state corresponds to an 
Anderson insulator. Our interest is in determining the phase 
boundary that separates the two phases in the situation where 
disorder and interaction compete. We have calculated it in 
the $U/W{-}$plane. Specifically, we can establish that 
at $W{>}0$ the Anderson-insulator survives as long as the interaction 
does not exceed a critical value, $U>U^{*}(W)>0$. 

\subsection{Conclusion and Outlook}
As it is typical with higher-order methods, the computational bottleneck 
restricts the feasible system sizes. In our applications, we found it practical 
to work with a single particle Hilbert space containing $N{\sim}50{-}100$ states.
Our preliminary tests indicate that substantially bigger system sizes 
of a few hundred states are realistically accessible, $N{\sim} 200{-}400$,
after additional improvements in the code performance have been
implemented.
It is only the limit of very large values of $N$, though, where the superior scaling of 
\efrg{} will become effective, so that the method becomes favorable as compared 
to other well established techniques, such as CCSD or quantum-Monte-Carlo. 
Whether these system sizes actually can be reached, future research will tell. 

At present, \efrg{} is readily applicable to models of interacting fermions in low dimensions, 
which includes Hubbard models with spin and (attractive) interactions
at different filling fractions, but also, e.g., 
small molecules.

\subsection{Organization of this paper}
The paper is organized in the following way. In section \ref{sec:methodology} we give 
the main formalism including the formul{\ae} needed to reconstruct physical 
observables, in particular densities and occupation numbers. 
Also the formul{\ae} for the finite-temperature 
formalism are given there, so that also, e.g., the effect of heat could be studied. 
Section \ref{sec:impl} provides the computational details of our specific implementation 
of the main formul\ae. In the consecutive section \ref{sec:verification} 
we test this implementation on 1D- and 2D-model systems of 
disordered fermions against numerically exact results 
from exact diagonalization and the density matrix renormalization group (DMRG)
for small system sizes. 

To illustrate the potential of \efrg{}, we present in section \ref{sec:results} 
an application to the disordered, spinless 2D-Hubbard model. 
We will calculate and discuss the phase boundary between 
the Anderson-insulator and the Mott-phase in the plane 
spanned by the disorder and interaction strength.

\section{General Methodology of \efrg{}}
\label{sec:methodology}

In this section we will develop our \efrg{}-scheme. We will assume that it is practical to diagonalize the non-interacting Hamiltonian exactly,
\begin{equation}
 \hat H_0 \ket{\alpha} = \epsilon_{\alpha} \ket{\alpha},
\end{equation}
yielding eigenstates $\{\ket{\alpha}\}$ with corresponding eigenenergies $\{\epsilon_\alpha\}$. This allows us to rewrite the full Hamiltonian
in terms of the non-interacting eigenbasis,
\begin{equation}
 \hat H = \sum_{\alpha} \epsilon_{\alpha} \mathrm{\hat c}_\alpha^\dagger \mathrm{\hat c}_\alpha
   + \frac{1}{4} \sum_{\alpha\beta\gamma\delta} U_{\alpha\beta\gamma\delta}
        \mathrm{\hat c}_\alpha^\dagger \mathrm{\hat c}_\beta^\dagger \mathrm{\hat c}_\delta \mathrm{\hat c}_\gamma.
\end{equation}
Here, $U_{\alpha\beta\gamma\delta}$ are the anti-symmetrized bare interaction matrix elements in the non-interacting eigenbasis.

As discussed in Ref.~\onlinecite{FrgReviewMetzner}, the FRG is a means to solve this interacting problem by introducing a cutoff into the bare propagator
of the system. As the systems we want to study are inhomogeneous in nature, and hence the single-particle states are not easily classified systematically,
we introduce a cutoff in frequency space (as opposed to momentum space), see Eq.~(57) in Ref.~\onlinecite{FrgReviewMetzner},
\begin{equation}
 \mathcal{G}^{0,\Lambda}(\ii\omega) = \frac{\Theta^\Lambda(\omega)}{\ii\omega - H_0 + \muchem},
 \label{eq:definition:G0}
\end{equation}
where $\Theta^\Lambda(\omega)$ vanishes at $\Lambda\to\infty$ and approaches $1$ at $\Lambda\to 0$; see below for a discussion of our
choice for $\Theta^\Lambda(\omega)$.

As a consequence of introducing the infrared cutoff, $\Lambda$, all other quantities of the system
depend on $\Lambda$. If we take the limit of $\Lambda\to\infty$, it can be shown (see Eq~(31) in Ref.~\onlinecite{FrgReviewMetzner})
that the self-energy vanishes and the effective interaction vertex $\Gamma$ is given by the matrix elements of the bare interaction, $U_{\cdot\cdot\cdot\cdot}$.
On the other hand, taking the limit of $\Lambda\to 0$, we recover the original system without the introduced cutoff.
There is now a continuous variable that connects the real system ($\Lambda\to 0$), where the physical quantities are
not known \emph{a priori}, with a trivial system ($\Lambda\to\infty$), where all quantities are known.

\subsection{Flow equations}

As is discussed in the literature\cite{FrgReviewMetzner,SalmhoferBook}, the derivatives of the vertex functions (self-energy, effective interaction, etc.)
yield a set of flow equations; a full derivation of their most generic form may be
found in Chapter.~4 of Ref.~\onlinecite{SalmhoferBook}.
Following Ref.~\onlinecite{FrgReviewMetzner} (Eq.~(50)), we will adopt the generic formulation of the flow equations,
\begin{equation}
 \ddLambda \Sigma^\Lambda(x',x)              = \sum_{y,y'} \mathcal{S}^\Lambda(y,y')\Gamma^\Lambda(x',y';x,y),
 \label{eq:fromref:flow:Sigma}
\end{equation}
and for the vertex, Ref.~\onlinecite{FrgReviewMetzner} (Eq.~(52)),
\begin{eqnarray}
 & & \ddLambda \Gamma^\Lambda(x_1',x_2';x_1,x_2)
  \nonumber \\ & & \hspace{2em}
     = \sum_{y_1,y_1'} \sum_{y_2,y_2'}
       \mathcal{G}^\Lambda(y_1,y_1') \mathcal{S}^\Lambda(y_2,y_2')
  \nonumber \\ & & \hspace{2em}
       \times\Big\{\Gamma^\Lambda(x_1',x_2';y_1,y_2) \Gamma^\Lambda(y_1',y_2';x_1,x_2)
  \nonumber \\ & & \hspace{2em}
       -\big[ \Gamma^\Lambda(x_1',y_2';x_1,y_1) \Gamma^\Lambda(y_1',x_2';y_2,x_2)
  \nonumber \\ & & \hspace{2em}
       \hphantom{\times}+(y_1 \leftrightarrow y_2, y_1' \leftrightarrow y_2')\big]
   \nonumber \\ & & \hspace{2em}
       +\big[ \Gamma^\Lambda(x_2',y_2';x_1,y_1)
              \Gamma^\Lambda(y_1',x_1';y_2,x_2)
  \nonumber \\ & & \hspace{2em}
       \hphantom{\times}+(y_1 \leftrightarrow y_2, y_1' \leftrightarrow y_2')\big]\Big\}
  \nonumber \\ & & \hspace{2em}
       -\sum_{y,y'} \mathcal{S}^\Lambda(y,y') \Gamma^{(6),\Lambda}(x_1',x_2',y';x_1,x_2,y).
   \label{eq:fromref:flow:Gamma}
\end{eqnarray}
Here, $x$ and $y$ are combined indices for space and time coordinates.
A diagrammatic representation of these equations is given in in Fig.~\ref{fig:flow_equations}.
Furthermore, we copy the definition of Ref.~\onlinecite{FrgReviewMetzner} (Eq.~(47)) for the
single-scale propagator,
\begin{equation}
 \mathcal{S}^\Lambda = - \mathcal{G}^\Lambda
                         \left[ \ddLambda \left( \mathcal{G}^{0,\Lambda} \right)^{-1} \right]
                         \mathcal{G}^\Lambda.
  \label{eq:definition:SingleScalePropagator}
\end{equation}
We next rewrite these quantities into our own nomenclature, where we  work in Matsubara space. Furthermore, we
separate the generic indices into Matsubara frequencies and Hilbert space indices, $x \rightarrow (\mu,\omega_n)$.
We also drop the term with $\Gamma^{(6),\Lambda}$ in accordance with the standard truncation scheme for
these equations,\cite{FrgReviewMetzner} where in the case of short-range interactions, power counting arguments
establish the scheme's validity.

Since energy is conserved, the self-energy, the single-particle Green's functions,
the single-scale propagator and the vertex include the corresponding $\delta$-function,
\begin{eqnarray}
 \Sigma^\Lambda_{\alpha\beta}(\omega_n;\omega_{n'}) & \to & T^{-1} \delta_{n,n'} \Sigma^\Lambda_{\alpha\beta}(\omega_n), \\
 \mathcal{G}^{0,\Lambda}_{\alpha\beta}(\omega_n;\omega_{n'}) & \to & T^{-1} \delta_{n,n'} \mathcal{G}^{0,\Lambda}_{\alpha\beta}(\omega_n), \\
 \mathcal{G}^\Lambda_{\alpha\beta}(\omega_n;\omega_{n'}) & \to & T^{-1} \delta_{n,n'} \mathcal{G}^\Lambda_{\alpha\beta}(\omega_n), \\
 \mathcal{S}^\Lambda_{\alpha\beta}(\omega_n;\omega_{n'}) & \to & T^{-1} \delta_{n,n'} \mathcal{S}^\Lambda_{\alpha\beta}(\omega_n), \\
 \Gamma^\Lambda_{\alpha\beta\gamma\delta}(\omega_n,\omega_{\tilde n};\omega_{n'},\omega_{\tilde n'}) & \to &
    T^{-1} \delta_{n+\tilde n,n'+\tilde n'}\times \nonumber \\
    & & \Gamma^\Lambda_{\alpha\beta\gamma\delta}(\omega_n,\omega_{\tilde n};\omega_{n'}).\hspace{1em}
\end{eqnarray}
Inserting this into Eq.~(\ref{eq:fromref:flow:Sigma}) yields
\begin{eqnarray}
\hspace{-1em} \ddLambda T^{-1} \delta_{n,n'} \Sigma^\Lambda_{\alpha\beta}(\omega_n) & = & T^2 \sum_{\omega_m\omega_{m'}} \sum_{\mu\nu}
  \mathcal{S}^\Lambda_{\mu\nu}(\omega_m)  \times \nonumber\\
 & &
    \Gamma^\Lambda_{\alpha\nu\beta\mu}(\omega_n,\omega_{m'};\omega_{n'}) \times \nonumber\\
 & &
     T^{-1} \delta_{m,m'} T^{-1}\delta_{n+m',n'+m},
\end{eqnarray}
and after evaluating the sum over the Matsubara frequency $\omega_{m'}$, one arrives at
\begin{eqnarray}
 \hspace{-1em}\ddLambda \Sigma^\Lambda_{\alpha\beta}(\omega_n) & = & T \sum_{\omega_m} \sum_{\mu\nu}
   \mathcal{S}^\Lambda_{\mu\nu}(\omega_m) \times \nonumber \\
   & & \Gamma^\Lambda_{\alpha\nu\beta\mu}(\omega_n,\omega_m;\omega_n). \label{eq:flow:SigmaWithOmegaAtFiniteT}
\end{eqnarray}
Here, we have used that a $\delta_{n,n'}$ appears on both sides and have multiplied the equation by $T$.

Proceeding in a similar way for the equation of the flow of the vertex, Eq.~(\ref{eq:fromref:flow:Gamma}), we arrive at
\begin{widetext}
\begin{eqnarray}
 \ddLambda \Gamma^\Lambda_{\alpha\beta\gamma\delta}(\omega_n,\omega_{\tilde n};\omega_{n'}) & = &
     T \sum_{\omega_m\omega_{\tilde m}}
       \sum_{\mu\nu\rho\sigma}
       \mathcal{G}^\Lambda_{\rho\mu}(\omega_m)
       \mathcal{S}^\Lambda_{\sigma\nu}(\omega_{\tilde m})
       \times \Big\{ 
 \nonumber \\ & & \hspace{-3em}
     \hphantom{+ \big[}  \Gamma^\Lambda_{\alpha\beta\rho\sigma}(\omega_n,\omega_{\tilde n};\omega_{m})
              \Gamma^\Lambda_{\mu\nu\gamma\delta}(\omega_m,\omega_{\tilde m};\omega_{n'})
              \delta^{(\text{c})}_{\tilde m}
 \nonumber \\ & & \hspace{-3em} 
          + \big[ \Gamma^\Lambda_{\beta\nu\gamma\rho}(\omega_{\tilde n},\omega_{\tilde m};\omega_{n'})
                  \Gamma^\Lambda_{\mu\alpha\sigma\delta}(\omega_{m},\omega_{n};\omega_{\tilde m})
                  \delta^{(\text{ph},1)}_{\tilde m}
 \nonumber \\ & & \hspace{-3em}
          \hphantom{+ \big[} + \Gamma^\Lambda_{\beta\mu\gamma\sigma}(\omega_{\tilde n},\omega_{m};\omega_{n'})
                  \Gamma^\Lambda_{\nu\alpha\rho\delta}(\omega_{\tilde m},\omega_{n};\omega_{m})
                  \delta^{(\text{ph},2)}_{\tilde m}
            \big]
 \nonumber \\ & & \hspace{-3em}
          - \big[ \Gamma^\Lambda_{\alpha\nu\gamma\rho}(\omega_{n},\omega_{\tilde m};\omega_{n'})
                  \Gamma^\Lambda_{\mu\beta\sigma\delta}(\omega_{m},\omega_{\tilde n};\omega_{\tilde m})
                  \delta^{(\text{ph},3)}_{\tilde m}
 \nonumber \\ & & \hspace{-3em}
            \hphantom{+ \big[} + \Gamma^\Lambda_{\alpha\mu\gamma\sigma}(\omega_{n},\omega_{m};\omega_{n'})
                  \Gamma^\Lambda_{\nu\beta\rho\delta}(\omega_{\tilde m},\omega_{\tilde n};\omega_{m})
                  \delta^{(\text{ph},4)}_{\tilde m}
            \big]
        \Big\},
         \label{eq:flow:GammaWithOmegaAtFiniteT}
\end{eqnarray}
\end{widetext}
where $\delta^{(\text{c})}_{\tilde m}$ and $\delta^{(\text{ph},\cdot)}_{\tilde m}$ reflect the energy conservation of the vertex, e.g.
$\delta^{\text{c}}_{\tilde m} = \delta_{n+\tilde n,m+\tilde m}.$

\subsection{Formalism at Zero Temperature}

For the most part, we will discuss the Formalism at $T = 0$. In that case, sums over Matsubara frequencies are replaced by integrals,
\begin{equation}
 T\sum_{\omega_n} \to (2\pi)^{-1} \int\diff\omega,
 \label{eq:Matsubara:T0-transition}
\end{equation}
and the Kronecker symbols will be replaced by $\delta$-functions,
\begin{equation}
 T^{-1} \delta_{n,n'} \to 2\pi\delta(\omega - \omega').
 \label{eq:Matsubara:T0-transition:Kronecker}
\end{equation}
Eqs.~(\ref{eq:flow:SigmaWithOmegaAtFiniteT},\ref{eq:flow:GammaWithOmegaAtFiniteT}) now read
\begin{widetext}
\begin{eqnarray}
 \ddLambda \Sigma^\Lambda_{\alpha\beta}(\omega) & = & \frac{1}{2\pi} \int\diff\bar\omega \sum_{\mu\nu}
   \mathcal{S}^\Lambda_{\mu\nu}(\bar\omega)
   \Gamma^\Lambda_{\alpha\nu\beta\mu}(\omega,\bar\omega;\omega), \label{eq:flow:SigmaWithOmegaAtT0} \\
 \ddLambda \Gamma^\Lambda_{\alpha\beta\gamma\delta}(\omega,\tilde\omega;\omega') & = &
     \frac{1}{2\pi} \int\diff\bar\omega\diff\bar\omega'
       \sum_{\mu\nu\rho\sigma}
       \mathcal{G}^\Lambda_{\rho\mu}(\bar\omega)
       \mathcal{S}^\Lambda_{\sigma\nu}(\bar\omega')
       \times \Big\{
   \Gamma^\Lambda_{\alpha\beta\rho\sigma}(\omega,\tilde\omega;\bar\omega)
              \Gamma^\Lambda_{\mu\nu\gamma\delta}(\bar\omega,\bar\omega';\omega')
              \delta^{(\text{c})}(\bar\omega')
 \nonumber \\ & & \hspace{-9em}
          + \big[ \Gamma^\Lambda_{\beta\nu\gamma\rho}(\tilde\omega,\bar\omega';\omega')
                  \Gamma^\Lambda_{\mu\alpha\sigma\delta}(\bar\omega,\omega;\bar\omega')
                  \delta^{(\text{ph},1)}(\bar\omega')
          + \Gamma^\Lambda_{\beta\mu\gamma\sigma}(\tilde\omega,\bar\omega;\omega')
                  \Gamma^\Lambda_{\nu\alpha\rho\delta}(\bar\omega',\omega;\bar\omega)
                  \delta^{(\text{ph},2)}(\bar\omega')
            \big]
 \nonumber \\ & & \hspace{-9em}
          - \big[ \Gamma^\Lambda_{\alpha\nu\gamma\rho}(\omega,\bar\omega';\omega')
                  \Gamma^\Lambda_{\mu\beta\sigma\delta}(\bar\omega,\tilde\omega;\bar\omega')
                  \delta^{(\text{ph},3)}(\bar\omega')
             + \Gamma^\Lambda_{\alpha\mu\gamma\sigma}(\omega,\bar\omega;\omega')
                  \Gamma^\Lambda_{\nu\beta\rho\delta}(\bar\omega',\tilde\omega;\bar\omega)
                  \delta^{(\text{ph},4)}(\bar\omega')
            \big]
        \Big\}, \label{eq:flow:GammaWithOmegaAtT0}
\end{eqnarray}
\end{widetext}
where again, $\delta^{(\text{c})}(\bar\omega')$ and $\delta^{(\text{ph},\cdot)}(\bar\omega')$ reflect the energy conservation of the
vertex, e.g., $\delta^{(\text{ph},1)}(\bar\omega') = \delta(\tilde\omega+\bar\omega'-\omega'-\bar\omega)$.

We now proceed to take the static limit, i.e. by replacing the frequency dependence of the vertex and the self-energy by their static
limit, e.g., $\Gamma(\omega,\omega';\bar\omega)\to\Gamma(0)$. For short-range interactions,
power counting of the flow equations demonstrates that the dominant contribution for small $\Lambda$ comes from zero frequencies and
states close to the Fermi energy. This approximation has been discussed extensively in Ref.~\onlinecite{FrgReviewMetzner}.

We arrive at
\begin{eqnarray}
 \ddLambda \Sigma^\Lambda_{\alpha\beta} & = & \frac{1}{2\pi} \int\diff\bar\omega \sum_{\mu\nu}
   \mathcal{S}^\Lambda_{\mu\nu}(\bar\omega)
   \Gamma^\Lambda_{\alpha\nu\beta\mu}, \label{eq:flow:SigmaWithOmegaStatic}  \\
 \ddLambda \Gamma^\Lambda_{\alpha\beta} & = & \frac{1}{2\pi} \int\diff\bar\omega
       \sum_{\mu\nu\rho\sigma} \Big\{
       \mathcal{G}^\Lambda_{\rho\mu}(\bar\omega)
       \mathcal{S}^\Lambda_{\sigma\nu}(-\bar\omega)
       \Gamma^\Lambda_{\alpha\beta\rho\sigma}
       \Gamma^\Lambda_{\mu\nu\gamma\delta}
    \nonumber \\ & & \hspace{-3em}
         + \mathcal{G}^\Lambda_{\rho\mu}(\bar\omega)
           \mathcal{S}^\Lambda_{\sigma\nu}(\bar\omega)
           \big[ \Gamma^\Lambda_{\beta\nu\gamma\rho}
                  \Gamma^\Lambda_{\mu\alpha\sigma\delta}
                + \Gamma^\Lambda_{\beta\mu\gamma\sigma}
                  \Gamma^\Lambda_{\nu\alpha\rho\delta}
           \big]
    \nonumber \\ & & \hspace{-3em}
         - \mathcal{G}^\Lambda_{\rho\mu}(\bar\omega)
           \mathcal{S}^\Lambda_{\sigma\nu}(\bar\omega)
           \big[ \Gamma^\Lambda_{\alpha\nu\gamma\rho}
                  \Gamma^\Lambda_{\mu\beta\sigma\delta}
                + \Gamma^\Lambda_{\alpha\mu\gamma\sigma}
                  \Gamma^\Lambda_{\nu\beta\rho\delta}
            \big]
        \Big\}. \label{eq:flow:GammaWithOmegaStatic}
\end{eqnarray}
Note that the vertex $\Gamma^\Lambda$ is antisymmetric under exchange of the first or the last pair of indices,
\begin{equation}
        \Gamma^{\Lambda}_{\alpha\beta\gamma\delta}
    = - \Gamma^{\Lambda}_{\beta\alpha\gamma\delta}
    = - \Gamma^{\Lambda}_{\alpha\beta\delta\gamma}
    =   \Gamma^{\Lambda}_{\beta\alpha\delta\gamma}.
 \label{eq:symmetry:Gamma} 
\end{equation}
Furthermore, one can easily show that in the static limit for finite system sizes the self-energy $\Sigma$ is hermitian.
To further simplify these equations, we choose our cutoff $\Theta^\Lambda(\omega)$ to be a simple step function,
\begin{equation}
 \Theta^\Lambda(\omega) = \Theta(|\omega| - \Lambda),
\end{equation}
such that its derivative is
\begin{equation}
 \ddLambda\Theta^\Lambda(\omega) = - \delta(|\omega| - \Lambda).
\end{equation}
Since by construction the self-energy is not frequency dependent, the frequency integrals may now be solved analytically. For
Eq.~(\ref{eq:flow:SigmaWithOmegaStatic}), we have to integrate
\[ \int\diff\bar\omega \mathcal{S}^\Lambda_{\mu\nu}(\bar\omega). \]
Inserting Dyson's equation into Eq.~(\ref{eq:definition:SingleScalePropagator}), we have
\begin{eqnarray}
 \mathcal{S} & = & - \mathcal{G}\left(\ddLambda\left[\mathcal{G}^{0}\right]^{-1}\right)\mathcal{G}
               =   - \mathcal{G}\left(\ddLambda\left[\mathcal{G}^{-1} + \Sigma\right]\right)\mathcal{G} \nonumber \\
             & = & \mathcal{\dot G} - \mathcal{G}\dot\Sigma\mathcal{G},
\end{eqnarray}
in matrix notation. We note that $\mathcal{G} = (\mathcal{Q} - \Theta\Sigma)^{-1}\Theta$, where we use the shorthand
$\Theta = \Theta(|\omega|-\Lambda)$ and $\mathcal{Q} = \ii\omega - H_0 + \muchem$.
Using $\ddLambda A^{-1}(\Lambda) = - A^{-1}(\Lambda) \dot A(\Lambda) A^{-1}(\Lambda)$,
simple algebra yields
\begin{equation}
 \mathcal{S} = - \delta\left(\mathds{1} + \frac{\Theta}{\mathcal{Q} - \Theta\Sigma} \Sigma\right) \frac{1}{\mathcal{Q} - \Theta\Sigma}.
 \label{eq:derivation:SingleScalePropagator:partialform}
\end{equation}
Since the $\delta$ and $\Theta$ functions are to be taken at the same argument, we employ Morris's Lemma\footnote{
$\delta(x)f(\Theta(x)) \to \delta(x)\int_0^1 f(t) \diff t$, see Ref.~\onlinecite{MorrisLemma}.}
to resolve this,
\begin{equation}
 \mathcal{S} = - \delta \int_0^1 \diff t \left(\mathds{1} + t \frac{1}{\mathcal{Q} - t\Sigma} \Sigma\right) \frac{1}{\mathcal{Q} - t\Sigma}.
 \label{eq:derivation:SingleScalePropagator:integral}
\end{equation}
Using the fact that
\[ \frac{\diff}{\diff t} \frac{1}{\mathcal{Q} - t\Sigma} = \frac{1}{\mathcal{Q} - t\Sigma} \Sigma \frac{1}{\mathcal{Q} - t\Sigma} \]
and partial integration, the second summand of the integral yields
\[
  - \left[ \frac{t}{\mathcal{Q} - t\Sigma} \right]_0^1 + \int_0^1 \diff t \frac{1}{\mathcal{Q} - t\Sigma},
\]
where it can be seen that the remaining integral cancels the first summand of the integral in
Eq.~(\ref{eq:derivation:SingleScalePropagator:integral}), so we arrive at
\begin{equation}
 \mathcal{S}^\Lambda(\omega) =
   - \frac{\delta(|\omega| - \Lambda)}{\ii\omega - H_0 + \muchem - \Sigma^\Lambda}.
\end{equation}
The frequency integral is now trivial, yielding
\begin{equation}
 \int\diff\bar\omega \mathcal{S}^\Lambda(\bar\omega) =
    - \sum_{\bar\omega=\pm\Lambda} \frac{1}{\ii\bar\omega - H_0 + \muchem - \Sigma^\Lambda}.
 \label{eq:result:SingleScalePropagatorAtT0}
\end{equation}
As the following quantity will appear also in the flow equation for the vertex, we will define
\begin{equation}
 P^\Lambda_{\mu\nu}(\bar\omega) := \left.\frac{1}{\ii\bar\omega - H_0 + \muchem - \Sigma^\Lambda}\right|_{\mu\nu}.
   \label{eq:definition:GenericPropagatorAtT0}
\end{equation}
Inserting Eqs.~(\ref{eq:result:SingleScalePropagatorAtT0},\ref{eq:definition:GenericPropagatorAtT0}) into
Eq.~(\ref{eq:flow:SigmaWithOmegaStatic}), the flow equation for the self-energy now reads
\begin{equation}
 \ddLambda\Sigma^\Lambda_{\alpha\beta} = - \frac{1}{2\pi} \sum_{\mu\nu}
     \underbrace{\left( P^{\Lambda}_{\mu\nu}(\Lambda) + P^{\Lambda}_{\mu\nu}(-\Lambda)\right)}_{=:\Pi^{\Sigma,\Lambda}_{\mu\nu}}
     \Gamma^\Lambda_{\alpha\nu\beta\mu}.
     \label{eq:flow:Sigma}
\end{equation}
When evaluating the flow equation for the vertex, Eq.~(\ref{eq:flow:GammaWithOmegaStatic}), one must take
care that the arguments for the $\delta$ and $\Theta$ functions coincide, so one may not simply take
the result derived for the single-scale propagator in the self-energy flow and apply it, but one rather
uses the same kind of treatment of the $\delta$ and $\Theta$ functions for the entire expression, on a
term by term basis. In the end, the flow equation for the vertex in the static limit reads,
\begin{eqnarray}
 \ddLambda \Gamma^\Lambda_{\alpha\beta\gamma\delta} & = & 
    - \frac{1}{2\pi} \sum_{\mu\nu\rho\sigma} \sum_{\bar\omega=\pm\Lambda} \Big\{
    \frac{1}{2} P^{\Lambda}_{\rho\mu}(-\bar\omega) P^{\Lambda}_{\sigma\nu}(\bar\omega)
   \Gamma^\Lambda_{\alpha\beta\rho\sigma} \Gamma^\Lambda_{\mu\nu\gamma\delta}
   \nonumber \\ &  &
   \hspace{-3em} + P^{\Lambda}_{\rho\mu}(\bar\omega) P^{\Lambda}_{\sigma\nu}(\bar\omega)
   \left[
     \Gamma^\Lambda_{\beta\nu\gamma\rho} \Gamma^\Lambda_{\alpha\mu\delta\sigma}
   - \Gamma^\Lambda_{\alpha\nu\gamma\rho} \Gamma^\Lambda_{\beta\mu\delta\sigma}
   \right] \Big\}
  \nonumber \\ & = & 
   - \frac{1}{2\pi} \sum_{\mu\nu\rho\sigma} \Big\{
           \Pi^{\text{c},\Lambda}_{\mu\nu\sigma\rho}
           \Gamma^\Lambda_{\nu\rho\gamma\delta} \Gamma^\Lambda_{\alpha\beta\sigma\mu}  
       \nonumber \\ & &
         + \Pi^{\text{ph},\Lambda}_{\mu\nu\rho\sigma}
          \left[ \Gamma^\Lambda_{\beta\nu\gamma\rho} \Gamma^\Lambda_{\alpha\sigma\delta\mu}
               - \Gamma^\Lambda_{\alpha\nu\gamma\rho} \Gamma^\Lambda_{\beta\sigma\delta\mu}
          \right], 
    \label{eq:flow:Gamma}
\end{eqnarray}
where we have used the symmetries of $\Gamma$ to simplify the equations and abberviated
\begin{eqnarray}
 \Pi^{\text{c},\Lambda}_{\mu\nu\sigma\rho} & := & P^{\Lambda}_{\mu\nu}(\Lambda) P^{\Lambda}_{\sigma\rho}(-\Lambda) \\
 \Pi^{\text{ph},\Lambda}_{\mu\nu\sigma\rho} & := & P^{\Lambda}_{\mu\nu}(\Lambda) P^{\Lambda}_{\rho\sigma}(\Lambda)
                                                 + P^{\Lambda}_{\mu\nu}(-\Lambda) P^{\Lambda}_{\rho\sigma}(-\Lambda).
\end{eqnarray}
The full derivation may be found in Appendix~\ref{app:T0StaticGamma}.

\subsubsection{Initial conditions}

The initial conditions at $\Lambda\to\infty$ are given by
\begin{equation}
 \Sigma^{\Lambda\to\infty}_{\alpha\beta}             = 0
 \hspace{1em}\text{and}\hspace{1em}
 \Gamma^{\Lambda\to\infty}_{\alpha\beta\gamma\delta} = U_{\alpha\beta\gamma\delta}.
 \label{eq:initcond:SigmaGamma:Infty}
\end{equation}
In order to solve the equations numerically, we need to choose an initial value $\Lambda_0$ that is still finite but larger than all other
energy scales in the system. For $\Lambda > \Lambda_0$ one may assume a form of $(\ii\omega)^{-1}\mathds{1}$ for the propagator,
allowing us to analytically integrate the flow equations from $\infty$ to $\Lambda_0$. In case of the flow equation for the
vertex, power counting in $U$ and $\Lambda_0$ immediately yields
\begin{equation}
 \Gamma^{\Lambda_0} - U \sim - \int_\infty^{\Lambda_0}
       U U \frac{1}{\Lambda^2} \diff\Lambda
       = \frac{1}{\Lambda_0} U U,
\end{equation}
and hence
\begin{equation}
 |\Gamma^{\Lambda_0}-U|/|U| \sim |U|/\Lambda_0.
\end{equation}
We therefore may simply use that $\Gamma^{\Lambda_0}$ does not differ from
$\Gamma^{\Lambda\to\infty}$ for large enough $\Lambda_0$ and arrive at
\begin{equation}
  \Gamma^{\Lambda}_{\alpha\beta\gamma\delta} = U_{\alpha\beta\gamma\delta}, \qquad \Lambda > \Lambda_0.
  \label{eq:initcond:Gamma}
\end{equation}
The same does not hold true for the flow equation for the self-energy, where the analytical
integral gives a non-negligible contribution,
\begin{eqnarray}
 \Sigma^{\Lambda_0}_{\alpha\beta} & = & - \frac{1}{2\pi} \sum_{\mu} U_{\alpha\mu\beta\mu}
    \lim_{\eta\to 0^{+}} \int_{\infty}^{\Lambda_0}
    \left( \frac{\e^{\ii\Lambda\eta}}{\ii\Lambda} - \frac{\e^{-\ii\Lambda\eta}}{\ii\Lambda} \right) \diff\Lambda \nonumber \\
   & = & - \frac{1}{\pi} \sum_{\mu} U_{\alpha\mu\beta\mu} \lim_{\eta\to 0^{+}} \eta
   \int_{\infty}^{\Lambda_0} \sinc(\eta\Lambda)\diff\Lambda \nonumber \\
   & = & \frac{1}{\pi} \sum_{\mu} U_{\alpha\mu\beta\mu} \lim_{\eta\to 0^{+}} \left[
     \int_0^\infty \sinc(x) \diff x - \mathcal{O}(\eta)
   \right] \nonumber \\
   & = & \frac{1}{2} \sum_{\mu} U_{\alpha\mu\beta\mu}.
   \label{eq:initcond:Sigma}
\end{eqnarray}
Here we have explicitly included the required convergence factor $\e^{\ii\omega 0^{+}}$ that appears in the
Green's function in imaginary frequency space.

\subsection{Systems with Spin}

In Eqs.~(\ref{eq:flow:Sigma},\ref{eq:flow:Gamma}), the indices represent generic states in the Hilbert space.
We will now discuss the case where the system is fully $\mathrm{SU}(2)$ symmetric. Here, it is convenient to
separate the orbital degrees of freedom from the spin degrees of freedom, $\alpha\to(\alpha,\sigma_1)$. Our
derivation will follow Ref.~\onlinecite{SalmhoferSpin}, but we will discuss the generic case without the
additional particle-hole symmetry. Single-particle quantities (self-energy, propagators) do not depend on
the spin degree of freedom,
\begin{eqnarray}
  \Sigma^\Lambda_{(\alpha,\sigma_1)(\beta,\sigma_2)}      & = & \Sigma^{\text{s},\Lambda}_{\alpha\beta} \delta_{\sigma_1\sigma_2}, \label{eq:definition:Sigma:with-spin} \\
  \mathcal{G}^\Lambda_{(\alpha,\sigma_1)(\beta,\sigma_2)} & = & \mathcal{G}^{\text{s},\Lambda}_{\alpha\beta} \delta_{\sigma_1\sigma_2},  \label{eq:definition:G:with-spin} \\
  \mathcal{S}^\Lambda_{(\alpha,\sigma_1)(\beta,\sigma_2)} & = & \mathcal{S}^{\text{s},\Lambda}_{\alpha\beta} \delta_{\sigma_1\sigma_2},  \label{eq:definition:S:with-spin} \\
  P^\Lambda_{(\alpha,\sigma_1)(\beta,\sigma_2)}           & = & P^{\text{s},\Lambda}_{\alpha\beta} \delta_{\sigma_1\sigma_2}. \label{eq:definition:P:with-spin}
\end{eqnarray}
The spin structure of the vertex is determined by the fact that two particles may either keep their spin or exchange it,
and may thus be decomposed into
\begin{eqnarray*}
  \Gamma^\Lambda_{(\alpha,\sigma_1),(\beta,\sigma_2),(\gamma,\sigma_3),(\delta,\sigma_4)}
  & = & \hphantom{+} c^{\text{I},\Lambda}_{\alpha\beta\gamma\delta} \delta_{\sigma_1\sigma_3} \delta_{\sigma_2\sigma_4} \nonumber\\
  &   &           +  c^{\text{II},\Lambda}_{\alpha\beta\gamma\delta} \delta_{\sigma_1\sigma_4} \delta_{\sigma_2\sigma_3},
\end{eqnarray*}
where $c^{\text{I}}$ and $c^{\text{II}}$ are the coefficients for each of these processes.

Using the antisymmetry of $\Gamma$, Eq.~(\ref{eq:symmetry:Gamma}), we may exchange $(\gamma,\sigma_3)$ with $(\delta,\sigma_4)$,
\begin{eqnarray*}
 & & \Gamma^\Lambda_{(\alpha,\sigma_1),(\beta,\sigma_2),(\gamma,\sigma_3),(\delta,\sigma_4)}
   = - \Gamma^\Lambda_{(\alpha,\sigma_1),(\beta,\sigma_2),(\delta,\sigma_4),(\gamma,\sigma_3)} \nonumber \\
  & & \hspace{4em} = - c^{\text{I},\Lambda}_{\alpha\beta\delta\gamma} \delta_{\sigma_1\sigma_4} \delta_{\sigma_2\sigma_3}
        - c^{\text{II},\Lambda}_{\alpha\beta\delta\gamma} \delta_{\sigma_1\sigma_3} \delta_{\sigma_2\sigma_4}.
\end{eqnarray*}
By comparing the coefficients of the Kronecker-$\delta$s, we may identify
\[ c^{\text{I},\Lambda}_{\alpha\beta\gamma\delta} = - c^{\text{II},\Lambda}_{\alpha\beta\delta\gamma}
   := - \Gamma^{\text{s},\Lambda}_{\alpha\beta\delta\gamma}, \]
and hence write the vertex as
\begin{eqnarray}
  \Gamma^\Lambda_{(\alpha,\sigma_1),(\beta,\sigma_2),(\gamma,\sigma_3),(\delta,\sigma_4)}
  & = & \hphantom{-} \Gamma^{\text{s},\Lambda}_{\alpha\beta\gamma\delta} \delta_{\sigma_1\sigma_4} \delta_{\sigma_2\sigma_3} \nonumber\\
  &   &           -  \Gamma^{\text{s},\Lambda}_{\alpha\beta\delta\gamma} \delta_{\sigma_1\sigma_3} \delta_{\sigma_2\sigma_4}.
  \label{eq:definition:Gamma:with-spin}
\end{eqnarray}
Using the symmetry of $\Gamma^\Lambda$, one can see that $\Gamma^{\text{s},\Lambda}$ is still symmetric under exchange of both
pairs of indices,
\begin{equation}
 \Gamma^{\text{s},\Lambda}_{\alpha\beta\gamma\delta} = \Gamma^{\text{s},\Lambda}_{\beta\alpha\delta\gamma},
\end{equation}
but in general it is not antisymmetric with respect to the exchange of a single pair of indices. Instead, one may identify the
part of $\Gamma^{\text{s},\Lambda}$ that is antisymmetric under exchange of $\alpha$ and $\beta$ with the triplet channel of the
vertex, whereas the part that is symmetric under the exchange of $\alpha$ and $\beta$ represents the singlet channel.

Inserting Eqs.~(\ref{eq:definition:Sigma:with-spin},\ref{eq:definition:P:with-spin},\ref{eq:definition:Gamma:with-spin})
into Eq.~(\ref{eq:flow:Sigma}), we have
\begin{eqnarray}
 \ddLambda\Sigma^{\text{s},\Lambda}_{\alpha\beta} \delta_{\sigma_1\sigma_2} & = & - \frac{1}{2\pi} \sum_{\mu\nu} \sum_{\sigma_3}
     \Pi^{\Sigma,\text{s},\Lambda}_{\mu\nu} \big(
     \Gamma^{\text{s},\Lambda}_{\alpha\nu\beta\mu} \delta_{\sigma_1\sigma_3} \delta_{\sigma_3\sigma_2}
  \nonumber \\ & & \hspace{6em}
     - \Gamma^{\text{s},\Lambda}_{\alpha\nu\mu\beta} \delta_{\sigma_1\sigma_2} \delta_{\sigma_3\sigma_3}
     \Big) \nonumber \\
    & = & - \frac{1}{2\pi} \sum_{\mu\nu} \Pi^{\Sigma,\text{s},\Lambda}_{\mu\nu} \big( \Gamma^{\text{s},\Lambda}_{\alpha\nu\beta\mu}
          - 2 \Gamma^{\text{s},\Lambda}_{\alpha\nu\mu\beta} \big) \delta_{\sigma_1\sigma_2} \nonumber,
\end{eqnarray}
and hence
\begin{equation}
 \ddLambda\Sigma^{\text{s},\Lambda}_{\alpha\beta} =
    - \frac{1}{2\pi} \sum_{\mu\nu} \Pi^{\Sigma,\text{s},\Lambda}_{\mu\nu}
    \big( \Gamma^{\text{s},\Lambda}_{\alpha\nu\beta\mu}
        - 2 \Gamma^{\text{s},\Lambda}_{\alpha\nu\mu\beta} \big).
   \label{eq:flow:Sigma:WithSpin}
\end{equation}
Here, we have defined
\begin{equation}
  \Pi^{\Sigma,\text{s},\Lambda}_{\mu\nu} := P^{\text{s},\Lambda}_{\mu\nu}(\Lambda) + P^{\text{s},\Lambda}_{\mu\nu}(-\Lambda)
\end{equation}
in analogy to the definition in Eq.~\ref{eq:flow:Sigma}, as we will do with $\Pi^{\text{ph},\text{s},\Lambda}_{\mu\nu}$ and
$\Pi^{\text{c},\text{s},\Lambda}_{\mu\nu}$ in the following.
To obtain the flow equation for $\Gamma^{\Lambda,s}$, we must insert
Eqs.~(\ref{eq:definition:Sigma:with-spin},\ref{eq:definition:P:with-spin},\ref{eq:definition:Gamma:with-spin})
into Eq.~(\ref{eq:flow:Gamma}). To simplify our notation, we will use $\delta^{34}_{12} = \delta_{\sigma_1\sigma_2}\delta_{\sigma_3\sigma_4}$.
For the first term with $\Pi^{\text{c},\text{s},\Lambda}$, we have
\begin{eqnarray}
 & & - \frac{1}{2\pi} \sum_{\mu\nu\rho\sigma} \sum_{\sigma_5\sigma_6} 
     \Pi^{\text{c},\text{s},\Lambda}_{\mu\nu\sigma\rho}
  \times \nonumber \\ & & \hspace{1em}
     \Big( \Gamma^{\text{s},\Lambda}_{\nu\rho\gamma\delta} \delta^{63}_{54}
         - \Gamma^{\text{s},\Lambda}_{\nu\rho\delta\gamma} \delta^{64}_{53} \Big)
  \times \nonumber \\ & & \hspace{1em}
     \Big( \Gamma^{\text{s},\Lambda}_{\alpha\beta\sigma\mu} \delta^{26}_{15}
         - \Gamma^{\text{s},\Lambda}_{\alpha\beta\mu\sigma} \delta^{25}_{16} \Big).
\end{eqnarray}
Multiplying out the main product, there are four terms of combinations of $\Gamma^{\text{s},\Lambda}$ that appear,
\begin{eqnarray*}
 \sum_{\sigma_5\sigma_6}
    \Gamma^{\text{s},\Lambda}_{\nu\rho\gamma\delta}
    \Gamma^{\text{s},\Lambda}_{\alpha\beta\sigma\mu}
    \delta^{63}_{54}
    \delta^{26}_{15}
  & = &
    \Gamma^{\text{s},\Lambda}_{\nu\rho\gamma\delta}
    \Gamma^{\text{s},\Lambda}_{\alpha\beta\sigma\mu}
    \delta^{23}_{14}.
 \\
 \sum_{\sigma_5\sigma_6}
    - \Gamma^{\text{s},\Lambda}_{\nu\rho\delta\gamma}
    \Gamma^{\text{s},\Lambda}_{\alpha\beta\sigma\mu}
    \delta^{64}_{53}
    \delta^{26}_{15}
  & = &
    - \Gamma^{\text{s},\Lambda}_{\nu\rho\delta\gamma}
    \Gamma^{\text{s},\Lambda}_{\alpha\beta\sigma\mu}
    \delta^{24}_{13},
\end{eqnarray*}
\begin{eqnarray*}
 \sum_{\sigma_5\sigma_6}
    - \Gamma^{\text{s},\Lambda}_{\nu\rho\gamma\delta}
    \Gamma^{\text{s},\Lambda}_{\alpha\beta\mu\sigma}
    \delta^{63}_{54}
    \delta^{25}_{16}
  & = &
    - \Gamma^{\text{s},\Lambda}_{\nu\rho\gamma\delta}
    \Gamma^{\text{s},\Lambda}_{\alpha\beta\mu\sigma}
    \delta^{24}_{13},
 \\
 \sum_{\sigma_5\sigma_6}
    \Gamma^{\text{s},\Lambda}_{\nu\rho\delta\gamma}
    \Gamma^{\text{s},\Lambda}_{\alpha\beta\mu\sigma}
    \delta^{64}_{53}
    \delta^{25}_{16}
  & = &
    \Gamma^{\text{s},\Lambda}_{\nu\rho\delta\gamma}
    \Gamma^{\text{s},\Lambda}_{\alpha\beta\mu\sigma}
    \delta^{23}_{14}.
\end{eqnarray*}
On the other hand, the left hand side of the flow equation reads
\begin{equation}
 \ddLambda \left(
    \Gamma^{\text{s},\Lambda}_{\alpha\beta\gamma\delta} \delta_{14}^{23}
  - \Gamma^{\text{s},\Lambda}_{\alpha\beta\delta\gamma} \delta_{13}^{24}
 \right).
\end{equation}
We may thus look at the products that contain $\delta_{14}^{23}$ to obtain the first term of the flow equation for $\Gamma^{\text{s},\Lambda}$,
\begin{eqnarray}
 & & - \frac{1}{2\pi} \sum_{\mu\nu\rho\sigma}
     \Pi^{\text{c},\Lambda}_{\mu\nu\sigma\rho}
  \Big(
    \Gamma^{\text{s},\Lambda}_{\nu\rho\gamma\delta}
    \Gamma^{\text{s},\Lambda}_{\alpha\beta\sigma\mu}
   +
    \Gamma^{\text{s},\Lambda}_{\nu\rho\delta\gamma}
    \Gamma^{\text{s},\Lambda}_{\alpha\beta\mu\sigma}
  \Big) \delta_{14}^{23}. \label{eq:spinflow:cooper:1423part}
\end{eqnarray}
We may now proceed in doing the same for the particle-hole channel,
\begin{eqnarray}
 & & - \frac{1}{2\pi} \sum_{\mu\nu\rho\sigma} \sum_{\sigma_5\sigma_6} 
     \Pi^{\text{ph},\text{s},\Lambda}_{\mu\nu\rho\sigma}
  \times \nonumber \\ & & 
     \Big\{
     \big( \Gamma^{\text{s},\Lambda}_{\alpha\nu\gamma\rho} \delta^{16}_{53}
         - \Gamma^{\text{s},\Lambda}_{\alpha\nu\rho\gamma} \delta^{13}_{56} \big)
     \big( \Gamma^{\text{s},\Lambda}_{\beta\sigma\delta\mu} \delta^{25}_{64}
         - \Gamma^{\text{s},\Lambda}_{\beta\sigma\mu\delta} \delta^{24}_{65} \big)
     \nonumber \\ & &
   + \big( \Gamma^{\text{s},\Lambda}_{\beta\nu\gamma\rho} \delta^{26}_{53}
         - \Gamma^{\text{s},\Lambda}_{\beta\nu\rho\gamma} \delta^{23}_{56} \big)
     \big( \Gamma^{\text{s},\Lambda}_{\alpha\sigma\delta\mu} \delta^{15}_{64}
         - \Gamma^{\text{s},\Lambda}_{\alpha\sigma\mu\delta} \delta^{14}_{65} \big)
    \Big\}.~~
\end{eqnarray}
Of the eight products that appear, we again pick out those that appear with a $\delta_{14}^{23}$, where we use
that
\begin{eqnarray*}
  \sum_{\sigma_5\sigma_6} \delta^{16}_{53} \delta^{25}_{64} = \delta_{14}^{23},
  & \hspace{2em} &
  \sum_{\sigma_5\sigma_6} \delta^{23}_{56} \delta^{14}_{65} = 2 \delta_{14}^{23},
 \\
  \sum_{\sigma_5\sigma_6} \delta^{26}_{53} \delta^{14}_{65} = \delta_{14}^{23},
  & \hspace{2em} &
  \sum_{\sigma_5\sigma_6} \delta^{23}_{56} \delta^{15}_{64} = \delta_{14}^{23},
\end{eqnarray*}
so that we arrive at
\begin{eqnarray}
 & & \hspace{-2em} - \frac{1}{2\pi} \sum_{\mu\nu\rho\sigma}
     \Pi^{\text{ph},\text{s},\Lambda}_{\mu\nu\rho\sigma}
  \times \nonumber \\ & & \hspace{1em}
   \Big(
    2 \Gamma^{\text{s},\Lambda}_{\beta\nu\rho\gamma}
      \Gamma^{\text{s},\Lambda}_{\alpha\sigma\mu\delta}
     +
      \Gamma^{\text{s},\Lambda}_{\alpha\nu\gamma\rho}
      \Gamma^{\text{s},\Lambda}_{\beta\sigma\delta\mu}
  \nonumber \\ & & \hspace{1em}
     -
      \Gamma^{\text{s},\Lambda}_{\beta\nu\rho\gamma}
      \Gamma^{\text{s},\Lambda}_{\alpha\sigma\delta\mu}
     -
      \Gamma^{\text{s},\Lambda}_{\beta\nu\gamma\rho}
      \Gamma^{\text{s},\Lambda}_{\alpha\sigma\mu\delta}
   \Big) \delta_{14}^{23}.~~~~ \label{eq:spinflow:ph:1423part}
\end{eqnarray}
Adding Eq.~\eqref{eq:spinflow:cooper:1423part} and Eq.~\eqref{eq:spinflow:ph:1423part}, the flow equation for $\Gamma^{\text{s},\Lambda}$ now reads
\begin{eqnarray}
  & & \ddLambda \Gamma^{\text{s},\Lambda}_{\alpha\beta\delta\gamma} = - \frac{1}{2\pi} \sum_{\mu\nu\rho\sigma} \Big\{
    \nonumber \\ & & \hspace{2em} \hphantom{+}
   \Pi^{\text{c},\text{s},\Lambda}_{\mu\nu\sigma\rho}   \big(
    \Gamma^{\text{s},\Lambda}_{\nu\rho\gamma\delta}
    \Gamma^{\text{s},\Lambda}_{\alpha\beta\sigma\mu}
   +
    \Gamma^{\text{s},\Lambda}_{\nu\rho\delta\gamma}
    \Gamma^{\text{s},\Lambda}_{\alpha\beta\mu\sigma}
  \big)
    \nonumber \\ & & \hspace{1em} +
   \Pi^{\text{ph},\text{s},\Lambda}_{\mu\nu\rho\sigma}  \big(
    2 \Gamma^{\text{s},\Lambda}_{\beta\nu\rho\gamma}
      \Gamma^{\text{s},\Lambda}_{\alpha\sigma\mu\delta}
     +
      \Gamma^{\text{s},\Lambda}_{\alpha\nu\gamma\rho}
      \Gamma^{\text{s},\Lambda}_{\beta\sigma\delta\mu}
    \nonumber \\ & & \hspace{5em}
     -
      \Gamma^{\text{s},\Lambda}_{\beta\nu\rho\gamma}
      \Gamma^{\text{s},\Lambda}_{\alpha\sigma\delta\mu}
     -
      \Gamma^{\text{s},\Lambda}_{\beta\nu\gamma\rho}
      \Gamma^{\text{s},\Lambda}_{\alpha\sigma\mu\delta}
  \big) \Big\}.
\end{eqnarray}

\subsection{Finite Temperature}

For completeness, we also derive the form of the flow equations at finite temperature. In this case, using a sharp $\Theta$-function is ill-suited. Instead, we
utilize the cutoff suggested in Ref.~\onlinecite{EnssImpurities}, hence we replace $\Theta(|\omega|-\Lambda)$ by $\chi^\Lambda(\omega_n)$, which is
given by
\begin{equation}
 \chi^\Lambda(\omega_n) = \left\{ \begin{array}{ll}
                                   0,                                                       & |\omega_n| \le \Lambda - \pi T, \\
                                   \frac{1}{2} + \frac{|\omega_n| - \Lambda}{2\pi T}        & \Lambda - \pi T \le |\omega_n| \le \Lambda + \pi T, \\
                                   1,                                                       & \Lambda + \pi T \le |\omega_n|,
                                   \end{array} \right.
\end{equation}
and its derivative with respect to $\Lambda$ is then given by
\begin{equation}
 -(\partial_\Lambda \chi^\Lambda(\omega_n)) = \left\{ \begin{array}{ll}
                                                       \frac{1}{2\pi T}      & \Lambda - \pi T \le |\omega_n| \le \Lambda + \pi T, \\
                                                       0                     & \text{otherwise}.
                                                      \end{array} \right.
\end{equation}
We note that $\chi^\Lambda(\omega_n) \to \Theta(|\omega|-\Lambda)$ as $T\to 0$. The full Green's function is now given by
\begin{equation}
 \mathcal{G}^\Lambda(\omega_n) = \frac{\chi^\Lambda(\omega_n)}{\ii\omega_n - H_0 + \muchem - \chi^\Lambda(\omega_n) \Sigma^\Lambda(\omega_n)},
\end{equation}
whereas the single-scale propagator, Eq.~\ref{eq:definition:SingleScalePropagator}, reads
\begin{eqnarray*}
 \mathcal{S}^\Lambda(\omega_n) & = & \frac{\partial_\Lambda \chi^\Lambda(\omega_n)}{\ii\omega_n - H_0 + \muchem - \chi^\Lambda(\omega_n) \Sigma^\Lambda(\omega_n)}
     \times \nonumber \\ & & \hspace{1.5em}
     \big( \ii\omega_n - H_0 + \muchem \big)
      \times \nonumber \\ & & \hspace{1.5em}
      \frac{1}{\ii\omega_n - H_0 + \muchem - \chi^\Lambda(\omega_n) \Sigma^\Lambda(\omega_n)}.
\end{eqnarray*}
With this form of a cutoff function, the Matsubara sums may be evaluated analytically. Since Matsubara frequencies have a distance of $2\pi T$ from
each other, the derivative of the cutoff is only nonzero for a two Matsubara frequencies, whose magnitude are that closest to the parameter $\Lambda$.
Any sum with a single derivative of $\chi^\Lambda$ may hence be evaluated as
\begin{equation}
 T \sum_{n} -(\partial_\Lambda \chi^\Lambda(\omega_n)) f(\omega_n) = \frac{1}{2\pi} \sum_{|\omega_n| \approx \Lambda} f(\omega_n).
\end{equation}
This structure is very similar to the situation at $T = 0$, where we have
\begin{equation}
 \frac{1}{2\pi} \int\diff\omega \delta(|\omega|-\Lambda) f(\omega) = \frac{1}{2\pi} \sum_{|\omega| = \Lambda} f(\omega).
\end{equation}
Again we adopt the static limit and define $P^{T,\Lambda}(\omega_n)$ as
\begin{equation}
 P^{T,\Lambda}(\omega_n) := \frac{1}{\ii\omega_n - H_0 + \muchem - \chi^\Lambda(\omega_n) \Sigma^\Lambda},
\end{equation}
and $P^{'T,\Lambda}, P^{''T,\Lambda}$ as
\begin{eqnarray}
 P^{'T,\Lambda}(\omega_n) & := & P^{T,\Lambda}(\omega_n) (\ii\omega_n - H_0 + \muchem)
                                                             P^{T,\Lambda}(\omega_n), \nonumber \\
    & & \\
 P^{''T,\Lambda}(\omega_n) & := & P^{T,\Lambda}(\omega_n) \chi^\Lambda(\omega_n),
\end{eqnarray}
the flow equation for the self-energy now reads
\begin{eqnarray}
 \ddLambda \Sigma^\Lambda_{\alpha\beta} & = & - \frac{1}{2\pi} \sum_{|\omega_n| \approx \Lambda} P^{'T,\Lambda}_{\mu\nu}(\omega_n)
                                                \Gamma^\Lambda_{\alpha\nu\beta\mu}.
  \label{eq:flow:Sigma:finiteT}
\end{eqnarray}
Setting all external frequencies to zero and dropping the frequency dependence of the vertex, its flow equation is now given by
\begin{eqnarray}
 \ddLambda \Gamma^\Lambda_{\alpha\beta\gamma\delta} & = & - \frac{1}{2\pi} \sum_{|\omega_n| \approx \Lambda} \sum_{\mu\nu\rho\sigma} \Big\{ 
                      P^{''T,\Lambda}_{\mu\nu}(\omega_n) P^{'T,\Lambda}_{\rho\sigma}(-\omega_n)
                  \times \nonumber \\  & & \hspace{2em}
                 \Gamma^\Lambda_{\alpha\beta\sigma\mu} \Gamma^\Lambda_{\nu\rho\gamma\delta} + \nonumber \\
                  & & \hspace{4em} P^{''T,\Lambda}_{\mu\nu}(\omega_n) P^{'T,\Lambda}_{\rho\sigma}(\omega_n) \times \nonumber \\
                  & & \hspace{1em} \big[ 
                       \Gamma^\Lambda_{\beta\nu\gamma\rho} \Gamma^\Lambda_{\alpha\mu\delta\sigma}
                       - \Gamma^\Lambda_{\alpha\mu\gamma\sigma} \Gamma^\Lambda_{\beta\nu\delta\rho}
                    \nonumber \\  & & \hspace{2em}
                       + \Gamma^\Lambda_{\beta\mu\gamma\sigma} \Gamma^\Lambda_{\alpha\nu\delta\rho}
                       - \Gamma^\Lambda_{\alpha\nu\gamma\rho} \Gamma^\Lambda_{\beta\mu\delta\sigma}
                    \big] \Big\}.
                  \label{eq:flow:Gamma:finiteT}
\end{eqnarray}
We note that $\chi^\Lambda(\omega_n) \to \frac{1}{2}$ for $\omega_n \to \Lambda$, so if taking the limit $T \to 0$ (and applying the symmetries
of the vertex) one recovers Eq.~(\ref{eq:flow:Gamma}).

\subsection{Observables and Correlators}

\subsubsection{Single-particle observables}

Single-particle observables may be expressed by the Green's function, which is given by
\begin{equation}
 \mathcal{G}(\ii\omega) = \frac{1}{\ii\omega - H_0 + \mu - \Sigma}\e^{\ii\omega 0^+}. \label{eq:obs:G}
\end{equation}
The convergence factor $\e^{\ii\omega 0^+}$ is explicitly required here. In the following we will summarize (trivial) statements that
follow from employing the static limit.
For example, the density matrix for the occupancy of single-particle states, $\rho_{ij}$, is given by
\begin{equation}
 \rho_{ij} = \sum_{\alpha\beta} V^{\mathrm{rn}}_{i\alpha} \left[ \frac{1}{2\pi} \int_{-\infty}^{\infty} \diff\omega\, \mathcal{G}_{\alpha\beta}(\ii\omega) \e^{\ii\omega 0^+} \right] V^{\mathrm{rn},-1}_{\beta j},
\end{equation}
where
\begin{equation}
 V^{\mathrm{rn}}_{i\alpha} = \braket{\alpha|i}
\end{equation}
and $\ket{i}$ is one out of $N$ basis-vectors spanning the single-particle Hilbert space $\mathcal{H}$.

The frequency integral may be calculated analytically by going into the basis where $\mathcal{G}$ is diagonal, i.e. the eigenbasis of $H_0 + \Sigma$.
We will denote indices in that basis by a tilde, e.g. $\tilde\mu$ and the eigenvalues of $H_0 + \Sigma$ with $\tilde\epsilon_{\tilde\mu}$. (As $\Sigma$ is hermitian in the static limit, $\tilde\epsilon_{\tilde\mu}$ are real.)
The basis transform from that basis into the basis chosen for observables will be denoted by $V^{\rm{ri}}_{i\tilde\mu}$. The integral may now be performed analytically,
closing the integration loop around the left half-plane,
\begin{eqnarray}
 \rho_{ij} & = & \sum_{\tilde\mu} V^{\mathrm{ri}}_{i\tilde\mu} \left[ \frac{1}{2\pi} \int_{-\infty}^{\infty} \diff\omega\, \frac{\e^{\ii\omega 0^+}}{\ii\omega - \tilde\epsilon_{\tilde\mu} + \muchem} \right] V^{\mathrm{ri},-1}_{\tilde\mu j} \nonumber \\
           & = & \sum_{\tilde\mu}^{\text{occ.}} V^{\mathrm{ri}}_{i\tilde\mu} V^{\mathrm{ri},-1}_{\tilde\mu j},
                 \label{eq:obs:density-matrix}
\end{eqnarray}
where the summation is now only performed over states below the chemical potential. (Occupied states.)

In order to obtain the result at finite temperature, $T > 0$, we must replace the integral by a Matsubara sum, performing the inverse of Eq.~(\ref{eq:Matsubara:T0-transition}).
The sum may be performed analytically, using the well-known relation
\begin{equation}
 T \sum_{\omega_n} \frac{1}{\ii\omega_n - \xi} = n_{\mathrm{F}}(\xi),
\end{equation}
where $n_{\mathrm{F}}$ is the Fermi function. We now obtain
\begin{eqnarray}
 \rho_{ij} & = & \sum_{\tilde\mu} V^{\mathrm{ri}}_{i\tilde\mu} \left[ T \sum_{\omega_n} \frac{1}{\ii\omega_n - \tilde\epsilon_{\tilde\mu} + \muchem} \right] V^{\mathrm{ri},-1}_{\tilde\mu j} \nonumber \\
           & = & \sum_{\tilde\mu} V^{\mathrm{ri}}_{i\tilde\mu} n_{\mathrm{F}}(\tilde\epsilon_{\tilde\mu} - \muchem) V^{\mathrm{ri},-1}_{\tilde\mu j},
           \label{eq:obs:density-matrix:finiteT}
\end{eqnarray}
which reproduces Eq.~(\ref{eq:obs:density-matrix}) for $T \to 0$.

Another single-particle quantity of interest is the (normalized) density of states (DOS), which may be calculated from the imaginary part of the retarded Green's function after Wick rotation.
As we work in the static limit for the self-eenergy, the Wick rotation is trivial and yields the following expression for the density of states at $T = 0$,
\begin{eqnarray}
 \rho(\epsilon) = - \frac{1}{2\pi N} \Im \sum_{\tilde\mu} \frac{1}{\epsilon - \tilde\epsilon_{\tilde\mu} + \muchem + \ii 0}.
\end{eqnarray}

Finally, in systems with spin rotational invariance the single-particle Green's function is diagonal in spin space and the previously discussed quantities simply
acquire a factor of 2.

\subsubsection{Correlator of Occupancy Numbers ($T = 0$)}

Two-particle observables may be rewritten in terms of single- and two-particle Green's functions. In the case of spinless Fermions the correlator of occupancy numbers,
$\mathcal{C}^{\mathrm{dd}}_{ij}$, may be rewritten as
\begin{eqnarray}
 \hspace{-1em}
  \mathcal{C}^{\mathrm{dd}}_{ij} & = & \big<\mathrm{\hat n}_i^{\vphantom{\dagger}} \mathrm{\hat n}_j^{\vphantom{\dagger}}\big>
                     =   \big<\mathrm{\hat c}_i^\dagger \mathrm{\hat c}_i^{\vphantom{\dagger}} \mathrm{\hat c}_j^\dagger \mathrm{\hat c}_j^{\vphantom{\dagger}}\big>
                     =   \big<\mathrm{\hat c}_j^\dagger \mathrm{\hat c}_i^\dagger \mathrm{\hat c}_i^{\vphantom{\dagger}} \mathrm{\hat c}_j^{\vphantom{\dagger}}\big>
                       + \big<\mathrm{\hat c}_i^\dagger \mathrm{\hat c}_i^{\vphantom{\dagger}}\big> \delta_{ij} \nonumber \\
                   & = & \mathcal{C}^{\mathrm{dd},(2)}_{ij} + \big<\mathrm{\hat n}_i^{\vphantom{\dagger}}\big> \big<\mathrm{\hat n}_j^{\vphantom{\dagger}}\big>
                                                - \big<\mathrm{\hat c}_i^\dagger \mathrm{\hat c}_j^{\vphantom{\dagger}}\big> \big<\mathrm{\hat c}_j^\dagger \mathrm{\hat c}_i^{\vphantom{\dagger}}\big>
                                                + \big<\mathrm{\hat n}_i^{\vphantom{\dagger}}\big> \delta_{ij},
\end{eqnarray}
where $\mathcal{C}^{\mathrm{dd},(2)}_{ij}$ is the part of the correlation function arising from the connected two-particle Green's function and thus the vertex.
In the case of spinful Fermions, the correlator includes a sum over the spin degrees of freedom,
\begin{equation}
 \mathcal{C}^{\mathrm{dd}}_{ij} = \sum_{\sigma\sigma'} \big<\mathrm{\hat n}_{i\sigma} \mathrm{\hat n}_{j\sigma'}\big>.
\end{equation}
For systems that obey the full $\mathrm{SU}(2)$ symmetry, it reads
\begin{eqnarray}
 \mathcal{C}^{\mathrm{dd}}_{ij} & = & \mathcal{C}^{\mathrm{dd},(2)}_{ij}
                                + 4 \big<\mathrm{\hat n}_{i\sigma}^{\vphantom{\dagger}}\big> \big<\mathrm{\hat n}_{j\sigma}^{\vphantom{\dagger}}\big>
                                - 2 \big<\mathrm{\hat c}_{i\sigma}^\dagger \mathrm{\hat c}_{j\sigma}^{\vphantom{\dagger}}\big> \big<\mathrm{\hat c}_{j\sigma}^\dagger \mathrm{\hat c}_{i\sigma}^{\vphantom{\dagger}}\big>
  \nonumber \\ & &
                                + 2 \big<\mathrm{\hat n}_{i\sigma}^{\vphantom{\dagger}}\big> \delta_{ij},
\label{eq:obs:ddcorr:spin}
\end{eqnarray}
where $\sigma$ is an arbitrary spin index that is not summed over, as the single-particle quantities are proportional to $\delta_{\sigma\sigma'}$.

We will first derive the expression for $\mathcal{C}^{\mathrm{dd},(2)}_{ij}$ for the spinless case at $T = 0$. Since we are looking at static quantities,
but our formalism is derived in Matsubara frequency space, we must perform a Fourier transform,
\begin{eqnarray}
 \mathcal{C}^{\mathrm{dd},(2)}_{ij} & = & \int\frac{\diff\omega_1}{2\pi} \int\frac{\diff\omega_2}{2\pi} \int\frac{\diff\omega_3}{2\pi} \int\frac{\diff\omega_4}{2\pi} \times
  \nonumber \\ & & 
                                          \mathcal{G}^{(2,c)}_{ijij}(\ii\omega_1, \ii\omega_2, \ii\omega_3, \ii\omega_4),
  \label{eq:TwoParticleGF:connected:fourier}
\end{eqnarray}
where $\mathcal{G}^{(2,c)}$ is the two-particle connected Green's function. Using the well-known relation between the two-particle connected Green's function
and the vertex,
\begin{equation}
 \raisebox{-2.3em}{\includegraphics[height=5em]{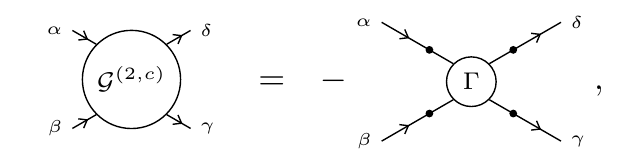}}
\end{equation}
we arrive at
\begin{eqnarray}
 & & \mathcal{G}^{(2,c)}_{ijij}(\ii\omega_1, \ii\omega_2, \ii\omega_3, \ii\omega_4) =
   - 2\pi \sum_{\alpha\beta\gamma\delta} \sum_{\alpha'\beta'\gamma'\delta'}
     V^{\mathrm{rn}}_{i\alpha'} V^{\mathrm{rn}}_{j\beta'}
 \times \nonumber \\ & & \hspace{1em}
   \mathcal{G}_{\alpha'\alpha}(\ii\omega_1) \mathcal{G}_{\beta'\beta}(\ii\omega_2)
   \Gamma_{\alpha\beta\gamma\delta} \delta(\ii\omega_1 + \ii\omega_2 - \ii\omega_3 - \ii\omega_4)
 \times \nonumber \\ & & \hspace{1em}
   \mathcal{G}_{\gamma\gamma'}(\ii\omega_3) \mathcal{G}_{\delta\delta'}(\ii\omega_4)
   V^{\mathrm{rn},-1}_{\gamma' i} V^{\mathrm{rn},-1}_{\delta' j}.
   \label{eq:TwoParticleGF:connected:frequency:initial}
\end{eqnarray}
In order to solve the frequency integral analytically, we again transform into the eigenbasis of $H_0 + \Sigma$.
Eq.~(\ref{eq:TwoParticleGF:connected:frequency:initial}) now reads
\begin{eqnarray}
 & & \mathcal{G}^{(2,c)}_{ijij}(\ii\omega_1, \ii\omega_2, \ii\omega_3, \ii\omega_4) =
   - 2\pi \sum_{\alpha\beta\gamma\delta} \sum_{\tilde\mu\tilde\nu\tilde\rho\tilde\sigma}
     V^{\mathrm{ri}}_{i\tilde\mu} V^{\mathrm{ri}}_{j\tilde\nu}
 \times \nonumber \\ & & \hspace{1em}
   \mathcal{G}_{\tilde\mu\tilde\mu}(\ii\omega_1) \mathcal{G}_{\tilde\nu\tilde\nu}(\ii\omega_2)
    V^{\mathrm{in}}_{\tilde\mu\alpha} V^{\mathrm{in}}_{\tilde\nu\beta}
 \times \nonumber \\ & & \hspace{1em}
   \Gamma_{\alpha\beta\gamma\delta} \delta(\ii\omega_1 + \ii\omega_2 - \ii\omega_3 - \ii\omega_4)
 \times \nonumber \\ & & \hspace{1em}
   V^{\mathrm{in},-1}_{\gamma\tilde\rho} V^{\mathrm{in},-1}_{\delta\tilde\sigma}
   \mathcal{G}_{\tilde\rho\tilde\rho}(\ii\omega_3) \mathcal{G}_{\tilde\sigma\tilde\sigma}(\ii\omega_4)
   V^{\mathrm{ri},-1}_{\tilde\rho i} V^{\mathrm{ri},-1}_{\tilde\sigma j}.
   \label{eq:TwoParticleGF:connected:frequency:ibasis}
\end{eqnarray}
For any given $\tilde\mu,\tilde\nu,\tilde\rho,\tilde\sigma$, we have for the frequency-dependent part
\begin{eqnarray}
 & & 2\pi \int\frac{\diff\omega_1}{2\pi} \int\frac{\diff\omega_2}{2\pi} \int\frac{\diff\omega_3}{2\pi} \int\frac{\diff\omega_4}{2\pi}
  \mathcal{G}_{\tilde\mu\tilde\mu}(\ii\omega_1) \mathcal{G}_{\tilde\nu\tilde\nu}(\ii\omega_2)
  \times \nonumber \\ & & \hspace{1em}
    \mathcal{G}_{\tilde\rho\tilde\rho}(\ii\omega_3) \mathcal{G}_{\tilde\sigma\tilde\sigma}(\ii\omega_4) \delta(\ii\omega_1 + \ii\omega_2 - \ii\omega_3 - \ii\omega_4)
   \nonumber \\ & = &
   \int\frac{\diff\omega_1}{2\pi} \int\frac{\diff\omega_2}{2\pi} \int\frac{\diff\omega_3}{2\pi} 
   \mathcal{G}_{\tilde\mu\tilde\mu}(\ii\omega_1) \mathcal{G}_{\tilde\nu\tilde\nu}(\ii\omega_2)
   \times \nonumber \\ & & \hspace{1em}
   \mathcal{G}_{\tilde\rho\tilde\rho}(\ii\omega_3) \mathcal{G}_{\tilde\sigma\tilde\sigma}(\ii(\omega_1 + \omega_2 - \omega_3)).
   \label{eq:TwoParticleGF:connected:frequency:nodelta}
\end{eqnarray}
Using the convention that $\tilde\epsilon_{\tilde\mu}$ is the $\tilde\mu$-th eigenvalue of $H_0 + \Sigma$, we may now write
\begin{equation}
 \mathcal{G}_{\tilde\mu\tilde\mu}(\ii\omega_1) = \frac{1}{\ii\omega_1 - \tilde\epsilon_{\tilde\mu} + \muchem} =: \frac{1}{\ii\omega_1 - \tilde\xi_{\tilde\mu}}.
 \label{eq:TwoParticleGF:Gdef}
\end{equation}
All occurring integrals are of similar form and may be solved by simply closing the integration loop around the left complex half-plane,
\begin{equation}
 \int \frac{\diff\omega}{2\pi} \frac{1}{\ii\omega - z} \frac{1}{\ii\omega - \xi} = \frac{g(z,\xi)}{z-\xi}.
\end{equation}
The exact result of the integral will depend on the position of each of the poles \{$z$, $\xi$\} relative to the integration loop. If they are
either both inside or both outside, the integral gives zero (either the residues cancel or there are no poles inside the loop), there is only a contribution
if there is just a single pole inside the loop. The residue is always $\pm(z-\xi)^{-1}$. Therefore, we define $g(z,\xi)$ to keep track of
the correct sign. It may be represented as
\begin{eqnarray}
 g(z,\xi) & = & - g(\xi,z) \nonumber \\
          & = & \Theta_{\Re}(-z)\Theta_{\Re}(\xi) - \Theta_{\Re}(z)\Theta_{\Re}(-\xi),~~
\end{eqnarray}
where $\Theta_{\Re}(z)$ is the Heaviside step function of the real part of $z$.

Performing the first integral over $\omega_1$, we have
\begin{eqnarray}
 &   & \int\frac{\diff\omega_1}{2\pi} \frac{1}{\ii\omega_1 - \tilde\xi_{\tilde\mu}} \frac{1}{\ii\omega_1 - (\tilde\xi_{\tilde\sigma} - \ii\omega_2 + \ii\omega_3)} \nonumber \\
 & = & \frac{g(\tilde\xi_{\tilde\mu}, \tilde\xi_{\tilde\sigma} + \ii(\omega_3 - \omega_2))}{\tilde\xi_{\tilde\mu} - \tilde\xi_{\tilde\sigma} + \ii\omega_2 - \ii\omega_3}.
\end{eqnarray}
The expression $g(\tilde\xi_{\tilde\mu}, \tilde\xi_{\tilde\sigma} + \ii(\omega_3 - \omega_2))$ may be simplified further, since for real $\omega_{2,3}$, it is equal to $g(\tilde\xi_{\tilde\mu}, \tilde\xi_{\tilde\sigma})$.\footnote{Note
that while closing the integrals over $\omega_{2,3}$, those frequencies may obtain an imaginary part, but since semi-circle contour parts have a vanishing contribution to
the integral itself, this may be ignored.}
Applying this result sequentially, the integral in Eq.~(\ref{eq:TwoParticleGF:connected:frequency:nodelta}) has the result
\begin{eqnarray}
 & & \frac{g(\tilde\xi_{\tilde\mu}, \tilde\xi_{\tilde\sigma}) g(\tilde\xi_{\tilde\sigma} - \tilde\xi_{\tilde\mu}, \tilde\xi_{\tilde\nu}) g(\tilde\xi_{\tilde\mu} + \tilde\xi_{\tilde\nu} - \tilde\xi_{\tilde\sigma}, \tilde\xi_{\tilde\rho})}{\tilde\xi_{\tilde\mu} + \tilde\xi_{\tilde\nu} - \tilde\xi_{\tilde\rho} - \tilde\xi_{\tilde\sigma}}.
\end{eqnarray}
Further simplification is possible: if $\Re \tilde\xi_{\tilde\mu} > 0$, then $\Re \tilde\xi_{\tilde\sigma}$ must be less than zero, or the contribution vanishes. In that case, it
follows that $\Re(\tilde\xi_{\tilde\sigma} - \tilde\xi_{\tilde\mu}) < 0$, and we may deduce in the same way that $\Re \tilde\xi_{\tilde\nu}$ should be greater than zero. Finally,
$\Re(\tilde\xi_{\tilde\mu} + \tilde\xi_{\tilde\nu} - \tilde\xi_{\tilde\sigma}) > 0$ leads to the conclusion that $\Re \tilde\xi_{\tilde\rho} < 0$. On the other hand, if
$\Re \tilde\xi_{\tilde\mu} < 0$, the analogous argument can be made with flipped inequalities. The only non-zero contributions arise from combinations where the real parts
of $\tilde\xi_{\tilde\mu}$ and $\tilde\xi_{\tilde\nu}$ have the same sign, but have the opposite sign to both $\tilde\xi_{\tilde\rho}$ and $\tilde\xi_{\tilde\sigma}$. Using
this result, Eq.~(\ref{eq:TwoParticleGF:connected:fourier}) now reads
\begin{eqnarray}
 \mathcal{C}^{\mathrm{dd},(2)}_{ij} & = & \sum_{\alpha\beta\gamma\delta} \left[
       \sum_{\substack{\tilde\mu,\tilde\nu\in\mathcal{H}_e\\\tilde\rho,\tilde\sigma\in\mathcal{H}_h}}
     - \sum_{\substack{\tilde\mu,\tilde\nu\in\mathcal{H}_h\\\tilde\rho,\tilde\sigma\in\mathcal{H}_e}}
    \right]
 \times \nonumber \\ & & \hspace{1em}
     V^{\mathrm{ri}}_{i\tilde\mu} V^{\mathrm{ri}}_{j\tilde\nu}
   \frac{1}{\tilde\epsilon_{\tilde\mu} + \tilde\epsilon_{\tilde\nu} - \tilde\epsilon_{\tilde\rho} - \tilde\epsilon_{\tilde\sigma}}
   V^{\mathrm{ri},-1}_{\tilde\rho i} V^{\mathrm{ri},-1}_{\tilde\sigma j}
 \times \nonumber \\ & & \hspace{1em}
   V^{\mathrm{in}}_{\tilde\mu\alpha} V^{\mathrm{in}}_{\tilde\nu\beta}
   \Gamma_{\alpha\beta\gamma\delta}
   V^{\mathrm{in},-1}_{\gamma\tilde\rho} V^{\mathrm{in},-1}_{\delta\tilde\sigma},
   \label{eq:TwoParticleGF:connected:T0:result}
\end{eqnarray}
where $\mathcal{H}_e$ is the subspace where $\tilde\epsilon - \muchem < 0$ (``electrons'') and $\mathcal{H}_h$ the subspace where $\tilde\epsilon + \muchem > 0$ (``holes'').

\subsubsection{Correlator of Occupancy Numbers ($T > 0$)}

At finite temperatures $T > 0$, the result is very similar. To derive it, we need to replace the integrals in Eq.~(\ref{eq:TwoParticleGF:connected:frequency:nodelta})
by Matsubara sums according to the inverse of Eqs.~(\ref{eq:Matsubara:T0-transition},\ref{eq:Matsubara:T0-transition:Kronecker}),
\begin{eqnarray}
 & & T\sum_{\omega_n} T\sum_{\omega_m} T\sum_{\omega_{n'}}
   \mathcal{G}_{\tilde\mu\tilde\mu}(\ii\omega_n) \mathcal{G}_{\tilde\nu\tilde\nu}(\ii\omega_m)
   \times \nonumber \\ & & \hspace{1em}
   \mathcal{G}_{\tilde\rho\tilde\rho}(\ii\omega_{n'}) \mathcal{G}_{\tilde\sigma\tilde\sigma}(\ii(\omega_n + \omega_m - \omega_{n'})).
   \label{eq:TwoParticleGF:connected:finiteT:frequency:nodelta}
\end{eqnarray}
Inserting Eq.~(\ref{eq:TwoParticleGF:Gdef}) into this expression, we may now perform the Matsubara sums analytically, which are of the form
\begin{equation}
 T\sum_{\omega_n} \frac{1}{\ii\omega_n - z} \frac{1}{\ii\omega_n - \xi} = \frac{n_{\mathrm{F}}(z) - n_{\mathrm{F}}(\xi)}{z - \xi},
\end{equation}
where $n_{\mathrm{F}}$ is the Fermi function. We note that due to its periodicity we have
$n_{\mathrm{F}}(\tilde\xi \pm \ii\omega_{n'}) = n_{\mathrm{F}}(\tilde\xi)$ if $\omega_{n'}$ is a Matsubara frequency, so we may simplify the
numerator again. Eq.~(\ref{eq:TwoParticleGF:connected:finiteT:frequency:nodelta}) is thus equal to
\begin{eqnarray}
  & & [n_{\mathrm{F}}(\tilde\xi_{\tilde\mu}) - n_{\mathrm{F}}(\tilde\xi_{\tilde\sigma})] [n_{\mathrm{F}}(\tilde\xi_{\tilde\sigma} - \tilde\xi_{\tilde\mu}) - n_{\mathrm{F}}(\tilde\xi_{\tilde\nu})]
 \nonumber \times \\ & &
 \qquad \frac{
    [n_{\mathrm{F}}(\tilde\xi_{\tilde\mu} + \tilde\xi_{\tilde\nu} - \tilde\xi_{\tilde\sigma}) - n_{\mathrm{F}}(\tilde\xi_{\tilde\rho})]
 }{\tilde\xi_{\tilde\mu} + \tilde\xi_{\tilde\nu} - \tilde\xi_{\tilde\rho} - \tilde\xi_{\tilde\sigma}}.
\end{eqnarray}
Therefore, we have
\begin{eqnarray}
  \mathcal{C}^{\mathrm{dd},(2)}_{ij} & = & \sum_{\alpha\beta\gamma\delta}\sum_{\tilde\mu\tilde\nu\tilde\rho\tilde\sigma}
     V^{\mathrm{ri}}_{i\tilde\mu} V^{\mathrm{ri}}_{j\tilde\nu}
   \frac{1}{\tilde\epsilon_{\tilde\mu} + \tilde\epsilon_{\tilde\nu} - \tilde\epsilon_{\tilde\rho} - \tilde\epsilon_{\tilde\sigma}}
 \times \nonumber \\ & & \hspace{1em}
   [n_{\mathrm{F}}(\tilde\xi_{\tilde\sigma}) - n_{\mathrm{F}}(\tilde\xi_{\tilde\mu})] [n_{\mathrm{F}}(\tilde\xi_{\tilde\sigma} - \tilde\xi_{\tilde\mu}) - n_{\mathrm{F}}(\tilde\xi_{\tilde\nu})]
 \times \nonumber \\ & & \hspace{1em}
   [n_{\mathrm{F}}(\tilde\xi_{\tilde\mu} + \tilde\xi_{\tilde\nu} - \tilde\xi_{\tilde\sigma}) - n_{\mathrm{F}}(\tilde\xi_{\tilde\rho})]
   V^{\mathrm{ri},-1}_{\tilde\rho i} V^{\mathrm{ri},-1}_{\tilde\sigma j}
 \times \nonumber \\ & & \hspace{1em}
   V^{\mathrm{in}}_{\tilde\mu\alpha} V^{\mathrm{in}}_{\tilde\nu\beta}
   \Gamma_{\alpha\beta\gamma\delta}
   V^{\mathrm{in},-1}_{\gamma\tilde\rho} V^{\mathrm{in},-1}_{\delta\tilde\sigma}.
   \label{eq:TwoParticleGF:connected:finiteT:result}
  \end{eqnarray}
For orbitals far away from the Fermi energy, $|\tilde\xi| \gg T$, this expression goes over into the expression for $T = 0$ and
we arrive at Eq.~(\ref{eq:TwoParticleGF:connected:T0:result}) again.

\subsubsection{Correlator of Occupancy Numbers (Systems with spin)}

In systems with spin we must also sum over two spin indices when calculating $\mathcal{C}^{\mathrm{dd},(2)}_{ij}$. We replace all orbital indices
in Eq.~(\ref{eq:TwoParticleGF:connected:finiteT:result}) by pairs of orbital and spin indices, $\alpha\to(\alpha,\sigma)$. For systems with $\mathrm{SU}(2)$
symmetry all single-particle quantities are diagonal in spin space, so after performing sums over all the relevant Kronecker-$\delta$s, we have
\begin{eqnarray}
  \mathcal{C}^{\mathrm{dd},(2)}_{ij} & = & \sum_{\sigma\sigma'} \sum_{\alpha\beta\gamma\delta}\sum_{\tilde\mu\tilde\nu\tilde\rho\tilde\sigma}
     V^{\mathrm{ri}}_{i\tilde\mu} V^{\mathrm{ri}}_{j\tilde\nu}
   \frac{1}{\tilde\epsilon_{\tilde\mu} + \tilde\epsilon_{\tilde\nu} - \tilde\epsilon_{\tilde\rho} - \tilde\epsilon_{\tilde\sigma}}
 \times \nonumber \\ & & \hspace{.5em}
   [n_{\mathrm{F}}(\tilde\xi_{\tilde\sigma}) - n_{\mathrm{F}}(\tilde\xi_{\tilde\mu})] [n_{\mathrm{F}}(\tilde\xi_{\tilde\sigma} - \tilde\xi_{\tilde\mu}) - n_{\mathrm{F}}(\tilde\xi_{\tilde\nu})]
 \times \nonumber \\ & & \hspace{.5em}
   [n_{\mathrm{F}}(\tilde\xi_{\tilde\mu} + \tilde\xi_{\tilde\nu} - \tilde\xi_{\tilde\sigma}) - n_{\mathrm{F}}(\tilde\xi_{\tilde\rho})]
   V^{\mathrm{ri},-1}_{\tilde\rho i} V^{\mathrm{ri},-1}_{\tilde\sigma j}
 \times \nonumber \\ & & \hspace{.5em}
   V^{\mathrm{in}}_{\tilde\mu\alpha} V^{\mathrm{in}}_{\tilde\nu\beta}
   \Gamma_{(\alpha,\sigma)(\beta,\sigma')(\gamma,\sigma)(\delta,\sigma')}
   V^{\mathrm{in},-1}_{\gamma\tilde\rho} V^{\mathrm{in},-1}_{\delta\tilde\sigma}.
   \label{eq:TwoParticleGF:connected:finiteT:spin:inserted}
\end{eqnarray}
Inserting Eq.~(\ref{eq:definition:Gamma:with-spin}), we may perform the summation over the remaining spin indices and arrive at
\begin{eqnarray}
  \mathcal{C}^{\mathrm{dd},(2)}_{ij} & = & \sum_{\alpha\beta\gamma\delta}\sum_{\tilde\mu\tilde\nu\tilde\rho\tilde\sigma}
     V^{\mathrm{ri}}_{i\tilde\mu} V^{\mathrm{ri}}_{j\tilde\nu}
   \frac{1}{\tilde\epsilon_{\tilde\mu} + \tilde\epsilon_{\tilde\nu} - \tilde\epsilon_{\tilde\rho} - \tilde\epsilon_{\tilde\sigma}}
 \times \nonumber \\ & & \hspace{.5em}
   [n_{\mathrm{F}}(\tilde\xi_{\tilde\sigma}) - n_{\mathrm{F}}(\tilde\xi_{\tilde\mu})] [n_{\mathrm{F}}(\tilde\xi_{\tilde\sigma} - \tilde\xi_{\tilde\mu}) - n_{\mathrm{F}}(\tilde\xi_{\tilde\nu})]
 \times \nonumber \\ & & \hspace{.5em}
   [n_{\mathrm{F}}(\tilde\xi_{\tilde\mu} + \tilde\xi_{\tilde\nu} - \tilde\xi_{\tilde\sigma}) - n_{\mathrm{F}}(\tilde\xi_{\tilde\rho})]
   V^{\mathrm{ri},-1}_{\tilde\rho i} V^{\mathrm{ri},-1}_{\tilde\sigma j}
 \times \nonumber \\ & & \hspace{.5em}
   V^{\mathrm{in}}_{\tilde\mu\alpha} V^{\mathrm{in}}_{\tilde\nu\beta}
   \big[ 2\Gamma^{\text{s}}_{\alpha\beta\gamma\delta} - 4\Gamma^{\text{s}}_{\alpha\beta\delta\gamma} \big]
   V^{\mathrm{in},-1}_{\gamma\tilde\rho} V^{\mathrm{in},-1}_{\delta\tilde\sigma}.
   \label{eq:TwoParticleGF:connected:finiteT:spin:result}
\end{eqnarray}
At $T = 0$, the result is analogously given by
\begin{eqnarray}
 \mathcal{C}^{\mathrm{dd},(2)}_{ij} & = & \sum_{\alpha\beta\gamma\delta} \left[
       \sum_{\substack{\tilde\mu,\tilde\nu\in\mathcal{H}_e\\\tilde\rho,\tilde\sigma\in\mathcal{H}_h}}
     - \sum_{\substack{\tilde\mu,\tilde\nu\in\mathcal{H}_h\\\tilde\rho,\tilde\sigma\in\mathcal{H}_e}}
    \right]
 \times \nonumber \\ & & 
     V^{\mathrm{ri}}_{i\tilde\mu} V^{\mathrm{ri}}_{j\tilde\nu}
   \frac{1}{\tilde\epsilon_{\tilde\mu} + \tilde\epsilon_{\tilde\nu} - \tilde\epsilon_{\tilde\rho} - \tilde\epsilon_{\tilde\sigma}}
   V^{\mathrm{ri},-1}_{\tilde\rho i} V^{\mathrm{ri},-1}_{\tilde\sigma j}
 \times \nonumber \\ & & 
   V^{\mathrm{in}}_{\tilde\mu\alpha} V^{\mathrm{in}}_{\tilde\nu\beta}
   \big[ 2\Gamma^{\text{s}}_{\alpha\beta\gamma\delta} - 4\Gamma^{\text{s}}_{\alpha\beta\delta\gamma} \big]
   V^{\mathrm{in},-1}_{\gamma\tilde\rho} V^{\mathrm{in},-1}_{\delta\tilde\sigma}.
   \label{eq:TwoParticleGF:connected:T0:spin:result}
\end{eqnarray}

\subsubsection{Spin-Spin Correlator}

In contrast to the expectation value of $\varvec{S}_i$, the expectation value of $\varvec{S}_i \cdot \varvec{S}_j$ does not automatically
vanish in systems with $\mathrm{SU}(2)$ symmetry. Using
\begin{equation}
 \varvec{\hat S}_i = \sum_{\sigma\sigma'} \mathrm{\hat c}^{\dagger}_{i\sigma} \vec\tau_{\sigma\sigma'} \mathrm{\hat c}^{\vphantom{\dagger}}_{i\sigma'},
\end{equation}
where $\vec\tau$ are the Pauli matrices and the identity
\begin{equation}
 \sum_{k=0}^{3} \tau^k_{\sigma\sigma'} \tau^k_{\bar\sigma\bar\sigma'} = 2 \delta_{\sigma\bar\sigma'} \delta_{\sigma'\bar\sigma},
\end{equation}
we may write
\begin{eqnarray}
 \mathcal{C}^{\mathrm{ss}}_{ij} 
   & := & \big< \varvec{S}_i \cdot \varvec{S}_j \big> \nonumber \\
   &  = & \sum_k \sum_{\sigma\sigma'} \sum_{\bar\sigma\bar\sigma'} \tau^k_{\sigma\sigma'} \tau^k_{\bar\sigma\bar\sigma'} 
          \big<  \mathrm{\hat c}^{\dagger}_{i\sigma}     \mathrm{\hat c}^{\vphantom{\dagger}}_{i\sigma'} 
                 \mathrm{\hat c}^{\dagger}_{j\bar\sigma} \mathrm{\hat c}^{\vphantom{\dagger}}_{j\bar\sigma'}\big> \nonumber \\
   &  = & 2 \sum_{\sigma\sigma'} \big< \mathrm{\hat c}^{\dagger}_{i\sigma}     \mathrm{\hat c}^{\vphantom{\dagger}}_{i\sigma'}
                                    \mathrm{\hat c}^{\dagger}_{j\sigma'}    \mathrm{\hat c}^{\vphantom{\dagger}}_{j\sigma} \big>
           - \big<\mathrm{\hat n}_i^{\vphantom{\dagger}} \mathrm{\hat n}_j^{\vphantom{\dagger}}\big> \nonumber \\
   &  = & 2 \sum_{\sigma\sigma'} \big< \mathrm{\hat c}^{\dagger}_{i\sigma} \mathrm{\hat c}^{\dagger}_{j\sigma'} 
                                       \mathrm{\hat c}^{\vphantom{\dagger}}_{j\sigma} \mathrm{\hat c}^{\vphantom{\dagger}}_{i\sigma'}  \big>
           - \big<\mathrm{\hat n}_i^{\vphantom{\dagger}} \mathrm{\hat n}_j^{\vphantom{\dagger}}\big>
           - 4 \delta_{ij} \big< \mathrm{\hat n}_{i}^{\vphantom{\dagger}} \big> \nonumber \\
   & = & \mathcal{C}^{\mathrm{ss},(2)}_{ij}
            - \big<\mathrm{\hat n}_i^{\vphantom{\dagger}} \mathrm{\hat n}_j^{\vphantom{\dagger}}\big>
            - 4 \delta_{ij} \big< \mathrm{\hat n}_{i}^{\vphantom{\dagger}} \big> \nonumber \\
   &   & + 2 \sum_{\sigma\sigma'} \Big[
              \big< \mathrm{\hat c}^{\dagger}_{j\sigma'} \mathrm{\hat c}^{\vphantom{\dagger}}_{j\sigma} \big>
              \big< \mathrm{\hat c}^{\dagger}_{i\sigma} \mathrm{\hat c}^{\vphantom{\dagger}}_{i\sigma'} \big>
         -    \big< \mathrm{\hat c}^{\dagger}_{i\sigma} \mathrm{\hat c}^{\vphantom{\dagger}}_{j\sigma} \big>
              \big< \mathrm{\hat c}^{\dagger}_{j\sigma'} \mathrm{\hat c}^{\vphantom{\dagger}}_{i\sigma'} \big>
           \Big] \nonumber \\
   & = & \mathcal{C}^{\mathrm{ss},(2)}_{ij}
         + 4 \big<\mathrm{\hat n}_{i\sigma}^{\vphantom{\dagger}}\big> \big<\mathrm{\hat n}_{j\sigma}^{\vphantom{\dagger}}\big>
         - 8 \big<\mathrm{\hat c}_{i\sigma}^\dagger \mathrm{\hat c}_{j\sigma}^{\vphantom{\dagger}}\big>
           \big<\mathrm{\hat c}_{j\sigma}^\dagger \mathrm{\hat c}_{i\sigma}^{\vphantom{\dagger}}\big> \nonumber \\
   &   & \hphantom{\mathcal{C}^{\mathrm{ss},(2)}_{ij}}
         - \big<\mathrm{\hat n}_i^{\vphantom{\dagger}} \mathrm{\hat n}_j^{\vphantom{\dagger}}\big>
         - 4 \delta_{ij} \big< \mathrm{\hat n}_{i}^{\vphantom{\dagger}} \big>.
\end{eqnarray}
Inserting Eq.~(\ref{eq:obs:ddcorr:spin}), several terms cancel and we arrive at
\begin{equation}
 \mathcal{C}^{\mathrm{ss}}_{ij} 
    =  \mathcal{C}^{\mathrm{ss},(2)}_{ij} - \mathcal{C}^{\mathrm{dd},(2)}_{ij}
          - 6 \Big( \big<\mathrm{\hat c}_{i\sigma}^\dagger \mathrm{\hat c}_{j\sigma}^{\vphantom{\dagger}}\big>
           \big<\mathrm{\hat c}_{j\sigma}^\dagger \mathrm{\hat c}_{i\sigma}^{\vphantom{\dagger}}\big>
          + \delta_{ij} \big< \mathrm{\hat n}_{i}^{\vphantom{\dagger}} \big> \Big).
\end{equation}
The expression for $\mathcal{C}^{\mathrm{ss},(2)}_{ij}$ may be derived in the same manner as the expression for $\mathcal{C}^{\mathrm{dd},(2)}_{ij}$. At
finite temperatures, it reads
\begin{eqnarray}
  \mathcal{C}^{\mathrm{ss},(2)}_{ij} & = & 2 \sum_{\alpha\beta\gamma\delta}\sum_{\tilde\mu\tilde\nu\tilde\rho\tilde\sigma}
     V^{\mathrm{ri}}_{i\tilde\mu} V^{\mathrm{ri}}_{j\tilde\nu}
   \frac{1}{\tilde\epsilon_{\tilde\mu} + \tilde\epsilon_{\tilde\nu} - \tilde\epsilon_{\tilde\rho} - \tilde\epsilon_{\tilde\sigma}}
 \times \nonumber \\ & & \hspace{.5em}
   [n_{\mathrm{F}}(\tilde\xi_{\tilde\sigma}) - n_{\mathrm{F}}(\tilde\xi_{\tilde\mu})] [n_{\mathrm{F}}(\tilde\xi_{\tilde\sigma} - \tilde\xi_{\tilde\mu}) - n_{\mathrm{F}}(\tilde\xi_{\tilde\nu})]
 \times \nonumber \\ & & \hspace{.5em}
   [n_{\mathrm{F}}(\tilde\xi_{\tilde\mu} + \tilde\xi_{\tilde\nu} - \tilde\xi_{\tilde\sigma}) - n_{\mathrm{F}}(\tilde\xi_{\tilde\rho})]
   V^{\mathrm{ri},-1}_{\tilde\rho i} V^{\mathrm{ri},-1}_{\tilde\sigma j}
 \times \nonumber \\ & & \hspace{.5em}
   V^{\mathrm{in}}_{\tilde\mu\alpha} V^{\mathrm{in}}_{\tilde\nu\beta}
   \big[ 4\Gamma^{\text{s}}_{\alpha\beta\gamma\delta} - 2\Gamma^{\text{s}}_{\alpha\beta\delta\gamma} \big]
   V^{\mathrm{in},-1}_{\gamma\tilde\rho} V^{\mathrm{in},-1}_{\delta\tilde\sigma}.
   \label{eq:sscorr:finiteT:spin:result}
\end{eqnarray}
As one is often interested in both the occupation number and spin correlators, we note that the expression for the difference between $\mathcal{C}^{\mathrm{ss},(2)}_{ij}$ and $\mathcal{C}^{\mathrm{dd},(2)}_{ij}$
simplifies slightly,
\begin{eqnarray}
 \mathcal{C}^{'\mathrm{ss},(2)}_{ij} & = & \mathcal{C}^{\mathrm{ss},(2)}_{ij} - \mathcal{C}^{\mathrm{dd},(2)}_{ij} \nonumber \\
 & = & 6 \sum_{\alpha\beta\gamma\delta}\sum_{\tilde\mu\tilde\nu\tilde\rho\tilde\sigma}
     V^{\mathrm{ri}}_{i\tilde\mu} V^{\mathrm{ri}}_{j\tilde\nu}
   \frac{1}{\tilde\epsilon_{\tilde\mu} + \tilde\epsilon_{\tilde\nu} - \tilde\epsilon_{\tilde\rho} - \tilde\epsilon_{\tilde\sigma}}
 \times \nonumber \\ & & \hspace{.5em}
   [n_{\mathrm{F}}(\tilde\xi_{\tilde\sigma}) - n_{\mathrm{F}}(\tilde\xi_{\tilde\mu})] [n_{\mathrm{F}}(\tilde\xi_{\tilde\sigma} - \tilde\xi_{\tilde\mu}) - n_{\mathrm{F}}(\tilde\xi_{\tilde\nu})]
 \times \nonumber \\ & & \hspace{.5em}
   [n_{\mathrm{F}}(\tilde\xi_{\tilde\mu} + \tilde\xi_{\tilde\nu} - \tilde\xi_{\tilde\sigma}) - n_{\mathrm{F}}(\tilde\xi_{\tilde\rho})]
   V^{\mathrm{ri},-1}_{\tilde\rho i} V^{\mathrm{ri},-1}_{\tilde\sigma j}
 \times \nonumber \\ & & \hspace{.5em}
   V^{\mathrm{in}}_{\tilde\mu\alpha} V^{\mathrm{in}}_{\tilde\nu\beta}
   \Gamma^{\text{s}}_{\alpha\beta\gamma\delta}
   V^{\mathrm{in},-1}_{\gamma\tilde\rho} V^{\mathrm{in},-1}_{\delta\tilde\sigma}.
   \label{eq:sscorr:finiteT:spin:result:v2}
\end{eqnarray}
At $T = 0$, the expression reads
\begin{eqnarray}
 \mathcal{C}^{'\mathrm{ss},(2)}_{ij} & = & 6 \sum_{\alpha\beta\gamma\delta} \left[
       \sum_{\substack{\tilde\mu,\tilde\nu\in\mathcal{H}_e\\\tilde\rho,\tilde\sigma\in\mathcal{H}_h}}
     - \sum_{\substack{\tilde\mu,\tilde\nu\in\mathcal{H}_h\\\tilde\rho,\tilde\sigma\in\mathcal{H}_e}}
    \right]
 \times \nonumber \\ & & \hspace{1em}
     V^{\mathrm{ri}}_{i\tilde\mu} V^{\mathrm{ri}}_{j\tilde\nu}
   \frac{1}{\tilde\epsilon_{\tilde\mu} + \tilde\epsilon_{\tilde\nu} - \tilde\epsilon_{\tilde\rho} - \tilde\epsilon_{\tilde\sigma}}
   V^{\mathrm{ri},-1}_{\tilde\rho i} V^{\mathrm{ri},-1}_{\tilde\sigma j}
 \times \nonumber \\ & & \hspace{.5em}
   V^{\mathrm{in}}_{\tilde\mu\alpha} V^{\mathrm{in}}_{\tilde\nu\beta}
   \Gamma^{\text{s}}_{\alpha\beta\gamma\delta}
   V^{\mathrm{in},-1}_{\gamma\tilde\rho} V^{\mathrm{in},-1}_{\delta\tilde\sigma}.
   \label{eq:sscorr:T0:spin:result}
\end{eqnarray}

\subsection{Reducting the Hilbert space size: Active-space approximation (\asa)}
\label{sec:orbital-reduction}

The flow equations for the self-energy and the vertex, even in their simplest form
Eqs.~(\ref{eq:flow:Sigma}, \ref{eq:flow:Gamma}),
are still computationally challenging.  
In translationally invariant systems simplifications arise, 
because the vertex only depends on three momenta,
the fourth given by momentum conservation. 
Moreover, one only tracks momenta near the Fermi surface:  
The Brillouin zone is divided into patches each containing
a single tracked momentum and the interaction vertex $\Gamma$ is only
calculated at these momenta. Whenever it needs to be evaluated for other
momenta, the other momentum is replaced by the tracked 
one located within the same patch (coarse graining).\cite{FRGHubbard}
In the absence of periodicity, this kind of patching is not possible, 
since there is no well-defined concept of a Fermi surface.

\begin{figure}[tbp]
   \centering
   \includegraphics[width=.95\linewidth]{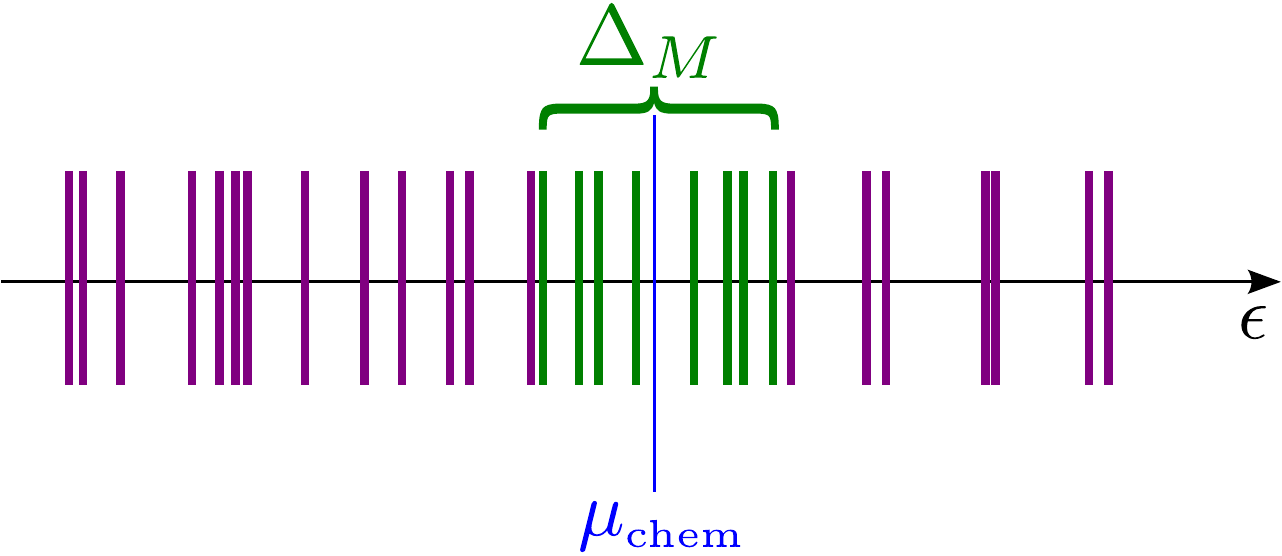}
   \caption[Selection of orbitals for which to renormalize the vertex]{Active space: Selection of $M$ orbitals (green)
around the chemical potential, $\muchem$,
   for which the vertex will be renormalized in the active space approximation (\asa). 
   The self-energy will still be renormalized
   for all $N$ orbitals, including the remaining (purple) ones.}
   \label{fig:state_selection}
\end{figure}

For systems without translational symmetries, we 
here propose an  approach alternative to Fermi-surface patching 
for reducing the number of explicit degrees of freedom.
Similar to the patching scheme, we define an 
``active space'' $\mathcal{H}_M$ of (effective) single-particle 
states near the chemical potential that are kept. 
In our case $\mathcal{H}_M$ simply contains the $M$ 
orbitals closest to the chemical potential, $\muchem$, (e.g. half above and half
below); see Fig.~\ref{fig:state_selection} for details.
We will refer to this approach in 
a loose manner of speaking as ``active-space approximation'' (\asa). 

Within \asa{} external indices of the flow equation for the vertex,
Eq.~(\ref{eq:flow:Gamma}), only refer to a reduced number of
states. In summations over the full single-particle Hilbert space, 
we adopt the approximation scheme
\begin{equation}
 \Gamma^\Lambda_{\alpha\beta\gamma\delta} \rightarrow \left\{
       \begin{array}{ll}
         \Gamma^\Lambda_{\alpha\beta\gamma\delta} & \{\alpha,\beta,\gamma,\delta\} \subseteq \mathcal{H}_M \\
         U_{\alpha\beta\gamma\delta}              & \text{otherwise}
       \end{array}
     \right., \label{eq:definition:replacement-gamma}. 
\end{equation}
To simplify the notation, in the following we label states from
the active space $\mathcal{H}_M$ with barred indices, e.g. $\bar\alpha$, 
whereas states from the full set of orbitals are denoted without
bars, e.g. $\alpha$.

We comment on the choice for $M$ at a given system size. 
As long as mostly the states close to the Fermi energy are important for screening
(as is also assumed in applications of the FRG for translationally invariant systems), 
we can argue that the number of states necessarily kept in $\mathcal{H}_M$, $M$, 
should grow sub-linearly with the total number of orbitals, $N$.
We remind ourselves that in a translationally invariant system, the Fermi 
surface has dimensionality $(d-1)$ within the
$d$-dimensional Brillouin zone. Since the number of states in the Brillouin zone 
grows as $L^d$, but the number of states on a surface
within that space grows as $L^{d-1}$, we suggest the number of states required 
should be proportional to $L^{d-1}$, which can be rewritten as
$L^{d-1} = (L^d)^{(d-1)/d} = N^{1-1/d}$. To the extent that $M$ 
scales the same also for generic systems, we have $M\sim N^{1-1/d}$, 
implying $M\sim N^{1/2}$ in 2D.

In Sec.~\ref{sec:verification} we will establish the efficacy of 
the \asa{} and also revisit the system size scaling.

\subsection{Runaway Flow}
\label{sec:frg:runaway-flow}

At present, one of the main applications of \kfrg{} is the study of phase diagrams, 
because an unbiased view of competing instabilities of
the system is provided. In parameter regimes where the system shows a phase transition, 
the instabilities pertaining to the new phase lead to
``runaway flow'': at a critical scale, $\Lambda_{\text{c}}$, 
the integration of the RG-equations exhibits matrix elements of the interaction 
vertex that diverge. The physical nature of the instability 
reveals itself in what matrix element actually shows 
the strongest divergence. This property of the FRG has been used very
successfully to study the phase diagram of a multitude of systems, 
for an overview see Ref.~\onlinecite{FrgReviewMetzner}.
With \efrg{} one needs to keep in mind that the eigenstate representation is not based on plane waves. 
Therefore, the physics of individual vertex-elements may not be as transparent 
as it is in the clean case. 
Hence, it can be helpful to calculate two-particle correlators at $\Lambda_{\text{c}}$ 
to support interpretations of the precise nature of the instability. 

We mention that cases exist in which competing order parameters influence each other 
(such as antiferromagnetism and $d$-wave superconductivity). 
Strategies how to deal with this situation have been developed within \kfrg{}.
Ideally, one should continue the flow to $\Lambda \to 0$ to obtain information 
about the true phase diagram of the system. 
This may be done in principle, e.g., by introducing an infinitesimal
symmetry-breaking 
term that grows under the RG-flow, as has been done for
superconductivity\cite{SalmhoferBrokenSymmetry}.
Alternatively, one may calculate the flow for the combined Bose-Fermi
system, where fermions were decoupled via a Hubbard-Stratonovich 
transformation.\cite{WetterichHubbard} 

\section{Implementation}
\label{sec:impl}

We implement the FRG procedure in C++, using the Eigen linear algebra library\cite{EigenLibrary} for matrix
products and the HDF5 file format \cite{Hdf5Library} for storage. We employ the OpenMP 3.1 standard \cite{OpenMP3.1}
for parallelization.

\subsection{Computational Details}
\label{sec:impl:efficient-traces}

The computational complexity of the self-energy flow, Eqs.~(\ref{eq:flow:Sigma},\ref{eq:flow:Sigma:WithSpin},\ref{eq:flow:Sigma:finiteT}),
is given by $\OO{N^4}$ -- two loops for each of the outer indices, two loops for the contraction with the non-diagonal
single-scale propagator. At first glance the flow of the vertex, e.g., Eq.~(\ref{eq:flow:Gamma}) appears to have a
complexity of $\OO{N^8}$. However, one may define intermediate products, $I^{\text{c},\pm}, I^{\text{ph},\pm}$,
\begin{eqnarray}
 I^{\text{c},+}_{\mu\rho\bar\gamma\bar\delta} & = & \sum_{\nu} P^{\Lambda,s}_{\mu\nu}(\Lambda) \Gamma^\Lambda_{\nu\rho\bar\gamma\bar\delta} \label{eq:definition:intermediate:cooper:plus} \\
 I^{\text{c},-}_{\bar\alpha\bar\beta\rho\mu} & = & \sum_{\sigma} P^{\Lambda,s}_{\rho\sigma}(-\Lambda) \Gamma^\Lambda_{\bar\alpha\bar\beta\sigma\mu} \label{eq:definition:intermediate:cooper:minus} \\
 I^{\text{ph},\pm}_{\bar\alpha\nu\bar\gamma\sigma} & = & \sum_{\rho} \Gamma^\Lambda_{\bar\alpha\nu\bar\gamma\rho} P^{\Lambda,s}_{\rho\sigma}(\pm\Lambda), \label{eq:definition:intermediate:particle-hole}
\end{eqnarray}
where each of these partial diagrams has a complexity of $\OO{N^5}$. The flow equation for the vertex now reads
\begin{eqnarray}
  \ddLambda \Gamma^\Lambda_{\bar\alpha\bar\beta\bar\gamma\bar\delta} & = & - \frac{1}{2\pi} \sum_{\mu\rho} \left\{
    I^{\text{c},+}_{\mu\rho\bar\gamma\bar\delta} I^{\text{c},-}_{\bar\alpha\bar\beta\rho\mu} 
\right.\nonumber\\ & & \left. \hspace{2em}
 + I^{\text{ph},+}_{\bar\alpha\mu\bar\gamma\rho} I^{\text{ph},+}_{\bar\beta\rho\bar\delta\mu}
 + I^{\text{ph},-}_{\bar\alpha\mu\bar\gamma\rho} I^{\text{ph},-}_{\bar\beta\rho\bar\delta\mu}
\right.\nonumber\\ & & \left. \hspace{2em}
 - I^{\text{ph},+}_{\bar\beta\mu\bar\gamma\rho}  I^{\text{ph},+}_{\bar\alpha\rho\bar\delta\mu}
 - I^{\text{ph},-}_{\bar\beta\mu\bar\gamma\rho}  I^{\text{ph},-}_{\bar\alpha\rho\bar\delta\mu}
\right\}, \label{eq:flow:Gamma:with-intermediates}
\end{eqnarray}
with a computational complexity of $\OO{N^6}$. In the case of $M < N$, using the replacement in Eq.~(\ref{eq:definition:replacement-gamma}),
this reduces to $\OO{N^3 N^3}$ for the calculation of the intermediates and to $\OO{N^2 M^4}$ for the trace.

Repeating our argument from Sec.~\ref{sec:orbital-reduction} that $M \propto \sqrt{N}$, we expect a scaling of $\OO{N^4}$ for two-dimensional
systems.

\subsubsection{Efficient Trace Evaluation}

In order to evaluate the temporary products for the flow of the vertex,
Eqs.~(\ref{eq:definition:intermediate:cooper:plus},\ref{eq:definition:intermediate:cooper:minus},\ref{eq:definition:intermediate:particle-hole}),
it is advantageous to rewrite the expression in terms of a matrix product, e.g.
\begin{equation}
 I^{\text{c},+}_{\mu,(\rho\bar\gamma\bar\delta)} = \sum_{\nu} P^{\Lambda,s}_{\mu\nu}(\Lambda) \Gamma^\Lambda_{\nu,(\rho\bar\gamma\bar\delta)},
\end{equation}
where we interpret $(\rho\bar\gamma\bar\delta)$ as a single index, because modern generic matrix-matrix multiplication (GEMM) kernels are
highly optimized and perform far better than a simple sum. For the cases where we calculate the renormalization of the vertex for all states,
this is trivial. Note that for some equations one needs to retain a copy of the vertex with transposed indices to be able to do this. Since
our implementation is typically not constrained by the available memory but rather the available processing power, this tradeoff is advantageous.
\begin{figure}[tbp]
  \centering
  \includegraphics[height=6cm]{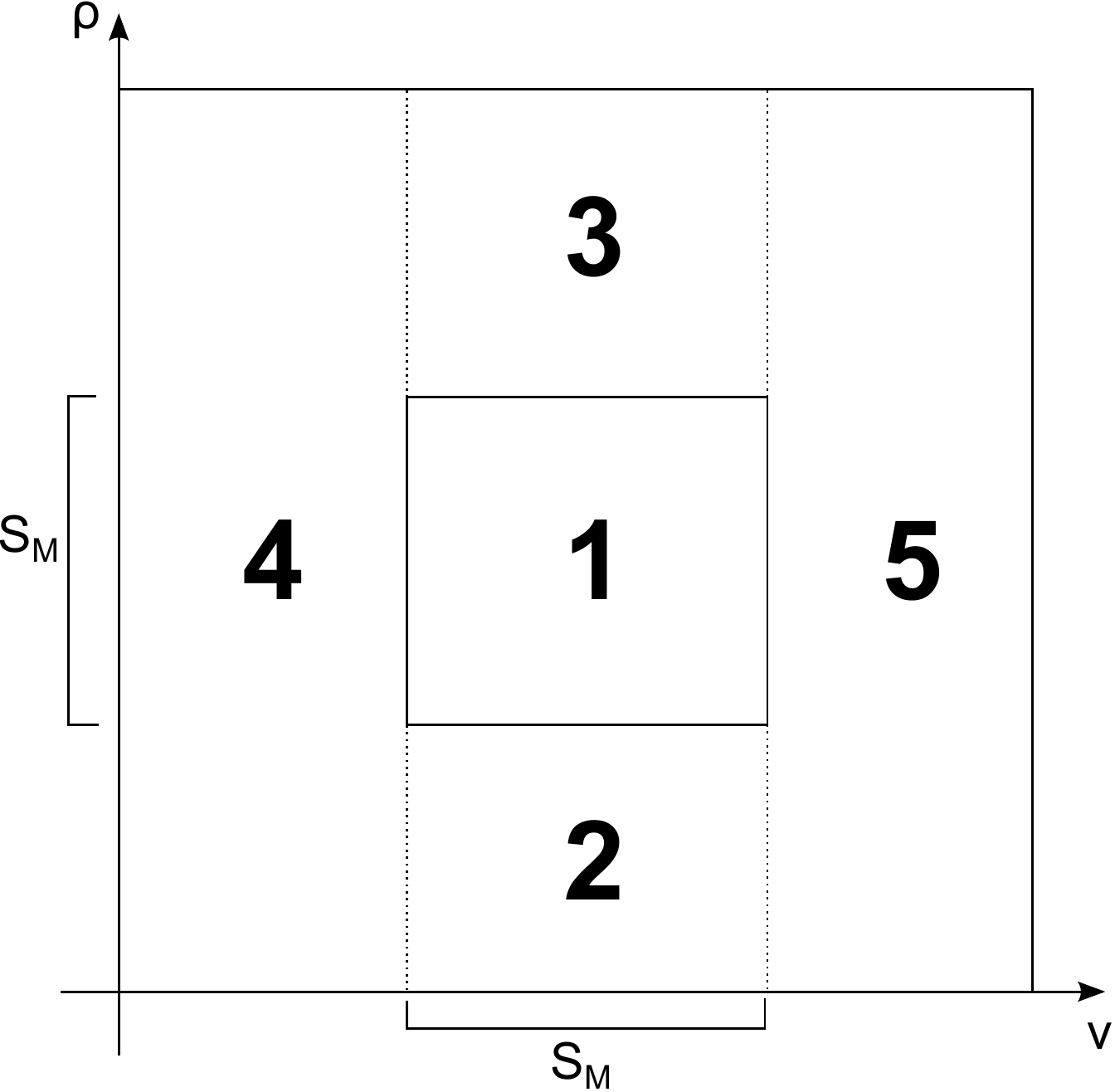}
  \caption[GEMM kernel subdivision for intermediate products]{\label{fig:impl:gemm-statesel} The subdivision of the GEMM kernel for the
  intermediate product $I^{\text{c},+}$ in the $\nu$ and $\rho$ indices. The regions one to five in the diagram correspond to the terms
  of Eqs.~(\ref{eq:impl:gemmsplit:tV},\ref{eq:impl:gemmsplit:tU0s},\ref{eq:impl:gemmsplit:tU0l},\ref{eq:impl:gemmsplit:tUs},\ref{eq:impl:gemmsplit:tUl}),
  respectively.}
\end{figure}
It is trickier to approximate the vertex according to Eq.~\eqref{eq:definition:replacement-gamma}. Instead of rewriting the entire expression
in terms of a GEMM kernel, we need to perform the loop on the external indices explicitly. We may then split the resulting matrix product
into five parts. Taking for example Eq.~\eqref{eq:definition:intermediate:cooper:plus} and using that $\mathcal{H}_M$ is the subset of
states for which the vertex is renormalized, we have
\begin{eqnarray}
 I^{\text{c},+}_{\mu\rho\bar\gamma\bar\delta} & = &
    \hphantom{+} \hspace{-1em}\sum_{\bar\nu\in \mathcal{H}_M} P^{\Lambda,s}_{\mu\bar\nu}(\Lambda) \Gamma^\Lambda_{\bar\nu\rho\bar\gamma\bar\delta}  \hspace{1em}[\rho\in \mathcal{H}_M] \label{eq:impl:gemmsplit:tV} \\
           & & \hspace{-1em} + \sum_{\bar\nu\in \mathcal{H}_M} P^{\Lambda,s}_{\mu\bar\nu}(\Lambda) U_{\bar\nu\rho\bar\gamma\bar\delta} \hspace{1em}[\rho < \min(\mathcal{H}_M)] \label{eq:impl:gemmsplit:tU0s} \\
           & & \hspace{-1em} + \sum_{\bar\nu\in \mathcal{H}_M} P^{\Lambda,s}_{\mu\bar\nu}(\Lambda) U_{\bar\nu\rho\bar\gamma\bar\delta} \hspace{1em}[\rho > \max(\mathcal{H}_M)] \label{eq:impl:gemmsplit:tU0l} \\
           & & \hspace{-1em} + \sum_{\nu < \min(\mathcal{H}_M)}  P^{\Lambda,s}_{\mu\nu}(\Lambda) U_{\nu\rho\bar\gamma\bar\delta} \label{eq:impl:gemmsplit:tUs} \\
           & & \hspace{-1em} + \sum_{\nu > \max(\mathcal{H}_M)}  P^{\Lambda,s}_{\mu\nu}(\Lambda) U_{\nu\rho\bar\gamma\bar\delta}. \label{eq:impl:gemmsplit:tUl}
\end{eqnarray}
We assume here that the non-interacting states are ordered in energy. The five subexpressions may then be written in terms of GEMM kernels with
rectangular blocks of the matrices $P^{\Lambda,s}$ and $U_{\cdot\cdot\bar\gamma\bar\delta}$. Figure~\ref{fig:impl:gemm-statesel} shows the
division into these terms in the plane of $\nu$ and $\rho$ indices.

There are no standard kernels for trace evaluation, e.g. Eq.~\eqref{eq:flow:Gamma:with-intermediates}, hence we implement that directly in terms
of a loop.

\subsubsection{Parametrization of the Flow Equations}

We use an exponential parametrization for the flow equations, Eqs.~(\ref{eq:flow:Sigma},\ref{eq:flow:Gamma:with-intermediates}),
\begin{equation}
 \Lambda = \Lambda_0 \mathrm{e}^{-l \Delta s}, \hspace{2em} l \in \mathds{N},
\end{equation}
where $\Lambda_0$ is the initial $\Lambda$ at which the flow starts and $l$ is our discretizing iteration number. This parametrization has
the advantage that it captures the physics close to the Fermi energy well, as the integration mesh gets denser, while still being
relatively fast in reaching that point. Both flow equations are of the form
\begin{equation}
  \ddLambda A(\Lambda) = -\frac{1}{2\pi} B(\Lambda).
\end{equation}
allowing for a trivial discretization,
\begin{equation}
 A(\Lambda(l+1)) = A(\Lambda(l)) + \frac{\Lambda(l)\Delta s}{2\pi} B(\Lambda(l)),
\end{equation}
assuming that $\Delta s$ is sufficiently small. In the following calculations we have chosen
the parameters $\Delta s = 0.02$ and $\Lambda_0 = 40$. Unless we encounter a divergence in the flow, we stop as soon as
$\Lambda < 10^{-4}$ (giving a total of $l_{\text{max}} = 645$ iterations).

\subsection{Chemical Potential}

We would like to keep the number of particles fixed to study the system at a given filling fraction. Since our flow modifies the real part of the self-energy,
we need to constantly adjust the chemical potential during the renormalization procedure.

At $T = 0$ we diagonalize the matrix $H_0 + \Sigma^\Lambda$ to obtain the updated quasi-particle energies for a given $\Lambda$ (including the initial $\Lambda_0$,
since $\Sigma^{\Lambda_0} \neq 0$). We choose our chemical potential to be
\begin{equation}
 \muchem^\Lambda = \frac{1}{2}(\tilde\epsilon^\Lambda_{N_{\text{e}}+1} + \tilde\epsilon^\Lambda_{N_{\text{e}}}),
 \label{eq:impl:muchem:T0}
\end{equation}
where $\tilde\epsilon^\Lambda_{N_{\text{e}}}$
is the energy of the highest occupied quasi-particle state and $\tilde\epsilon^\Lambda_{N_{\text{e}}}$ the energy of the lowest unoccupied quasi-particle state.

At $T > 0$ the value of $\muchem^\Lambda$ follows as usual from the solution to the equation
\begin{equation}
 N_{\text{e}} = \sum_{\tilde\epsilon^\Lambda_{\tilde\alpha} < \muchem^\Lambda} n_{\mathrm{F}}(\tilde\epsilon_{\tilde\alpha} - \muchem^\Lambda),
 \label{eq:impl:muchem:finiteT:tosolve}
\end{equation}
where $N_{\text{e}}$ is the number of electrons and $\tilde\epsilon^\Lambda_{\tilde\alpha}$ are the quasi-particle energies for
a given $\Lambda$, i.e. the eigenvalues of $H_0 + \Sigma^\Lambda$.

\subsection{Correlators}
\label{sec:impl:correlators}

Starting from  Eq.~\eqref{eq:TwoParticleGF:connected:T0:result}, we first transform the vertex into the $\Lambda$-dependent
quasi-particle basis,
\begin{equation}
 \tilde\Gamma^\Lambda_{\tilde\mu\tilde\nu\tilde\rho\tilde\sigma} = \sum_{\alpha\beta\gamma\delta}
 V^{\mathrm{in}}_{\tilde\mu\alpha} V^{\mathrm{in}}_{\tilde\nu\beta}
   \Gamma^{\Lambda}_{\alpha\beta\gamma\delta}
   V^{\mathrm{in},-1}_{\gamma\tilde\rho} V^{\mathrm{in},-1}_{\delta\tilde\sigma}.
\end{equation}
We exploit fast matrix multiplication routines to perform these basis transforms. As these routines require us to group either
the three left- or rightmost indices together, we first transform the vertex in $\alpha$ and $\delta$, then transpose it to
have $\beta$ as the first index and $\gamma$ as the last index, and apply the final pair of transformations, yielding the
following sequence of steps:
\begin{eqnarray}
 \Gamma^{\Lambda,(1)}_{\tilde\mu\beta\gamma\delta} & = &
     \sum_{\alpha} V^{\mathrm{in}}_{\tilde\mu\alpha} \Gamma^{\Lambda}_{\alpha\beta\gamma\delta} \\
 \Gamma^{\Lambda,(2)}_{\tilde\mu\beta\gamma\tilde\sigma} & = &
     \sum_{\delta} \Gamma^{\Lambda,(1)}_{\alpha\beta\gamma\delta} V^{\mathrm{in},-1}_{\delta\tilde\sigma} \\
 \Gamma^{\Lambda,(3)}_{\beta\tilde\mu\tilde\sigma\gamma} & = &
     \Gamma^{\Lambda,(2)}_{\tilde\mu\beta\gamma\tilde\sigma} \\
 \Gamma^{\Lambda,(4)}_{\tilde\nu\tilde\mu\tilde\sigma\gamma} & = &
     \sum_{\tilde\nu} V^{\mathrm{in}}_{\tilde\nu\beta} \Gamma^{\Lambda,(3)}_{\beta\tilde\mu\tilde\sigma\gamma} \\
 \tilde\Gamma^\Lambda_{\tilde\mu\tilde\nu\tilde\rho\tilde\sigma} & = &
     \sum_{\tilde\rho} \Gamma^{\Lambda,(4)}_{\tilde\nu\tilde\mu\tilde\sigma\gamma} V^{\mathrm{in},-1}_{\gamma\tilde\rho}
\end{eqnarray}
We do not need to transpose the final result because of the symmetry of $\Gamma$. 
If our ``active space'' approximation (\asa) is used,
Eq.~\eqref{eq:definition:replacement-gamma}, we employ rectangular submatrices of the
$V^{\mathrm{in}}$, since $\Gamma$ is only of size $\mathds{C}^{M^4}$ but $\tilde\Gamma$ needs to be of size
$\mathds{C}^{N^4}$.

Within \asa{} a decomposition similar to the one used in the flow equations,
Eqs.~(\ref{eq:impl:gemmsplit:tV}-\ref{eq:impl:gemmsplit:tUl}),
is not useful here, as a single matrix multiplication already decomposes into 5 products. Instead, we transform the entire bare
interaction, $U$, in the full Hilbert space, and additionally transform $\Gamma - U$ in the activate space and add the results
together in the end.

We then proceed to multiply the transformed vertex by the energy denominator of Eq.~\eqref{eq:TwoParticleGF:connected:T0:result},
\begin{equation}
 \tilde\Gamma^{\Lambda,\text{div}}_{\tilde\mu\tilde\nu\tilde\rho\tilde\sigma} =
 \tilde\Gamma^\Lambda_{\tilde\mu\tilde\nu\tilde\rho\tilde\sigma}
 \frac{1}{\tilde\epsilon_{\tilde\mu} + \tilde\epsilon_{\tilde\nu} - \tilde\epsilon_{\tilde\rho} - \tilde\epsilon_{\tilde\sigma}}.
 \label{eq:impl:ddcorr:div}
\end{equation}

Finally, we need to transform to the target basis and select the proper orbitals. At $T = 0$, we have
\begin{eqnarray}
 \mathcal{C}^{\mathrm{dd-pre},(2)}_{i\tilde\nu\tilde\sigma} & = &
    \left[ \sum_{\substack{\tilde\mu\in\mathcal{H}_e\\\tilde\rho\in\mathcal{H}_h}} - \sum_{\substack{\tilde\mu\in\mathcal{H}_h\\\tilde\rho\in\mathcal{H}_e}} \right]
    V^{\mathrm{ri}}_{i\tilde\mu}
    \tilde\Gamma^{\Lambda,\text{div}}_{\tilde\mu\tilde\nu\tilde\rho\tilde\sigma}
    V^{\mathrm{ri},-1}_{\tilde\rho i},
    \label{eq:impl:ddcorr:pre-final} \\
 \mathcal{C}^{\mathrm{dd},(2)}_{ij} & = &
    \left[ \sum_{\substack{\tilde\nu\in\mathcal{H}_e\\\tilde\sigma\in\mathcal{H}_h}} - \sum_{\substack{\tilde\nu\in\mathcal{H}_h\\\tilde\sigma\in\mathcal{H}_e}} \right]
    V^{\mathrm{ri}}_{j\tilde\nu}
       \mathcal{C}^{\mathrm{dd-pre},(2)}_{i\tilde\nu\tilde\sigma}
    V^{\mathrm{ri},-1}_{\tilde\sigma j}.
    \nonumber \\ & &
    \label{eq:impl:ddcorr:final}
\end{eqnarray}

Because we transform into the basis of the quasi-particles for a given $\Lambda$, the transformation matrices
$V^{\mathrm{in}}$ are $\Lambda$-dependent and the contribution 
from the bare interaction, $U$, cannot be calculated just once initially. This means that
for each $\Lambda$ the density-density correlator incurs a cost of $\OO{N^5}$. Eq.~\eqref{eq:impl:ddcorr:div} has a complexity
of $\OO{N^4}$ and Eq.~\eqref{eq:impl:ddcorr:pre-final} a complexity of $\OO{N^5}$. This cannot be simplified further without
additional approximations, making it the most expensive object to calculate.

Fortunately, the density-density-correlator is not actually required for the flow of the vertex or the self-energy. Therefore,
unless we see a divergence in
our flow in $\Lambda$, we calculate it only once at the very end of the flow. In case a divergence
is seen, we perform a backtracking procedure: while we don't store the vertex for all iteration steps, we do keep it for
the last $n_{\text{bt}}$ iterations. Once we detect a divergence, we reset the system
to the current iteration minus $n_{\text{bt}}$ steps (typically 10) and calculate the density-density correlator at that
iteration step and proceed to the next iteration again. This is performed for a total of $n_{\text{dv}} \le n_{\text{bt}}$
iterations (typically 1 or 2), where we don't need to recalculate the flow but can just use the known self-energy and the vertex.

\section{Verification - Tests on the spinless Hubbard model}
\label{sec:verification}

\begin{figure}[tbp]
 \includegraphics[width=.95\linewidth]{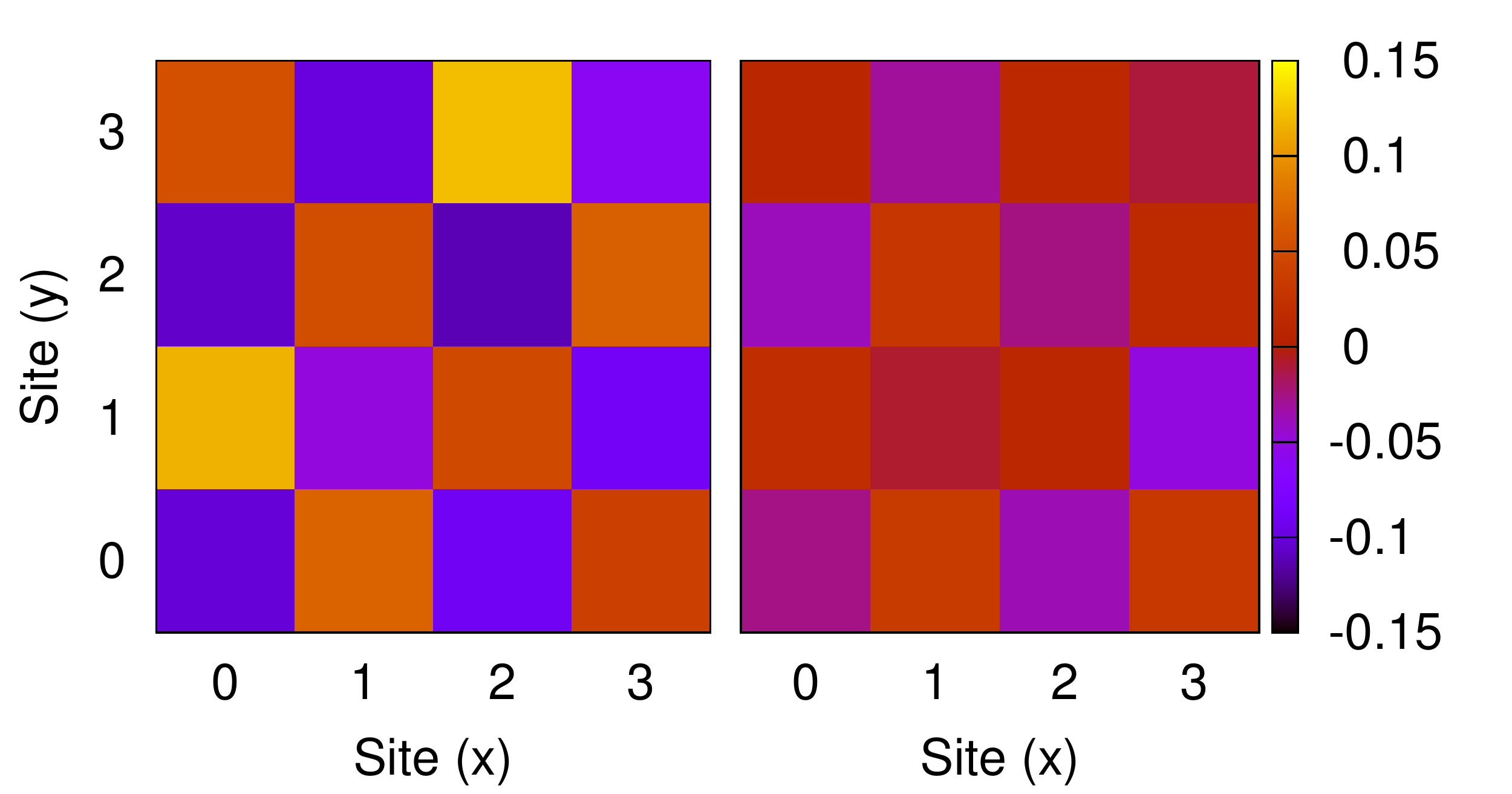}
 \caption[Real space density compared with ED]{\label{fig:verification:w0.1-u0.01:density:cmp}
 Comparison of the particle density, $n(\varvec r)$, calculated from ED ($n_\text{ED}$) and 
 \efrg{} ($n_\text{\efrg}$) for a single disorder realization at $U = 0.01$ and $W = 0.1$.
 Left: Normalized relative deviation $(n_\text{ED}-n_0)/n_0U$, where $n_0$ 
 denotes the density for the same disorder realization at $U{=}0$. 
 Right: $(n_\text{ED}-n_\text{\efrg})/n_0U$.
 }
\end{figure}

In this section we test our implementation applying it to disordered spinless Hubbard model.
We compare results from \efrg{} for the quasiparticle energies and the particle density to
the exact diagonalization (ED) in 2D and to the density matrix renormalization group (DMRG) in 1D.

The corresponding Hamiltonian reads
\begin{eqnarray}
 \hat H & = & -t \sum_{<ij>} \mathrm{\hat c}_i^\dagger \mathrm{\hat c}_j
   + \sum_{i} \delta\epsilon_i \mathrm{\hat n}_i
   + U \sum_{<ij>} \mathrm{\hat n}_i \mathrm{\hat n}_j,
   \label{eq:verification:hamiltonian}\label{e124}
\end{eqnarray}
where $t$ is the hopping parameter, $U$ the interaction strength and the $\delta\epsilon_i$ the on-site energies, which are
chosen at random from a box distribution with width $W$ centered around $\epsilon = 0$.
\footnote{In 2D this model could be realized in terms of a strongly screened two-dimensional electron gas with a strong in-plane
magnetic field. This would polarize all of the spins due to the Zeemann effect, but have no orbital contribution.}
In all calculations we will be working at half-filling. All energies will be measured in units of $t$.

\subsection{\efrg{} vs. ED for square lattices}
\label{sec:disorder:2d}

In this section we test our implementation of the 
\efrg{} equations. To this end, we work with 
small systems, so ED is feasible and there is no need to 
apply the \asa{}. Specifically, we consider the model Hamiltonian of
Eq.~\eqref{eq:verification:hamiltonian} 
on a $4{\times}4$ square lattice with $N{=}16$
sites and periodic boundary conditions at half filling, $\nu=1/2$. 
The details of our ED-implementation are given in App.~\ref{app:ED}. 
\paragraph{Density}

\begin{figure}[tbp]
  \includegraphics[width=.95\linewidth]{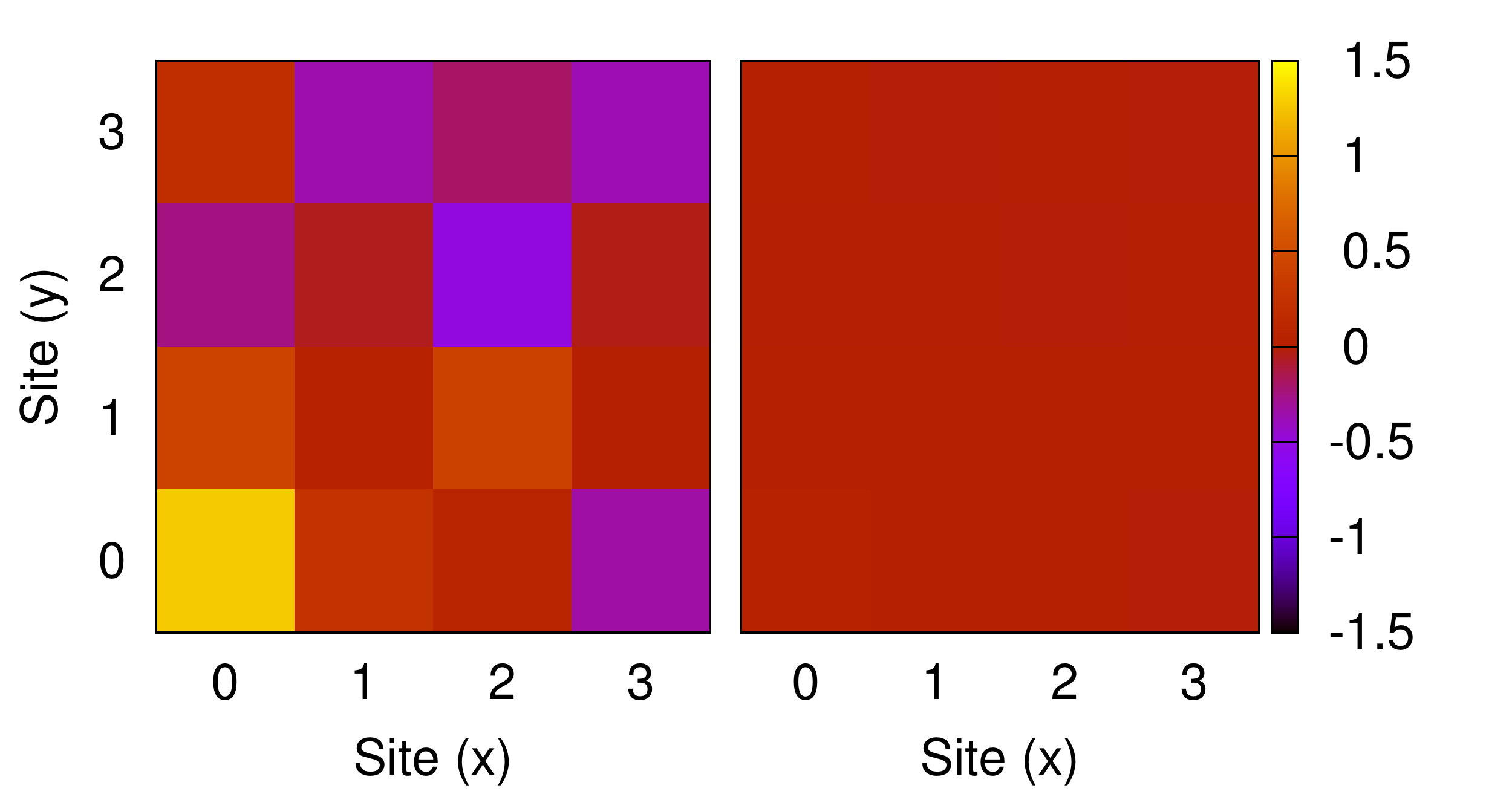}
  \caption[Comparison with ED at stronger disorder]{\label{fig:verification:w5-u0.1:density:cmp}\label{f5}
 Plot similar to Fig. \ref{fig:verification:w0.1-u0.01:density:cmp}
 with $U = 0.1$ and $W = 5$.}
\end{figure}

Fig.~\ref{fig:verification:w0.1-u0.01:density:cmp} (left) displays the interaction induced 
shift of the particle density as it is obtained for a typical disorder realization
at very weak interactions and disorder $U{=}0.01, W{=}0.1$. 
To highlight the density response, we have divided the relative displacement 
by $U$. We obtain a checkerboard pattern
that we interpret as a precursor to the system ordering in a
charge-density wave (CDW). In the absence of disorder there is a two-fold 
degeneracy associated with the placement of the wave. The pattern 
is visible in 
our calculation due to the disorder which breaks this symmetry.
As seen from Fig.~\ref{fig:verification:w0.1-u0.01:density:cmp} (right) 
the density response to very small values of $U$ is reproduced by 
the \efrg{} reasonably well with a typical error of about 30\%.

A comparison at stronger interaction and disorder is given in Fig.
\ref{f5} where $U{=}0.1$ and where the disorder potential of the previous
realization has been recycled, but multiplied with a factor of fifty
corresponding to $W=5.0$.

\paragraph{Quasiparticle energies}

\begin{figure}[tbp]
 \centering
 \includegraphics[width=.8\linewidth]{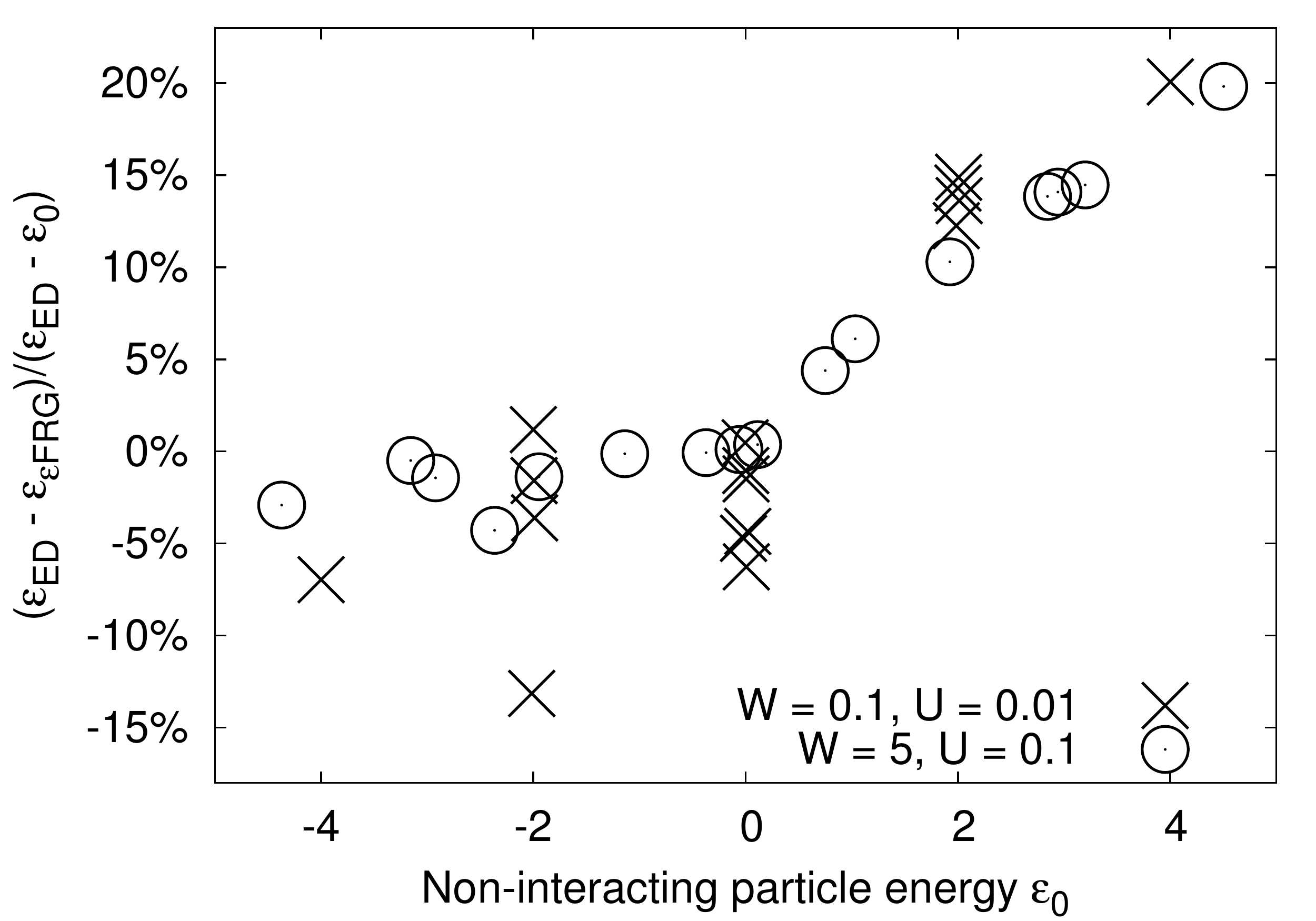}
 \caption[Quasiparticle energies compared with ED]{\label{fig:verification:qpe}
 Comparison of the quasiparticle energies obtained with \efrg{} and ED, normalized by the interaction-induced
 shift, $(\epsilon_{\text{ED}} - \epsilon_{\text{\efrg}})/(\epsilon_{\text{ED}} - \epsilon_0)$, for the same
 systems as in Fig.~\ref{fig:verification:w0.1-u0.01:density:cmp} (crosses) and Fig.~\ref{f5} (circles), respectively.}
\end{figure}

We also compare the spectral properties, i.e. the quasiparticle energies, for both systems,
see Fig.~\ref{fig:verification:qpe}.
\footnote{To obtain the quasiparticle energies in the ED case, we calculate the spectral function
utilizing the truncated Chebyshev expansion discussed in App.~\ref{app:ED}, where we have kept $10^5$ Chebyshev moments. With an
artifical broadening ($2\cdot 10^{-3}t$) to ensure the validity of the truncation of the expansion, the resulting density of states has
been fitted against Lorentzians (with a maximum relative error of the position always below $10^{-6}t$ for each peak).}
The ordinate shows the energies of the corresponding non-interacting system, 
i.e. of $\hat H_0$. At low disorder, $W =
0.1$, the
degeneracies of the clean system are only slightly lifted, 
hence the crosses in Fig.~\ref{fig:verification:qpe} appear
in groups.
The vertical spreading of these groups is seen to be larger 
than for the case with stronger disorder, $W{=}5$. 
We attribute the larger error for the near-degenerate 
situation to the fact that our formulation of the \efrg{} assumes that $\hat H_0$ 
is non-degenerate and becomes singular, otherwise.

We observe that the normalized deviations between \efrg{} and ED 
are approximately independent of the interaction strength $U$.
For the occupied states below the chemical potential, 
$\muchem\approx 0$, 
the error depends very weakly on energy with a typical error smaller 
than 5\%. In contrast, the deviations keep growing for the 
unoccupied levels reaching values of 20\% near the band edge.

\subsection{\hTrunc (\asa)}

As has been discussed in Sec.~\ref{sec:orbital-reduction}, we will consider the renormalized vertex within an active space of $M < N$
states. In this section we test the sensitivity of $n(\varvec r)$ and
the spectral function to variation of $M$.
To this end we will use a $6{\times}6$ square lattice with periodic boundary conditions,
so $N{=}36$. Each system is calculated twice, with the full $M{=}36$, and with $M{=}16$.

\begin{figure}[bp]
  \includegraphics[width=.95\linewidth]{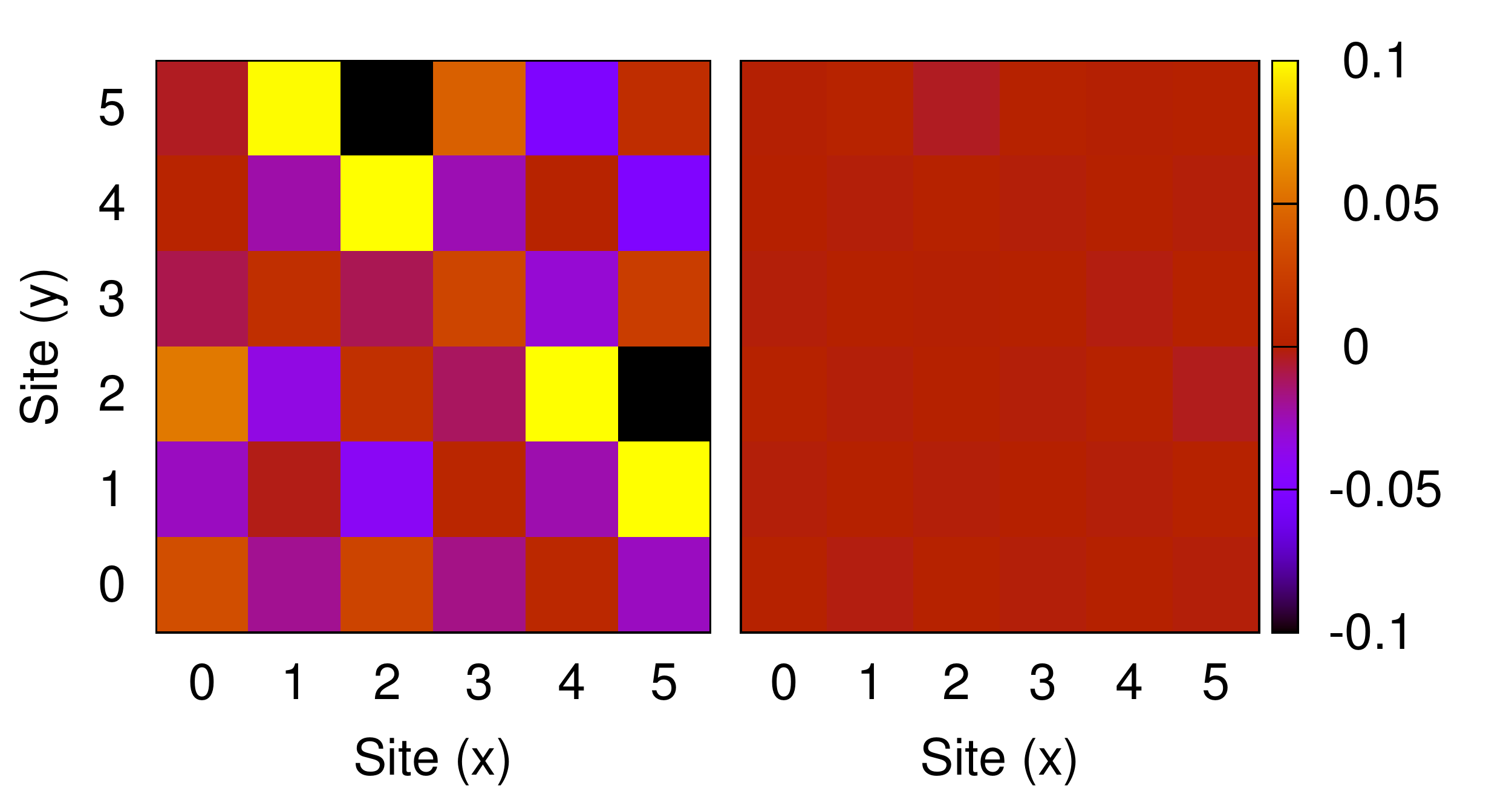}
  \caption[Real space density comparison between FRG variants]{\label{fig:verification:36:density}\label{f6}
  Testing the \asa{} via density calculations at $N{=}36$ with $U{=}0.1$ 
  for a given disorder realization at $W=1.0$. 
  Left: $(n_\text{\efrg}{-}n_0)/U n_0$. Right: $(n_\text{\efrg}{-}n_\text{\asa})/U n_0$ where $M=16$
  has been used in the \asa{}-calculation. 
  }
\end{figure}

The real space density at $U = 0.01$ for a specific disorder realization at $W=1$
is shown in Fig.~\ref{fig:verification:36:density}.
We see that there is a very good agreement between the density profiles of both methods, 
validating our approach at least for small system sizes and moderate interaction strengths.  

\begin{figure}[t]
  \centering
  \includegraphics[width=0.85\linewidth]{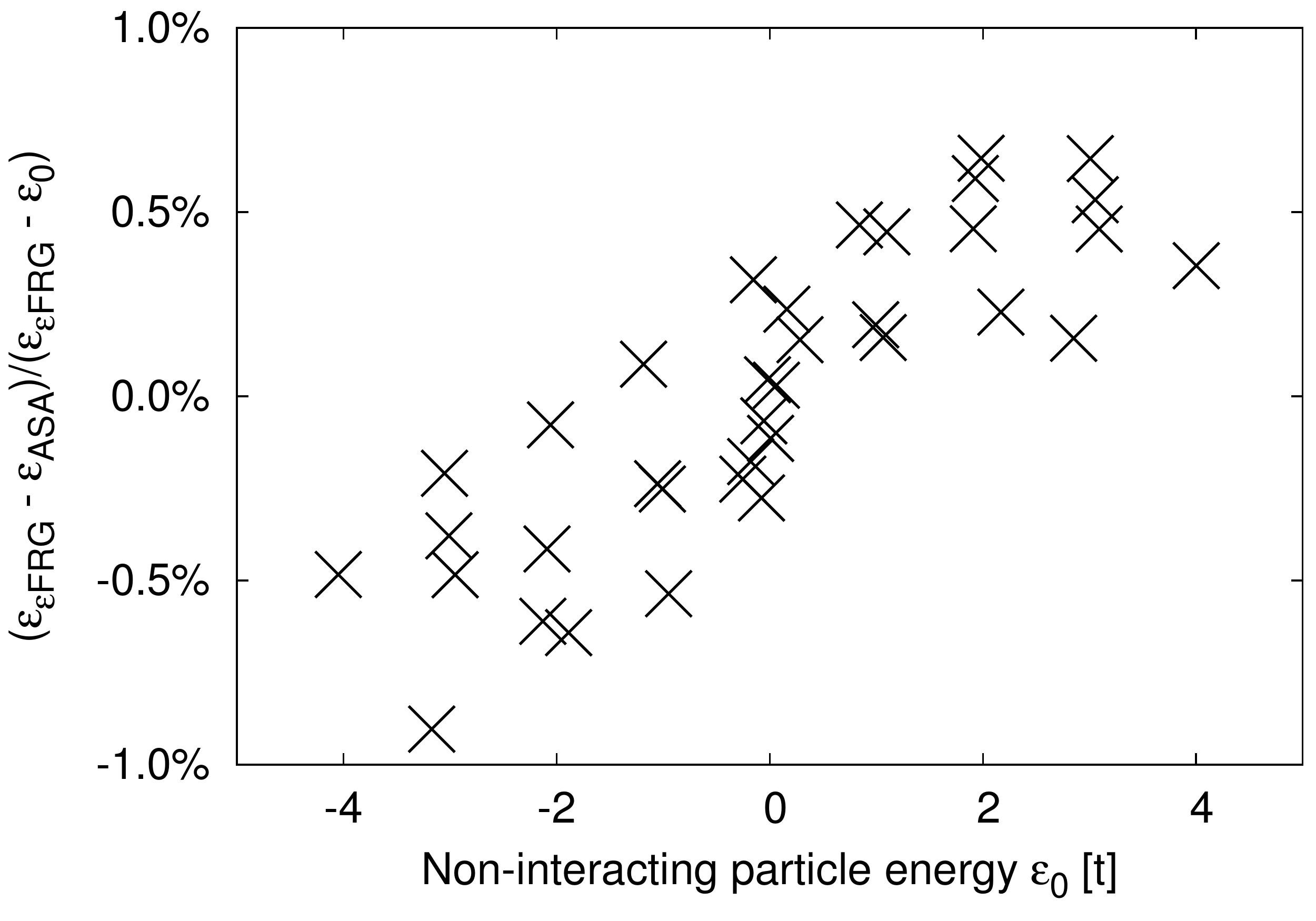}
  \caption[Spectral function comparison]{\label{fig:verification:36:w1-u0.01:dos}\label{f7}
  Difference of the quasi-particle energies obtained with \asa{} ($\epsilon_\text{\asa}$, $M{=}16$) and without
($\epsilon_\text{\efrg}$, $N{=}M{=}36$)
  normalized by the interaction induced shift:
$(\epsilon_\text{\efrg}-\epsilon_\text{\asa})/(\epsilon_\text{\efrg}{-}\epsilon_0)$. The same sample was used  
  as in the previous Fig. \ref{fig:verification:36:density}.
 }
\end{figure}

We also compare the quasi-particle energies as obtained from \efrg{} for both 
choices of $M{=}36, 16$. Fig.~\ref{fig:verification:36:w1-u0.01:dos} shows 
the normalized difference of both spectra. As can be seen, the overall performance 
of \asa{} is acceptable with a relative error of about 0.5\% for quasi-particle 
energies close to the Fermi level. Remarkably, the error does not exceed 
1\% even for states outside of the active space.
\subsection{\efrg{} vs. DMRG for chains}

As a second, independent line of testing we  also compare the 
results from \efrg{} with DMRG calculations. 
To this end we consider the same Hamiltonian
\eqref{eq:verification:hamiltonian}
as before, but now the geometry represents a short chain of $L = 16$ sites. 
In the \efrg{} we keep $N{=}16{=}M$. 
At a given, fixed disorder configuration with $W = 0.2$
we compare the particle density for two different interaction strengths, $U = 0.2$ and
$U = 1.5$.

\begin{figure}[b]
  \centering
  \includegraphics[width=.85\linewidth]{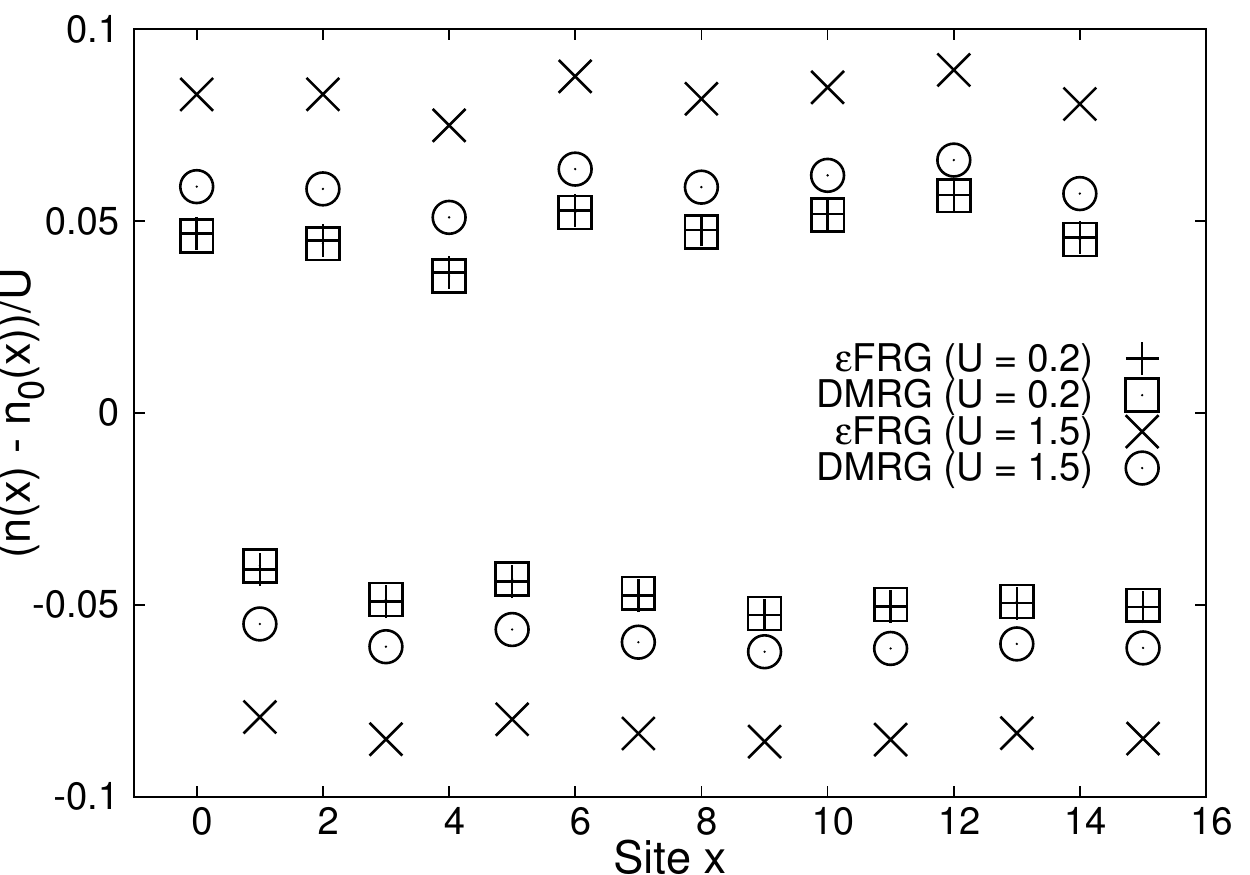}
  \caption[Real space density comparison between FRG and DMRG]{\label{fig:verification:1d:lowU}\label{f8}
  Normalized interaction induced density response, $(n_\text{X}-n_0)/U$, of a $16$ site chain obtained with DMRG
  (empty squares an circles) and FRG (crosses) at $U = 0.2$ and $U = 1.5$; $n_0$ denotes the non-interacting density.
}
\end{figure}

Fig.~\ref{fig:verification:1d:lowU} displays the response of the density 
when switching on $U$ as obtained with both methods. 
At smaller interaction values, $U = 0.2$, 
the \efrg{} reproduces the DMRG results quantitatively
with errors in the percent-regime.
When the interaction reaches values of the order of the band-width, $2t$, 
larger deviations occur reaching values of up to 50\%. 
The systematic overshooting that is observed in the data, we 
tentatively attribute to a lack of screening related to the static approximation.

\subsection{Detecting the CDW state with \efrg{} and \asa{}}
\label{sec:results:cdw-frg}

\begin{figure}[b]
  \centering
  \includegraphics[width=.9\linewidth]{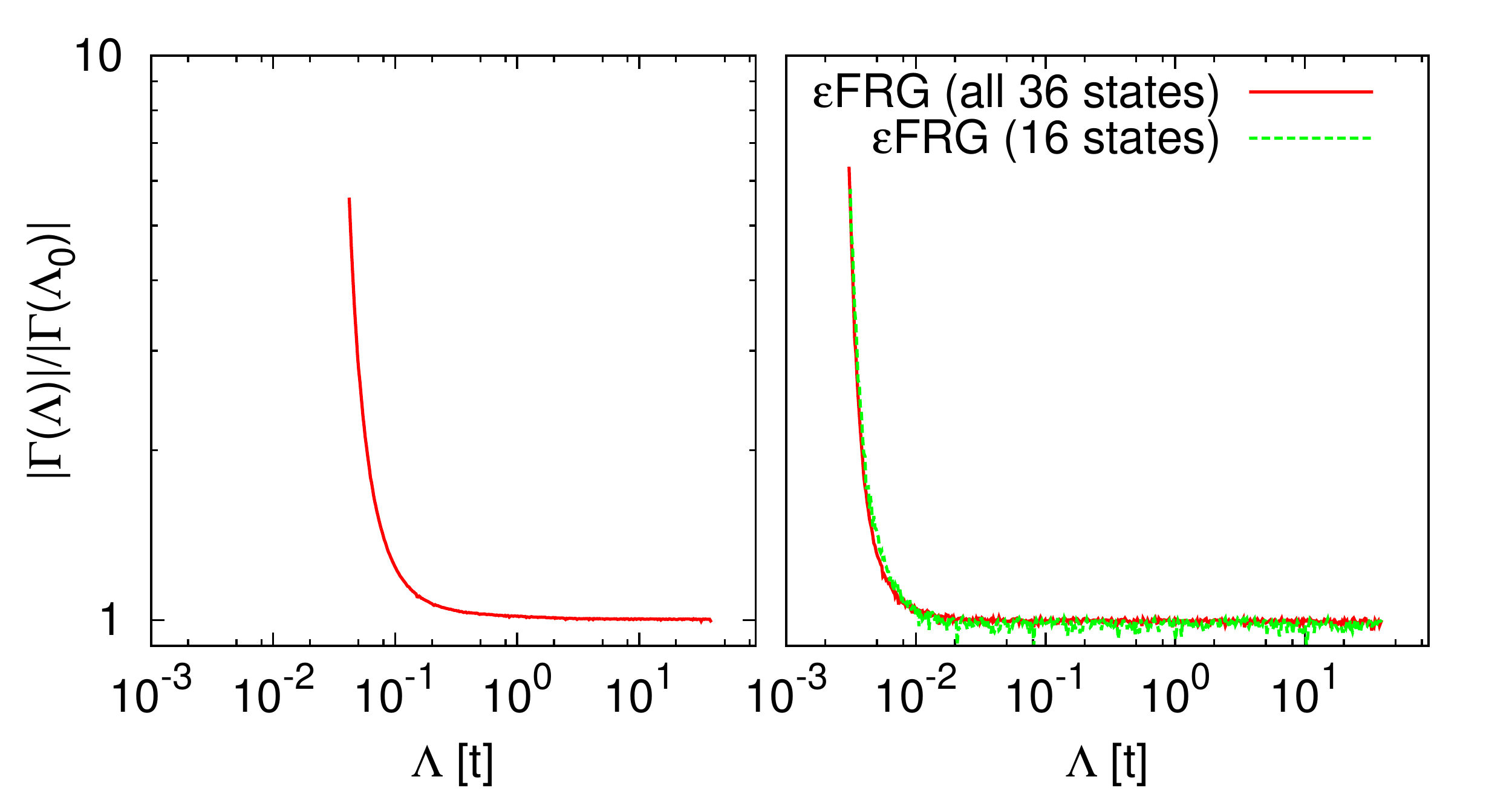}
   \caption[Divergent FRG flow]{\label{fig:verification:w0.1-u0.1:flow-frg}\label{f9} 
  RG flow of the norm of $\Gamma^\Lambda$ for $U{=}W{=}0.1$ on a lattice $4\times 4$ (left) 
  and $U{=}0.01, W{=}0.001, N{=}36$ on a lattice $6{\times}6$ (right). 
  The right panel shows in addition to full \efrg{} ($M=36$) also \asa-data with 
  $M{=}16$ demonstrating that the critical value $\Lambda_c$ is a very robust indicator 
  of runaway flow.}
\end{figure}

As we pointed out in section \ref{sec:frg:runaway-flow}, the \frg{} formalism signalizes the 
presence of an instability of the Fermi-liquid via runaway flow of certain
elements of the interaction vertex $\Gamma$. Therefore, a matrix-norm, e.g., 
\begin{equation}
 |\Gamma^\Lambda| = M^{-4} \sqrt{\sum_{\bar\alpha\bar\beta\bar\gamma\bar\delta}
(\Gamma^\Lambda_{\bar\alpha\bar\beta\bar\gamma\bar\delta})^2},
\end{equation}
is a reliable indicator of a nearby instability. 
Fig.~\ref{fig:verification:w0.1-u0.1:flow-frg} shows how this norm flows 
under the action of the RG. It is seen to diverge, 
e.g., at $\Lambda_{\text{c}}\approx 0.04$ for $U{=}0.1$.

\begin{figure}[t]
  \centering
  \includegraphics[width=.45\linewidth]{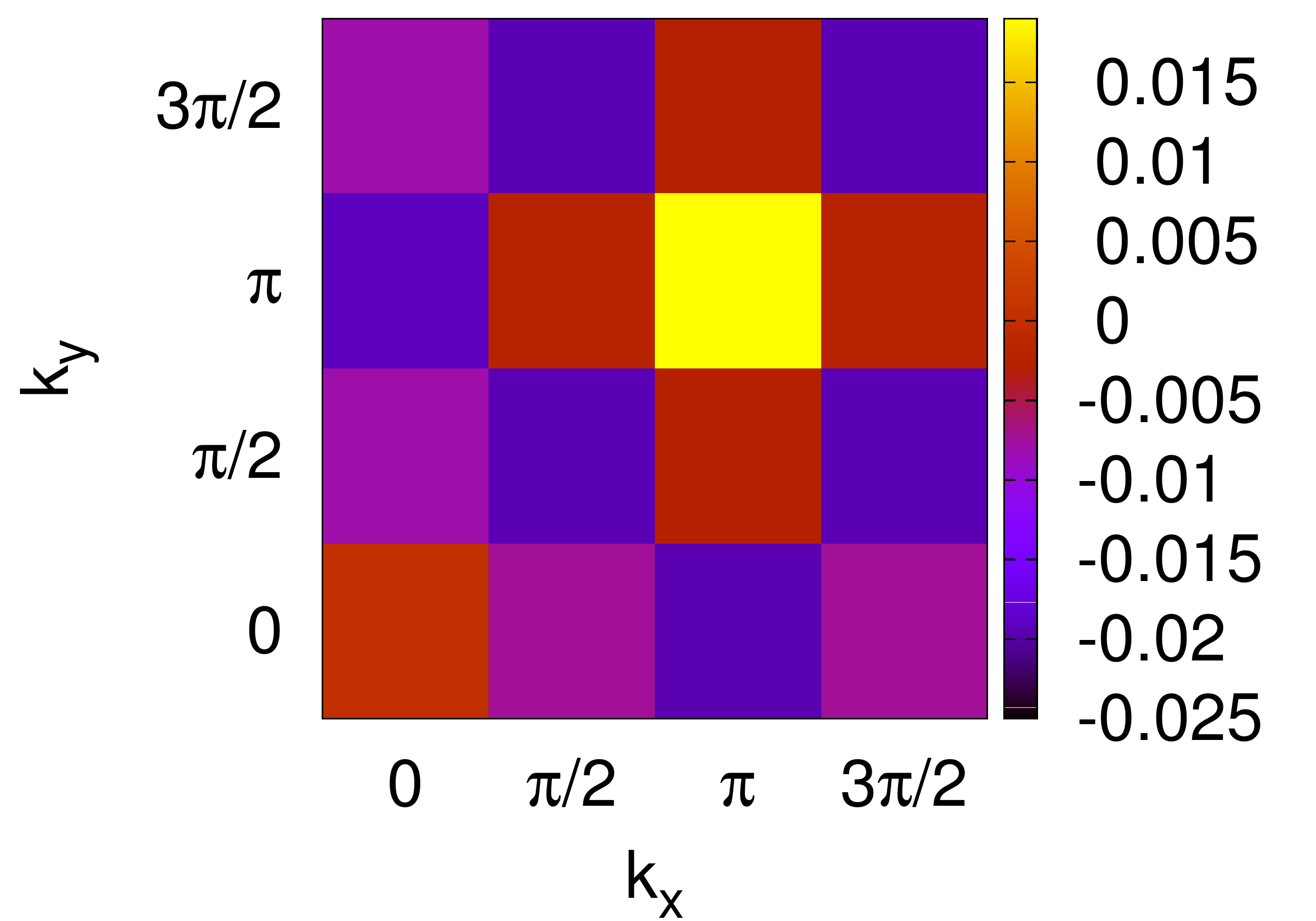}
  \includegraphics[width=.45\linewidth]{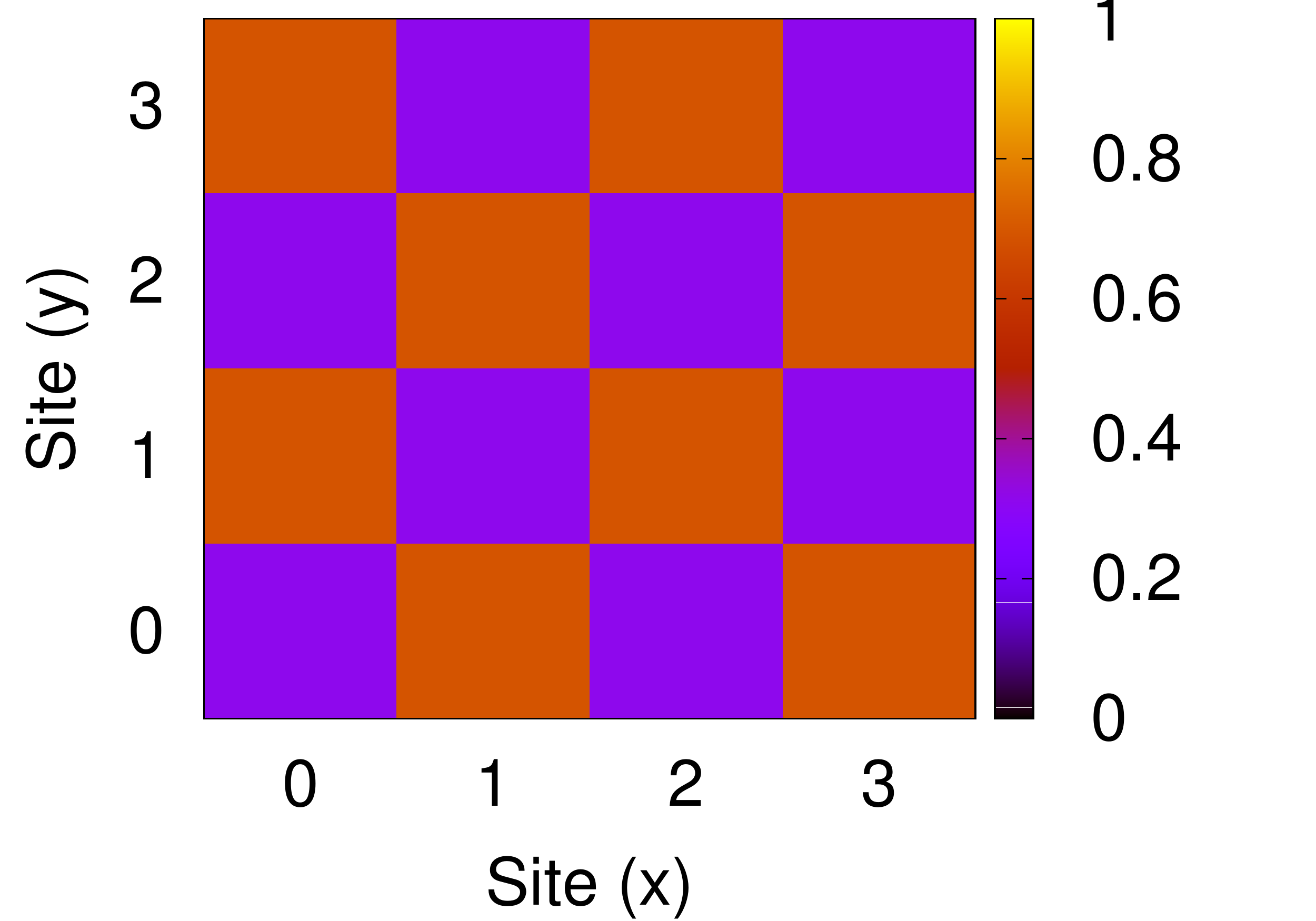}
  \includegraphics[width=.45\linewidth]{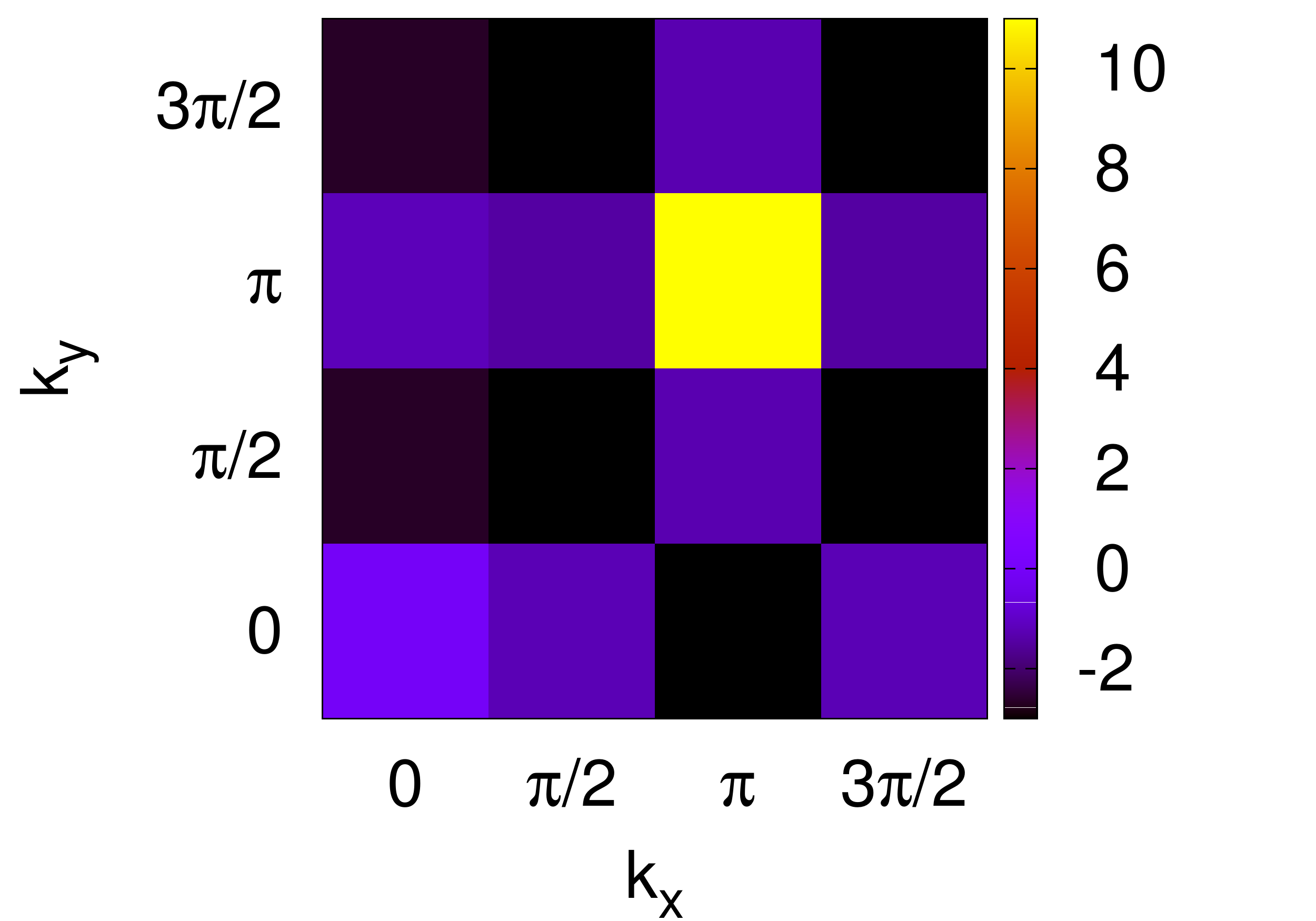}
  \includegraphics[width=.45\linewidth]{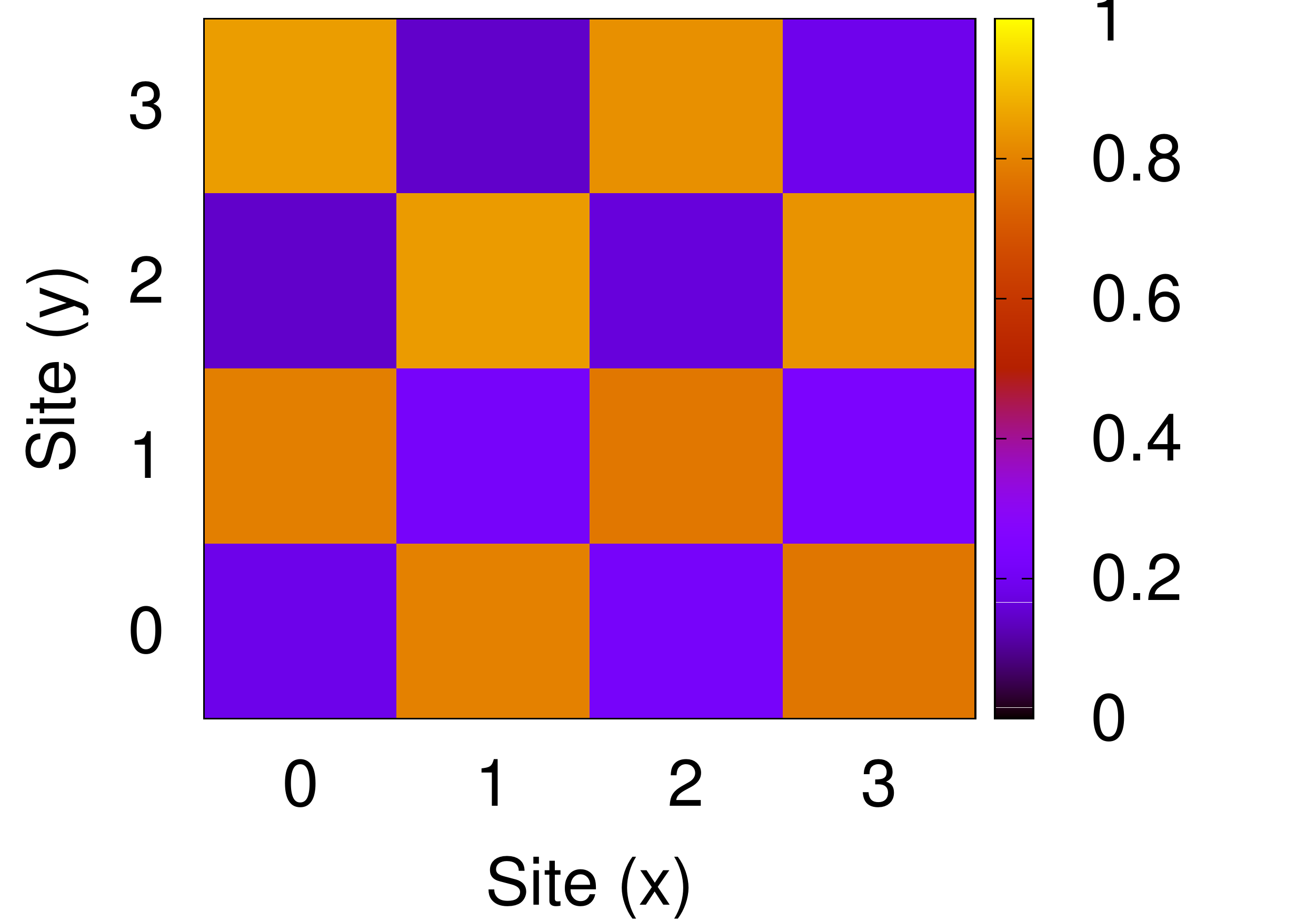}
  \caption[FRG density-density correlator at $\Lambda_{\text{c}}$]{
  \label{fig:verification:w0.01-u0.01:ddcorr-frg}\label{f10} 
  Left column: The density-density correlator as defined in Eq.~\eqref{eq:verification:ddcorr:ft} 
  calculated at $\Lambda_{\text{c}}$ at $U{=}W{=}0.01$ (upper row, system in Fig.~\ref{f9}) and 
  at $U{=}W{=}5$ (lower row). The peak indicates the CDW instability with wave-vector $(\varvec Q{=}\pi,\pi)$.
  Right column: Respective densities $n(\varvec r)$ from exact diagonalization (ED) exhibiting 
  the correspondig pinned CDW.}
\end{figure}

Ideally, to pinpoint the nature of the instability, 
one would investigate which one of the matrix elements 
of $\Gamma$ diverges so as to predict the nature of the instability. 
Since we here expect a CDW, we omit this step and just check that 
this interpretation is indeed consistent with the \efrg{} results. 
At first sight one might suspect that it would be sufficient to this end 
calculating the particle density $n(\varvec r)$ and ensuring that it indeed exhibits the 
checkerboard pattern. 
However, this perspective is slightly misleading. In the presence 
of runaway flow, we cannot evaluate the density at $\Lambda=0$, 
but only at $\Lambda\gtrsim \Lambda_c$ where 
the ground-state does not yet fully exhibit the broken symmetry.
Therefore, instead of calculating $n({\bf r})$ one rather 
evaluates the density-correlator at $\Lambda = \Lambda_{\text{c}}$,
\begin{equation}
 \mathcal{D}({\varvec k}) = N^{-1} \sum_{\varvec x \varvec x'} \mathrm{e}^{\ii \varvec k (\varvec x - \varvec x')} \mathcal{C}^{\mathrm{dd},(2)}_{i=(x,y),j=(x',y')},
 \label{eq:verification:ddcorr:ft}
\end{equation}
where $\mathcal{C}^{\mathrm{dd},(2)}$ may be calculated according to Eq.~(\ref{eq:TwoParticleGF:connected:T0:result}).
The result for $\mathcal{D}$ is displayed in Fig.~\ref{f10} (left column)
for two different values of interactions and disorder.
The peak at the correct ordering wavenumber $(\varvec Q{=}\pi,\pi)$ of the density response 
is already clearly visibly foreshadowing the upcoming ordered phase. 

To give further evidence of the correct prediction of charge ordering, 
we also calculate the real space density. 
Since due to runaway flow this cannot be done with \efrg{}, we again employ the ED. 
As expected, the resulting densities -- shown in Fig. \ref{f10} (right column) -- 
exhibit the checkerboard pattern. 
\begin{figure}[b]
  \centering
  \includegraphics[width=.9\linewidth]{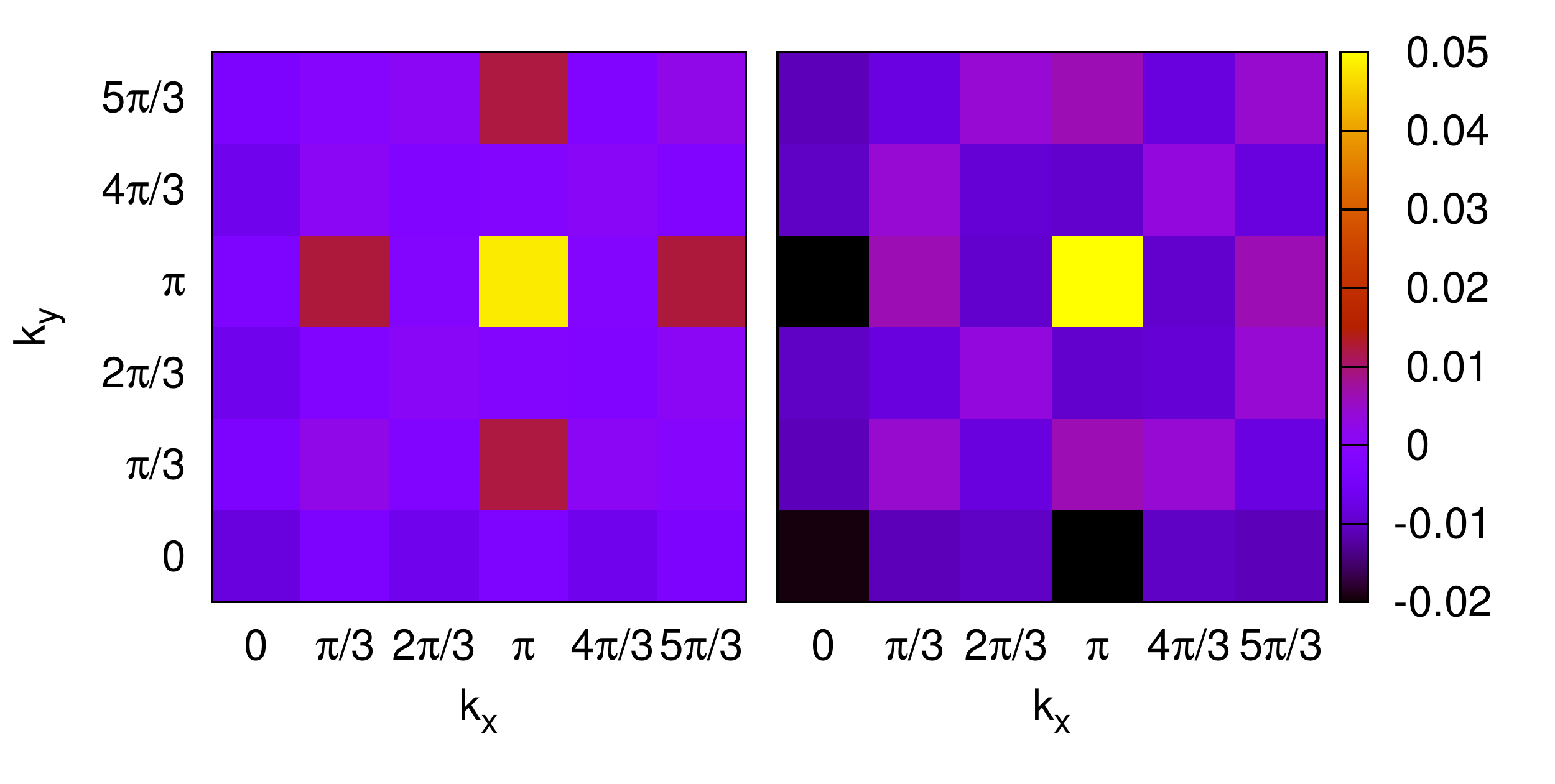}
  \caption[Density-density correlator comparison between FRG variants]{\label{fig:verification:36:ddcorr:36}\label{f11}
    {\label{fig:verification:36:ddcorr}
  The density-density correlator as obtained from \efrg{} at $\Lambda_{\text{c}}$ with (left, $M{=}16$) and without (right, $N{=}M{=}36$)
  \asa{} for $W = 0.001$, $U = 0.01$ on a $6{\times}6$-lattice.
  The peak is well exposed in both plots, so the CDW-nature of the ordering phase is 
  reliably reproduced by \asa. 
  }}
\end{figure}

We have already demonstrated that $\Lambda_\text{c}$ 
is properly reproduced within \asa{}. As a final 
step in this section we show that this is also the 
case for the density response $\mathcal{D}$. 
In Fig. \ref{f11} we compare two calculations with 
full \efrg{}, $N{=}M{=}36$ and with \asa{} ($M{=}16$)
for a system with very weak disorder and interaction. 
As is seen there, the ordering peak is quantitatively 
reproduced by the active-space approximation to the 
\efrg{}. 

\section{Application -- Phase-diagram of spinless disordered Hubbard model}
\label{sec:results}

As a relevant application of our method, 
we determine the phase diagram of the spinless Hubbard 
model on a square lattice with periodic boundary
conditions.
For two limiting cases the phases of the model are well known. 
In the absence of disorder, $W{=}0$, the ground state exhibits 
the charge-density wave (CDW) at any finite value of $U{>}0$;\cite{RGShankar}
it already made its appearance in the previous section. 
On the other hand, in the absence of interaction, $U{=}0$,
the system becomes an Anderson insulator (AI) 
for any finite disorder $W > 0$.\cite{AndersonScaling}
The purpose of this investigation is to determine the phase-boundary in the general case, 
$U,W > 0$, as is indicated in the \efrg{} by runaway flow.

Our tests on small systems so far have indicated, that with 
disorder, $W>0$, a minimum value of the interaction, $U^*(W)$, is required
for the system to form a CDW ground state. 
This is in contrast to the clean case 
where for any $U>0$ a charge density order is established, 
at least for large enough systems. 
We evaluate  $U^{*}(W)$ with the \efrg{}.

\begin{figure}[tbp]
  \centering
  \includegraphics[width=.9\linewidth]{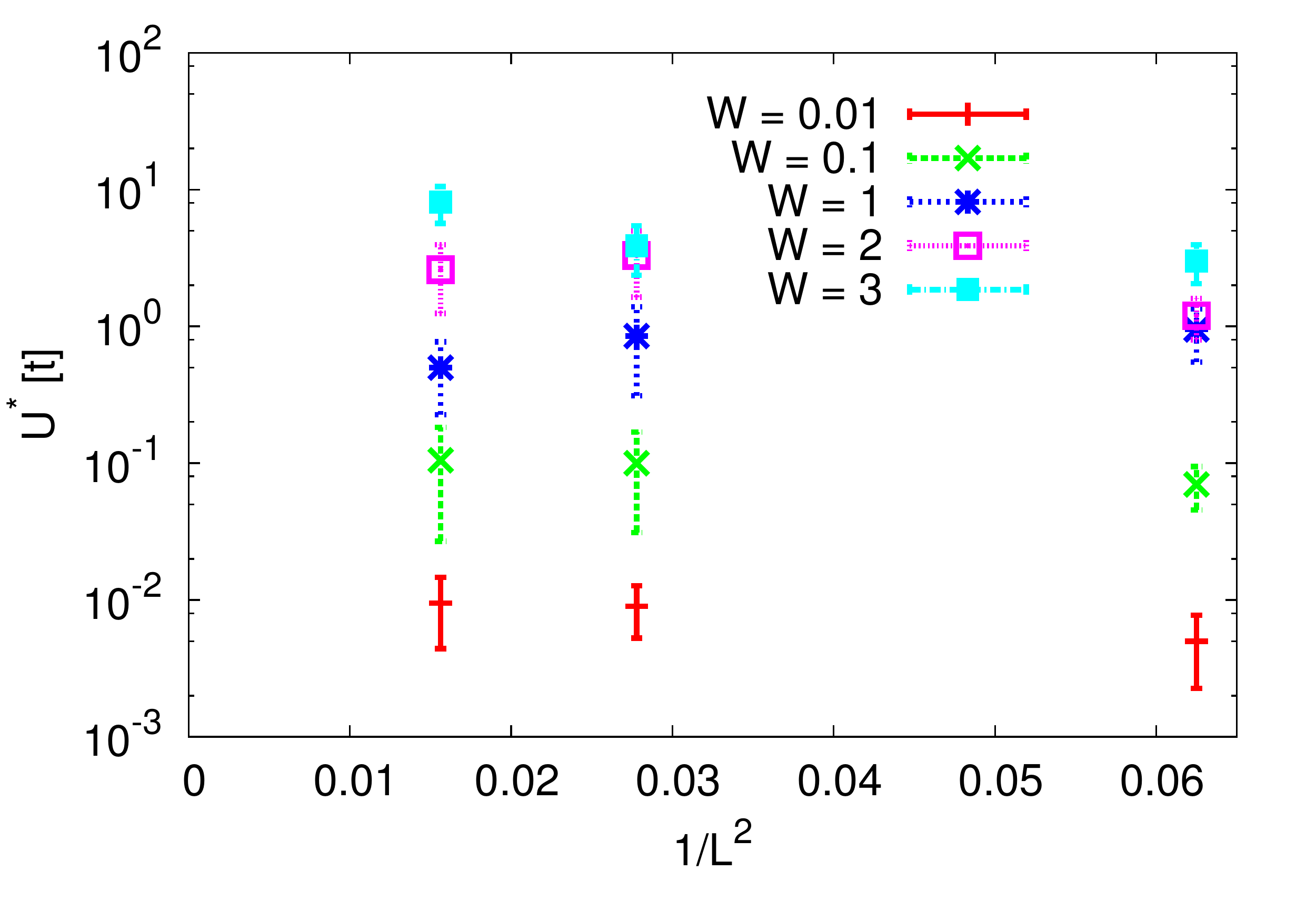}
  \caption[$U^{*}$ for different system sizes $L$ and disorder strengths $W$]{\label{fig:ucrit-plot}\label{f12}
  Critical interaction $U^{*}$ beyond which \efrg{} predicts CDW-ordering plotted over the inverse  system size
  $1/L^2$ for multiple different values of the disorder
  strength $W$. The results have been averaged over $5$ disorder configurations and $1\sigma$-error bars are 
  given.
  }
\end{figure}

Note, that $U^{*}$ will somewhat vary between different disorder realization and 
may, in addition, exhibit a dependency on the system size $L$.
To deal with this, we apply the following strategy: for a fixed system size and disorder
realization, we scan over $U$ and thus obtain $U^{*}$ for this specific sample.  
We repeat the run for more samples with different disorder realizations 
keeping the same disorder strength $W$
thus finding the average $U^{*}(W,L)$. Finally, to account for finite size effects 
we analyze the behavior of $U^{*}(L,W)$ for varying system sizes. 

\subsection{Results -- Phase diagram}

\begin{figure}[tbp]
  \centering
  \includegraphics[width=.85\linewidth]{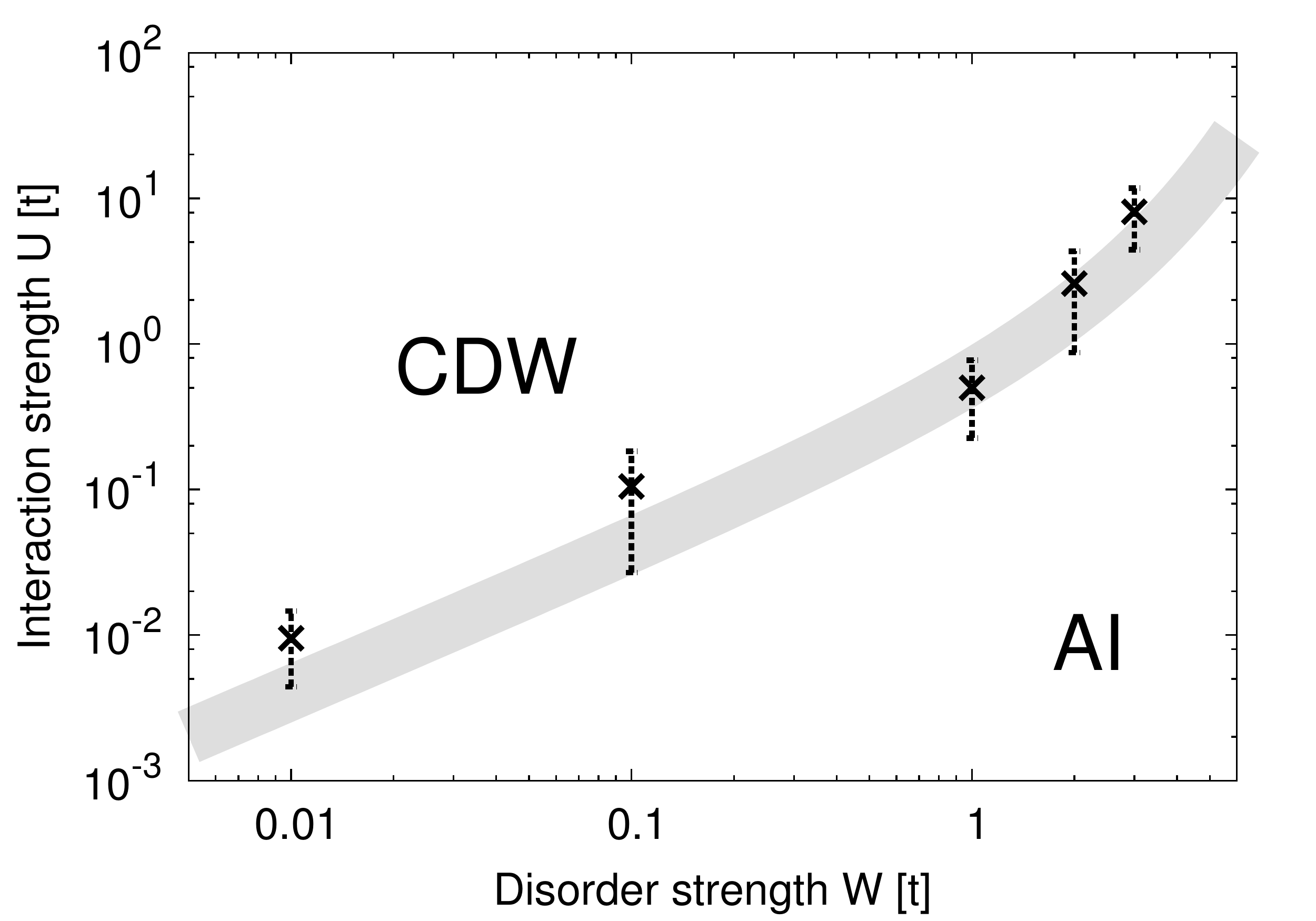}
  \caption[Phase diagram: results]{\label{fig:pdres}\label{f13}
  The phase-diagram for the spinless disordered Hubbard model in 2D
  as calculated with \efrg{}. 
  }
\end{figure}

Fig.~\ref{fig:ucrit-plot} displays $U^{*}(W,L)$ after averaging over five 
disorder configurations for $L{\times} L$-lattices with $L = 4,6,8$. 
For $L=6,8$ we have used \asa{} with $M = 16$ states in both cases.
Our data indicates that except at very large disorder values, $W{=}3$, 
$U^{*}$ appears to remain largely insensitive 
to variations of the (lateral) system size by a factor of two. 
We take this as an indication that $U^{*}$ will indeed remain finite even at large 
system sizes. Thus encouraged we take the data at $L{=}8$ as an estimate 
for the phase boundary $U^{*}(W)$ at $L\to\infty$. 
Fig.~\ref{fig:pdres} shows the resulting phase diagram. 

\paragraph*{Computational details.}
We found it practical 
to work with a single particle Hilbert space containing $N{\sim}50{-}100$ states.
For example, with a single-particle Hilbert space consisting of $N = 64$ states
and the active space consisting of $M{=}32$ states, 
a single calculation on 8 CPU cores takes less than 24 hours.

\subsection{Discussion}

\subsubsection{Stability arguments and quasistatic approximation} 
Due to the quasi-static approximation the \efrg{}-self-energy
is hermitian and energy-independent. On this level of 
approximation, the interaction is dealt with by replacing 
the non-interacting Hamiltonian $H_0$ with an effective 
quasi-particle (qp) Hamiltonian $H_\text{qp}$. 
The latter deviates from $H_0$ by a renormalized kinetic 
energy term, a renormalized effective potential that can, 
in general, carry off-diagonal entries. 

After these preliminaries, one expects that the 
Anderson-localized phase (at $U{=}0$) is seen to be stable 
within quasi-static \efrg{} against introducing 
a small repulsive interaction.
\footnote{We tacitly assume here that the 
short-range Hubbard term does not introduce long-range 
correlations in the matrix elements of $H_\text{qp}$.} 
After all, the renormalized Hamiltonian $\hat H_\text{qp}$ 
is still a generic representative of the orthogonal symmetry class 
and hence should exhibit conventional behavior.  

A similar stability argument also applies to the ordered phase: 
the leading effect of weak disorder is pinning of the charge-density 
wave (CDW). The wave is destroyed only when strong fluctuations of the 
local potential allow for lattice defects, 
where two neighboring lattice sites are occupied. 
For box-distributed on-site potentials, isolated defects can occur only when 
$W\sim U$. As a consequence, one expects  $U^{*}(W)\sim W$ 
at weak disorder $W$, which is consistent with the phase-boundary seen in 
Fig.~\ref{f13}. 
Remarkably, at interaction strengths comparable to the band-width, $U\gtrsim 1$, 
the disorder strength necessary to destroy the CDW appears to be considerably smaller than $U$. 
We hypothesize that we here witness the onset of a collective effect in 
which several particles can optimize their energy with respect to the disorder 
potential at the expense of very few particles that built up a defect line thus 
producing a phase-separation. 

\paragraph*{Physics beyond the quasi-static approximation.}
The quasi-static approximation ignores the energy exchange between 
the quasi-particles that of course also is included in the model 
Hamiltonian, Eq.~\eqref{e124}. Effects of dephasing and 
many-body localization\cite{nandkishore15} are beyond its scope. 
Therefore, we consider it likely that the phase seen as 
(conventional) Anderson-insulator by the (quasi-static) \efrg{}
is missing aspects of dynamical physics that dominate essential  
properties of the phase at non-vanishing temperature.
What implications this may have on the (zero-temperature) 
phase-boundary between the CDW and the Anderson insulator remains 
to be seen. 

\subsubsection{Relation to earlier work}
The spin$-1/2$ Hubbard model enjoyed considerable attention 
in recent years, because physical realizations can be found not 
only within condensed matter systems but also in cold atomic gases, 
see Refs.~\onlinecite{gemelke16,cocchi16} for very recent results. 
In principle, also the spinless model, Eq.~\eqref{e124},
that we deal with in this work could find a 
cold-atom realization which, however, would require the application 
of a strong homogeneous in-plane magnetic field. 
This could be one reason, why the spinless model, Eq.~\eqref{e124}, has 
received considerably less attention over the years. 

Numerical investigations of the spinless model have been concentrating 
on its quantum glass variant that deviates from Eq.~\eqref{e124} replacing 
the short-range interaction by a long-range Coulomb interaction.
\cite{vojta98,benenti99,berkovits01}
An analytical treatment of the model, Eq.~\eqref{e124}, has been given 
by Vlaming et al., Refs.~\onlinecite{vlaming92,uhrig93}. The authors employed 
the Bethe lattice where an exact solution can be given in the 
limit of infinite branching number. The physical picture 
developed there for the zero temperature limit is in qualitative 
agreement with our own findings. More recently, Foster and Ludwig 
studied the model, Eq.~\eqref{e124}, with (complex) off-diagonal disorder 
focussing on the effect of interactions on the Gade-fixpoint.\cite{foster08} 
In that case the non-interacting reference state is not an insulator
but a (critical) metal that -- according to perturbative RG --
is unstable against weak repulsive interactions.

\section{Conclusion and Outlook}
\label{sec:outlook}

The main purpose of this work was a methodological one: 
to develop, implement and test a variant of the traditional 
functional renormalization group (FRG) method 
that is applicable to generic systems, such as molecules 
or disordered metal grains, which are lacking translational invariance. 
Within the new approach (\efrg), the renormalization of the interaction 
vertex occurs only for matrix elements with single-particle states 
that are situated in an energy shell around the Fermi-energy ({\em active space}). 
The method is computationally efficient provided this shell 
can be taken smaller than the (non-interacting) bandwidth. 
We argue that the scaling with the size of the 
single-particle Hilbert space $N$ should be $N^4$ 
for 2D-lattice systems which compares favorably well 
with the typical $N^6$ scaling of competing methods, 
such as CCSD(T). Specifically, calculations with $N{=}64$ 
and an active space of size $M{=}32$ require less than 
24h on 8 CPU-cores.

An explicit implementation of \efrg{} has been coded 
for the spinless Hubbard model in 1D and 2D in the presence of 
on-site disorder. A comparison to 
(numerically exact) calculations employing the diagonalization 
of small systems suggests that the accuracy of \efrg{} 
concerning quasiparticle energies typically is below $20\%$ in relative
error to the interaction-induced shift, as compared to the non-interacting
system.
Similarly, the interaction induced shift in the ground-state 
density is recovered quantitatively at small interaction strength $U$
with an error that increases to $\sim$50\% if $U$ reaches the 
band-width. 

At its current development stage, the \efrg{} is readily applicable 
to models of interacting fermions in low dimensions, 
which includes Hubbard clusters with spin and (attractive) interactions
at different filling fractions, but also, e.g., 
small molecules. 
Our preliminary tests suggests that with the current formalism 
system sizes of, e.g., $N{=}256$ are already within reach. 
Significantly bigger system sizes might be attainable, 
after additional improvements in the code performance have been implemented.
As an example we mention the numerical integration of the flow-equations 
that at present is done in the simplest possible discretization scheme.
Also, the flow equations are well-suited for parallelization on distributed
memory systems, allowing for a significant increase in the number of CPU cores
used in a single calculation.
To give a perspective, we mention that the molecules in the GW100 test set 
have been described with a QZVP-basis set requiring ca. 800 basis function 
for the biggest species, the amino-acids Guanin and Adenin.\cite{vanSetten15}

We hope that this work helps paving the way for electronic-structure 
calculations beyond the present paradigm of GW-BSE. 
Admittedly before the envisioned applications to real systems, an efficient 
\efrg{}-implementation should be installed that is also prepared for dealing with long-range 
interactions. Here, we see at present the biggest bottleneck to be overcome 
in future research. Perhaps additional motivation to overcome this obstacle could come 
from the fact that we have also given formul{\ae} for the finite-temperature 
formalism in this work, so that the effect of heat could be included.

\section*{Acknowledgements}
We thank  S. Bera,  A. D. Mirlin,  J. Reuther, J. Schmalian, M. van Setten 
and P. W{\"o}lfle for inspiring discussions. We are indebted to A. D. Mirlin 
for supporting our project in an early stage.
Support has also been received from the DFG under grants EV30/7-1, EV30/11-1 and EV30/12-1.
and from the Landesgraduiertenf\"orderung of the state of Baden-W\"urttemberg. 
The DMRG results shown here have been provided by F. Weiner using the 
Schmitteckert-code. We acknowledge the support provided by computational resources
of the Institute of Nanotechnology (INT) and the Steinbuch Centre for Computing (SCC),
both at the Karlsruhe Institute of Technology (KIT).

\appendix
\section{Flow equations for $\Gamma$ in the static limit}
\label{app:T0StaticGamma}

Here we will derive the flow equation for $\Gamma$ in the static limit, Eq.~(\ref{eq:flow:Gamma}), analogous to the derivation for the
self-energy. Starting at Eq.~(\ref{eq:flow:GammaWithOmegaStatic}), looking at the first term,
\[ \int\diff\bar\omega \sum_{\mu\nu\rho\sigma}
   \mathcal{G}^\Lambda_{\rho\mu}(\bar\omega)
   \mathcal{S}^\Lambda_{\sigma\nu}(-\bar\omega)
   \Gamma^\Lambda_{\alpha\beta\rho\sigma}
   \Gamma^\Lambda_{\mu\nu\gamma\delta},
\]
it can be seen that by exchanging all traced indices in both vertices that appear, and then renaming the summation indices,
the formula may be rewritten as 
\[ \int\diff\bar\omega \sum_{\mu\nu\rho\sigma}
   \mathcal{S}^\Lambda_{\rho\mu}(-\bar\omega)
   \mathcal{G}^\Lambda_{\sigma\nu}(\bar\omega)
   \Gamma^\Lambda_{\alpha\beta\rho\sigma}
   \Gamma^\Lambda_{\mu\nu\gamma\delta},
\]
which is just an exchange of both propagators. Utilizing this, we may write it formulated in terms of matrix products,
\[
 \frac{1}{2} \mathrm{tr}\, \int\diff\bar\omega \big[
     \mathcal{S'} \Gamma^{\mathrm{T}} \mathcal{G}^{\mathrm{T}}  \Gamma
     + 
     \mathcal{G}  \Gamma^{\mathrm{T}} \mathcal{S'}^{\mathrm{T}} \Gamma
 \big].
\]
We note that the frequency of the single-scale propagator is negative here, which we denote with prime for $\mathcal{Q}$ and $\Sigma$;
the $\Theta$ and $\delta$-functions only depend on the modulus.
Inserting Eq.~(\ref{eq:derivation:SingleScalePropagator:partialform}) and using the same representation for $\mathcal{G}$,
we can separate four terms,
\begin{eqnarray}
 \hspace{-1em} & & -\frac{\delta}{2} \frac{1}{\mathcal{Q'} - \Theta\Sigma'} \Gamma^{\mathrm{T}} \left( \frac{\Theta}{\mathcal{Q}     - \Theta\Sigma    } \right)^{\mathrm{T}} \Gamma,
     \label{eq:derivation:flow:Gamma:cooper:term1} \\
 \hspace{-1em} & & -\frac{\delta}{2} \frac{\Theta}{\mathcal{Q}     - \Theta\Sigma    } \Gamma^{\mathrm{T}} \left( \frac{1}{\mathcal{Q'} - \Theta\Sigma'} \right)^{\mathrm{T}} \Gamma,
     \label{eq:derivation:flow:Gamma:cooper:term2} \\
 \hspace{-1em} & & -\frac{\delta}{2} \frac{1}{\mathcal{Q'} - \Theta\Sigma'} \Sigma' \frac{\Theta}{\mathcal{Q'} - \Theta\Sigma'}
    \Gamma^{\mathrm{T}} \left( \frac{\Theta}{\mathcal{Q}     - \Theta\Sigma    } \right)^{\mathrm{T}} \Gamma,
     \label{eq:derivation:flow:Gamma:cooper:term3} \\
 \hspace{-1em} & & -\frac{\delta}{2} \frac{\Theta}{\mathcal{Q} - \Theta\Sigma}
    \Gamma^{\mathrm{T}} \left( \frac{\Theta}{\mathcal{Q'}    - \Theta\Sigma'   } \right)^{\mathrm{T}}
    \hspace{-.5em} \Sigma'^{\mathrm{T}} \hspace{-.3em} 
    \left( \frac{1}{\mathcal{Q'}    - \Theta\Sigma'   } \right)^{\mathrm{T}} \Gamma.
    \label{eq:derivation:flow:Gamma:cooper:term4}
\end{eqnarray}
Since all of these terms occur underneath an integral over $\int\diff\bar\omega \delta(|\omega|-\Lambda)$,
we may switch primes within each term, and we note for future use that the terms of
Eqs.~(\ref{eq:derivation:flow:Gamma:cooper:term1},\ref{eq:derivation:flow:Gamma:cooper:term2}) are equal to
each other.

We now apply Morris's Lemma again.
In both other terms, Eqs.~(\ref{eq:derivation:flow:Gamma:cooper:term3},\ref{eq:derivation:flow:Gamma:cooper:term4}), we can rewrite them in
terms of derivatives w.r.t. the integration variable $t$,
\begin{eqnarray}
 & & -\frac{\delta}{2} \int_0^1 t^2
   \left( \frac{\diff}{\diff t} \frac{1}{\mathcal{Q'} - t \Sigma'} \right)
   \Gamma^{\mathrm{T}}
   \left( \frac{1}{\mathcal{Q} - t \Sigma} \right)^{\mathrm{T}}
   \Gamma
 \diff t, \label{eq:derivation:flow:Gamma:cooper:term3:morris} \\
 & & -\frac{\delta}{2} \int_0^1 t^2
   \frac{1}{\mathcal{Q} - t \Sigma}
   \Gamma^{\mathrm{T}}
   \left( \frac{\diff}{\diff t} \frac{1}{\mathcal{Q'} - t \Sigma'} \right)^{\mathrm{T}}
   \Gamma
 \diff t. \label{eq:derivation:flow:Gamma:cooper:term4:morris}
\end{eqnarray}
Partial integration of Eq.~\ref{eq:derivation:flow:Gamma:cooper:term3:morris} yields
\begin{eqnarray}
 & & - \frac{\delta}{2} \left[ t^2 \frac{1}{\mathcal{Q'} - t \Sigma'} \Gamma^{\mathrm{T}} 
       \left( \frac{1}{\mathcal{Q} - t \Sigma} \right)^{\mathrm{T}} \Gamma \right]_0^1 \nonumber \\
 & & + \frac{\delta}{2} \int_0^1 2t
       \frac{1}{\mathcal{Q'} - t \Sigma'} \Gamma^{\mathrm{T}} 
       \left( \frac{1}{\mathcal{Q} - t \Sigma} \right)^{\mathrm{T}} \Gamma
       \diff t \nonumber \\
 & & + \frac{\delta}{2} \int_0^1 t^2
   \frac{1}{\mathcal{Q} - t \Sigma}
   \Gamma^{\mathrm{T}}
   \left( \frac{\diff}{\diff t} \frac{1}{\mathcal{Q'} - t \Sigma'} \right)^{\mathrm{T}}
   \Gamma
 \diff t.
\end{eqnarray}
One sees that the second term cancels Eqs.~(\ref{eq:derivation:flow:Gamma:cooper:term1},\ref{eq:derivation:flow:Gamma:cooper:term2})
and the third term cancels Eq.~(\ref{eq:derivation:flow:Gamma:cooper:term4:morris}),
leaving the result
\begin{equation}
 - \frac{\delta}{2} \frac{1}{\mathcal{Q'} - \Sigma'} \Gamma^{\mathrm{T}} 
       \left( \frac{1}{\mathcal{Q} - \Sigma} \right)^{\mathrm{T}} \Gamma,
\end{equation}
which can be rewritten in terms of the index notation as
\begin{equation}
 - \frac{1}{2} \sum_{\bar\omega=\pm\Lambda} \sum_{\mu\nu\rho\sigma}
   P^{\Lambda}_{\rho\mu}(-\bar\omega) P^{\Lambda}_{\sigma\nu}(\bar\omega)
   \Gamma^\Lambda_{\alpha\beta\rho\sigma} \Gamma^\Lambda_{\mu\nu\gamma\delta}.
\end{equation}
We note that if one were to keep the frequency dependence of the vertex and the self-energy, two cases need to be distinguished:
for the case where all external frequencies are zero, the same derivation applies, so our result holds there. For the case
where at least some external frequencies are non-zero, the arguments for the $\delta$ and $\Theta$ functions differ, so one may
directly insert Eq.~\ref{eq:result:SingleScalePropagatorAtT0} into the flow equations for the vertex.

An analogous treatment is possible for the other four terms in Eq.~(\ref{eq:flow:GammaWithOmegaStatic}). The other terms may be
written as
\[
 \mathrm{tr}\, \int\diff\bar\omega \big[
     \mathcal{S}  \Gamma_{\alpha\cdot\delta\cdot} \mathcal{G} \Gamma_{\beta\cdot\gamma\cdot}
     + 
     \mathcal{G}  \Gamma_{\alpha\cdot\delta\cdot} \mathcal{S} \Gamma_{\beta\cdot\gamma\cdot}
     - [\alpha\leftrightarrow\beta]
 \big].
\]
Looking at the first two terms, they may be divided in the same mannger as in
Eqs.~(\ref{eq:derivation:flow:Gamma:cooper:term1},\ref{eq:derivation:flow:Gamma:cooper:term2},\ref{eq:derivation:flow:Gamma:cooper:term3},\ref{eq:derivation:flow:Gamma:cooper:term4}),
without the factor $1/2$, and with the same frequency for the single-scale and the regular propagator. This yields the result
\begin{equation}
 - \sum_{\bar\omega=\pm\Lambda} \sum_{\mu\nu\rho\sigma}
   P^{\Lambda}_{\rho\mu}(\bar\omega) P^{\Lambda}_{\sigma\nu}(\bar\omega)
   \Gamma^\Lambda_{\beta\nu\gamma\rho} \Gamma^\Lambda_{\alpha\mu\delta\sigma} + [\alpha\leftrightarrow\beta].
\end{equation}

Putting this all together, one arrives at Eq.~(\ref{eq:flow:Gamma}).

\section{Implementation Details}
\label{app:implementation}

\subsection{Chemical Potential for $T > 0$}

Our algorithm to solve this equation for $\muchem$ works in three stages: obtain an initial guess for $\muchem$, $\muchem^{(0)}$, (trivially) obtain a second guess, $\muchem^{(1)}$,
with $\sgn(N_{\text{e}}(\muchem^{\Lambda,(1)}) - N_{\text{e}}) = - \sgn(N_{\text{e}}(\muchem^{\Lambda,(0)}) - N_{\text{e}})$ and then use the secant algorithm \cite{Numerical77} to
iteratively find the final $\muchem$.

The initial guess is taken to be the same as for $T = 0$, Eq.~\eqref{eq:impl:muchem:T0}, since at low temperatures the value is a very
good approximation. We then calculate
$$
\muchem^{\Lambda,(0)} + \sgn(N_{\text{e}}(\muchem^{(0)}) - N_{\text{e}}) \frac{\Delta}{4} i,
$$
where $\Delta$ is the mean level spacing of the system and $i$ is an integer that starts at $1$ and is incremented until the condition
$\sgn(N_{\text{e}}(\muchem^{(1)}) - N_{\text{e}}) = - \sgn(N_{\text{e}}(\muchem^{(0)}) - N_{\text{e}})$ is satisfied. In practice $i = 1$
or $i = 2$ will already be sufficient, which is why $\Delta/4$ is a good empirical choice here.\footnote{We cut this
scheme off at $i = 10$, since it is only used to accelerate the convergence of the secant algorithm, which is likely to also work if
the second value does not satisfy the condition, albeit more slowly.}

Both initial guesses are then used as input for the secant algorithm. Since $N_{\text{e}}(\epsilon)$ is monotonous and the value searched
for is encompassed with both guesses, convergence will be quite fast ($10$ to $20$ iterations in practice). We consider the chemical
potential to be converged if the relative error of the number of electrons,
$$
\left|\frac{N_{\text{e}}(\muchem^{(i)}) - N_{\text{e}}}{N_{\text{e}}(\muchem^{(i)}) - N_{\text{e}}}\right|,
$$
is larger than the square root of the machine precision. While the smallest possible error here would be of the order of $\hat\epsilon N$,
with $\hat\epsilon$ being the machine precision and $N$ the number of orbitals in the system, the energies $\tilde\epsilon_{\tilde\alpha}$
only have a precision of $\sqrt{\hat\epsilon}$ due to the diagonalization procedure.

\subsection{Parallelization}

We will now discuss how we exploit parallelization in our implementation. We use a scheme based on a shared memory architecture, OpenMP \cite{OpenMP3.1}.
It is in principle possible to utilize distributed memory methods, such as MPI (Message Passing Interface, \cite{MPI}), which allow the usage of far more
processor cores for the same calculation.

The intermediate products offer a trivial way to parallelize: it is possible to use a parallel version of the GEMM kernel to calculate the matrix products.
In the case we track the renormalization of the entire vertex, this would likely be the most efficient avenue. In our case, however, the effective matrix
size that is fed into the GEMM kernel is relatively small (we want to calculate the vertex for as few states as possible), so it is unlikely that using a
parallel matrix product kernel will scale well even for a low amount of processors. Instead, we parallelize the loops over the two outer indices in the
intermediate products and perform serialized matrix products on each processor. This is trivially possible, since the calculations are independent of each
other for any given pair of external indices.

Similarly, for the evaluation of the trace, we parallelize the loops over all four external indices and have each processor evaluate the trace for a given
set of external indices serially.

\subsection{Restarting}

Calculations for larger systems may take a relatively long time. In case of technical difficulties, we implement a
restarting
procedure that allows us to continue a calculation at the point where it last stopped. We save the initial $\Lambda$,
the step
size, the number of selected states $M$, the chosen target $\Lambda$. Furthermore, we keep the last self-energy and
vertex
as well as the number of the last iteration to complete. These quantities suffice to reproduce the calculation at a
later point
in time.

\section{ED Implementation}
\label{app:ED}

In Sec.~\ref{sec:disorder:2d} we compare the FRG to exact diagonalization. In the following we provide edtails on how we implemented
ED as a reference method. In our implementation, we construct the full $N_{\text{e}}$-particle Hilbert space. Its dimension
is $\left( \begin{array}{c} N \\ N_{\text{e}} \end{array} \right)$ and grows exponentially with the number of orbitals $N$.
We systematically construct the basis states of that space and implement the action of the full many-body Hamiltonian on that
basis (we do \emph{not} explicitly construct the matrix elements of the Hamiltonian itself). An iterative eigensolver for
sparse problems is employed to calculate the full many-body ground state for a given system. We utilize the standard ARPACK
package \cite{Arpack} in direct mode.\footnote{The shift-inverse mode is not required, since the eigenvalues we are interested
in are taken from the spectrum edges, not the center.}

For simple observables, such as the density, we may then simply calculate expectation values with respect to the many-body
ground state,
\begin{equation}
 \left< \mathrm{\hat n}_i \right> = \bra{0} \mathrm{\hat c}_i^\dagger \mathrm{\hat c}_i \ket{0}.
\end{equation}
We also want to calculate the single-particle density of states, $\rho(\epsilon)$. This is given by the expectation value
\begin{eqnarray}
 \rho(\epsilon) & = & -\frac{1}{\pi} \Im\, \tr_{ij} \left< \mathrm{\hat c}_i \frac{1}{\epsilon - \hat H + E_0 + i\eta} \mathrm{\hat c}_j^\dagger \right>
   \nonumber \\ & & \hphantom{\frac{1}{\pi} \Im\, \tr_{ij}}
                    + \left< \mathrm{\hat c}_j^\dagger \frac{1}{\epsilon + \hat H - E_0 + i\eta} \mathrm{\hat c}_i \right>,
\end{eqnarray}
which we arrive at by Fourier transforming the definition of the retarded Green's function. This expressions contains the inverse
of a very large matrix, which needs to be done for every single energy at which the density of states is to be evaluated at. Furthermore,
directly inverting such a large matrix is only possible using iterative algorithms, which would again have to be applied for every single
energy. We therefore follow an alternative approach as outlined in the PhD thesis of Alexander Braun \cite{AlexBraunPhD}. One may
expand the denominator in terms of Chebyshev polynomials $T_n(x)$, such that we get
\begin{eqnarray}
c^{(+)}_{ij,n} & = & \bra{0} \mathrm{\hat c}_i T_n\big(a(\hat H - E_0 - b)\big) \mathrm{\hat c}_j^\dagger \ket{0}, \\
c^{(-)}_{ij,n} & = & \bra{0} \mathrm{\hat c}_i^\dagger T_n\big(a(\hat H - E_0 - b)\big) \mathrm{\hat c}_j \ket{0},
\end{eqnarray}
where $E_0$ is the ground state energy. The variables $a$ and $b$ are scaling factors that arise due to the fact that the
Chebyshev polynomials are only well-defined in the interval $[-1, 1]$, so the Hamiltonian needs to be scaled to fit into
that range. We note that since we are calculating expectations in the Hilbert spaces for $N_{\text{e}}+1$ and $N_{\text{e}}-1$
particles, we need to take into account the extremal eigenvalues of the Hamiltonian in those spaces. To make
sure we don't suffer from numerical artifacts, we scale the argument of the Chebyshev polynomials into the interval
$[-0.9, 0.9]$.\footnote{Using exactly $[-1,1]$ does not work, since the polynomials are fixed at the boundaries of the
interval. One needs to distance oneself at least by relative error in the eigenvalues from the boundary.} This gives us
\begin{eqnarray}
 \delta & = & 0.1,  \hspace{2em}\text{(distance to interval boundaries)} \nonumber \\
 a      & = & \frac{2 (1 - \delta)}{(\epsilon_{\text{max}} - E_0) - (\epsilon_{\text{min}} - E_0)}, \\
 b      & = & \frac{(\epsilon_{\text{max}} - E_0) + (\epsilon_{\text{min}} - E_0)}{2} - \delta,
\end{eqnarray}
where $\epsilon_{\text{min,max}}$ are the extremal many-body eigenvalues of the system with $N_{\text{e}}+1$ ($N_{\text{e}}-1$)
particles and $E_0$ is the ground state energy for $N_{\text{e}}$ particles.

We may then rewrite the single-particle retarded Green's function in terms of these coefficients,
\begin{eqnarray}
\mathcal{G}_{ij}(\omega) & = & a \sum_{n=0}^{\infty} \Big(
    \alpha_n^{+}\big( a(\omega + i\eta \mp b) \big) c^{(+)}_{ij,n}
  \nonumber \\ & & \hspace{3em}
  - \alpha_n^{-}\big( a(\omega + i\eta \mp b) \big) c^{(-)}_{ji,n} \Big).
\end{eqnarray}
The density of states is then given by the imaginary part of this expression traced over the real space indices, which is
why we only need to calculate the diagonal part of this expression. If we terminate the expansion at a finite $n$, the
formula remains only valid for finite $\eta$, with
\begin{equation}
 \eta \gtrsim \frac{1}{a n_{\text{max}}}. \label{eq:verification:ed-eta}
\end{equation}
For further discussion on this topic we would like to defer to Alexander Braun's thesis. \cite{AlexBraunPhD}

\bibliography{frg_p1}

\end{document}